\pdfoutput=1

\documentclass[onecolumn]{emulateapj}

\usepackage{amsmath}
\usepackage{graphicx}
\usepackage{color}
\usepackage[colorlinks]{hyperref}
\hypersetup{
    colorlinks,	
    citecolor=blue,
}

\newcommand{\be}{\begin{eqnarray}}
\newcommand{\ee}{\end{eqnarray}}
\newcommand{\rar}{\rightarrow}

\shorttitle{Reflection spectra from disk-like coronae}
\shortauthors{Riaz et al.}

\begin{document}

\title{Reflection spectra of accretion disks illuminated by disk-like coronae}

\author{Shafqat~Riaz\altaffilmark{1}, Askar~B.~Abdikamalov\altaffilmark{1,2,3}, Dimitry~Ayzenberg\altaffilmark{4}, Cosimo~Bambi\altaffilmark{1,\dag}, Haiyang~Wang\altaffilmark{1}, and Zhibo~Yu\altaffilmark{1}}

\altaffiltext{1}{Center for Field Theory and Particle Physics and Department of Physics, 
Fudan University, 200438 Shanghai, China. \email[\dag E-mail: ]{bambi@fudan.edu.cn}}
\altaffiltext{2}{Ulugh Beg Astronomical Institute, Tashkent 100052, Uzbekistan}
\altaffiltext{3}{Tashkent Institute of Irrigation and Agricultural Mechanization Engineers, Tashkent 100000, Uzbekistan}
\altaffiltext{4}{Theoretical Astrophysics, Eberhard-Karls Universit\"at T\"ubingen, D-72076 T\"ubingen, Germany}

\begin{abstract}
Relativistic reflection features in the X-ray spectra of black hole binaries and AGNs are thought to be produced through illumination of a cold accretion disk by a hot corona. In this work, we assume that the corona has the shape of an infinitesimally thin disk with its central axis the same as the rotational axis of the black hole. The corona can either be static or corotate with the accretion disk. We calculate the disk's emissivity profiles and iron line shapes for a set of coronal radii and heights. We incorporate these emissivity profiles into {\tt relxill\_nk} and we simulate some observations of a black hole binary with \textsl{NuSTAR} to study the impact of a disk-like coronal geometry on the measurement of the properties of the system and, in particular, on the possibility of testing the Kerr nature of the source. 
We find that, in general, the astrophysical properties of the accretion disk are recovered well even if we fit the data with a model employing a broken power-law or a lamppost emissivity profile, while it is more challenging to constrain the geometric properties of the black hole spacetime.
\end{abstract}


\section{Introduction}

Relativistic reflection features are commonly observed in the X-ray spectra of black hole binaries~\citep{Fabian:1989ej,Blum:2009ez,Fabian:2012kv,Miller:2013rca,Tomsick:2013nua,Xu:2017yrm} and AGNs~\citep{Tanaka:1995en,Nandra:1996vv,Nandra:2007rp,Walton:2012aw}. They are thought to be generated through illumination of a cold accretion disk by a hot corona~\citep{Fabian:1995qz,Zoghbi:2009wd,Risaliti:2013cga}. Thermal photons of the accretion disk inverse Compton scatter off free electrons in the corona. The resulting Comptonized photons have a power-law spectrum with an exponential high energy cut-off and can illuminate the accretion disk, generating a relativistic reflection spectrum. The most prominent features in the reflection spectrum are usually the iron K$\alpha$ complex in the soft X-ray band and the Compton hump peaked around 20-30~keV~\citep{George:1991jj,MZ95,Ross:2005dm,Garcia:2010iz}. In the presence of high-quality data and with the correct astrophysical model, the analysis of these reflection features in the X-ray spectra of accreting black holes can be a powerful tool to study the accretion process onto these objects, measure black hole spins~\citep{Brenneman:2006hw,Blum:2009ez,Fabian:2012kv,Miller:2013rca,Reynolds:2013qqa,Marinucci:2014ita,Reynolds:2019uxi}, and even test Einstein's theory of General Relativity in the strong field regime~\citep{Cao:2017kdq,Tripathi:2018lhx,Zhang:2019ldz,2020arXiv201013474T}.

The possibility of using X-ray reflection spectroscopy for precision measurements of accreting black holes depends, among other things, on the possibility of developing sufficiently sophisticated relativistic reflection models to limit systematic uncertainties related to simplifications in the theoretical model~\citep[see, e.g.,][]{2020arXiv201104792B}. Generally speaking, we can group these simplifications in the theoretical model into four classes: $(i)$ simplifications in the calculation of the reflection spectra at the emission point on the disk and in the rest-frame of the particles of the gas~\citep[e.g.,][]{jiang19a,jiang19b}, $(ii)$ simplifications in the description of the accretion flow~\citep[e.g.,][]{Reynolds:1997ek,Reynolds:2007rx,Svoboda:2012cy,Taylor:2017jep,2020arXiv200309663A,Cardenas-Avendano:2020xtw}, $(iii)$ simplifications in the description of the corona~\citep[e.g.,][]{Miniutti:2003yd,Dauser:2013xv,Wilkins:2015nfa,Wilkins:2014caa,2017MNRAS.471.4436W,Steiner17}, and $(iv)$ relativistic effects not taken into account~\citep[e.g.,][]{Niedzwiecki:2016ncz,Niedzwiecki:2018wtc,Riaz:2020zqb,Zhou:2019dfw}. Such simplifications might lead to modeling bias in the final measurements of the properties of a source. In part, systematic uncertainties can be limited by selecting the source and the observation. For example, theoretical models usually employ the Novikov-Thorne model for the description of the accretion disk and this would require to limit the analysis to sources with an Eddington-scaled disk luminosity in the range $\sim$5\% to $\sim$30\%, while we can easily get inaccurate black hole spin measurements if such a restriction is neglected~\citep{Riaz:2019kat,Riaz:2019bkv}.

Among the assumptions of the theoretical models, the geometry of the corona is thought to play quite an important role in the final measurement of the properties of a black hole. The geometry of the corona would determine the exact emissivity profile of the accretion disk. In the case of coronae of unknown geometry, it is common to model the emissivity profile with a power-law ($\varepsilon \propto 1/r^q$, where $q$ is the emissivity index) or a broken power-law ($\varepsilon \propto 1/r^{q_{\rm in}}$ for $r < r_{\rm br}$ and $\varepsilon \propto 1/r^{q_{\rm out}}$ for $r > r_{\rm br}$, where $q_{\rm in}$ and $q_{\rm out}$ are, respectively, the inner and the outer emissivity indices and $r_{\rm br}$ is the breaking radius). However, these two profiles are, at best, simple approximations of the actual emissivity profile and, especially in the presence of high-quality data, the measurements of the model parameters may be affected by undesirable systematic uncertainties.

If we consider a specific coronal geometry, we can calculate the emissivity profile in terms of some parameters that describe the corona. The most popular coronal geometry is currently the lamppost model~\citep{Dauser:2013xv}, where the corona is assumed to be a point-like source at a certain height $h$ along the black hole spin axis. Ring-like and disk-like coronae have been investigated in~\citet{Miniutti:2003yd}, \citet{Suebsuwong:2006mq}, and \citet{2012MNRAS.424.1284W}. \citet{Miniutti:2003yd} argue that the emissivity profile of a ring-like corona can be approximated well by a twice broken power-law, and in such a case we would find a very steep emissivity profile for the inner part of the accretion disk, an almost flat emissivity profile for the intermediate part, and an emissivity index slightly lower than 3 for the outer part. Extended coronae and moving coronae were discussed in \citet{Dauser:2013xv} and \citet{2012MNRAS.424.1284W}. 
The choice of the emissivity profile may be crucial for the estimate of some model parameters~\citep[see, e.g.,][]{Fabian:2012kv,Dauser:2013xv,2014MNRAS.439.2307F,Zhang:2019zsn}. Understanding the morphology of the corona is thus quite a relevant issue if we want to use X-ray reflection spectroscopy for precision measurements of accreting black holes.

In the present paper, we extend previous work in the literature and we discuss the case of a corona with the shape of an infinitesimally thin disk above the black hole and the accretion disk. Our coronae are described by two parameters: their height above the equatorial plane, $H$, and their radius, $R_{\rm disk}$. We consider the possibility that the corona is either static or corotating with the accretion disk. With ray-tracing calculations, we determine the emissivity profile of the accretion disk for different values of $H$, $R_{\rm disk}$, and for static/corotating coronae. We calculate the iron line shapes generated by similar emissivity profiles and we compare the results with those expected from a power-law emissivity profile and an emissivity profile of a lamppost corona. We incorporate the emissivity profiles of disk-like coronae in our reflection model {\tt relxill\_nk}~\citep{Bambi:2016sac,Abdikamalov:2019yrr}, which permits us to have angle-resolved calculations of the reflection spectrum\footnote{Due to relativistic light bending, the emission angle of the photons (i.e. the angle between the photon trajectory and the normal to the disk) changes over the surface of the accretion disk. Past reflection modeling employed angle-averaged calculations by assuming that the emission angle was equivalent to the inclination angle of the disk (i.e. the angle between the line of sight of the observer and the normal to the disk). Recent reflection models using the angle-dependent {\tt xillver} table take this difference into account \citep[see, for instance,][for an analysis on the differences of the measurement of the properties of a source between the two approaches.]{2020MNRAS.498.3565T}. {For further details on this issue, the readers are referred to \citet{2009AA5071S}, \citet{Svoboda:2012cy}, and \citet{Garcia:2013lxa} } }. Last, we simulate some observations of a black hole binary with \textsl{NuSTAR}~\citep{Harrison:2013md} to estimate the systematic uncertainties on the measurements of the properties of the system if the source has a disk-like corona and we fit the data assuming a broken power-law emissivity profile or a lamppost geometry.

The content of our paper is as follows. In Section~\ref{s-corona}, we present the disk-like coronal geometry of our work and we calculate the resulting emissivity profile of the accretion disk. In Section~\ref{s-refl}, we implement the new emissivity profiles to our ray-tracing code and we present iron line profiles generated by the disk-like coronae. In Section~\ref{s-sim}, we simulate some observations of a bright black hole binary with \textsl{NuSTAR} assuming a disk-like corona and we fit the data with the broken power-law and lamppost models to see whether we can recover the input parameters of the system. In particular, we will focus on the impact of the coronal geometry on tests of the Kerr metric. Summary and conclusions are in Section~\ref{s-conclusions}. Throughout the paper, we use units in which $G_{\rm N} = c = 1$ and a metric with signature $(-+++)$.



\begin{figure*}[t]
\begin{center}
\includegraphics[width=0.5\textwidth,trim={0cm 0cm 0cm 4.0cm},clip]{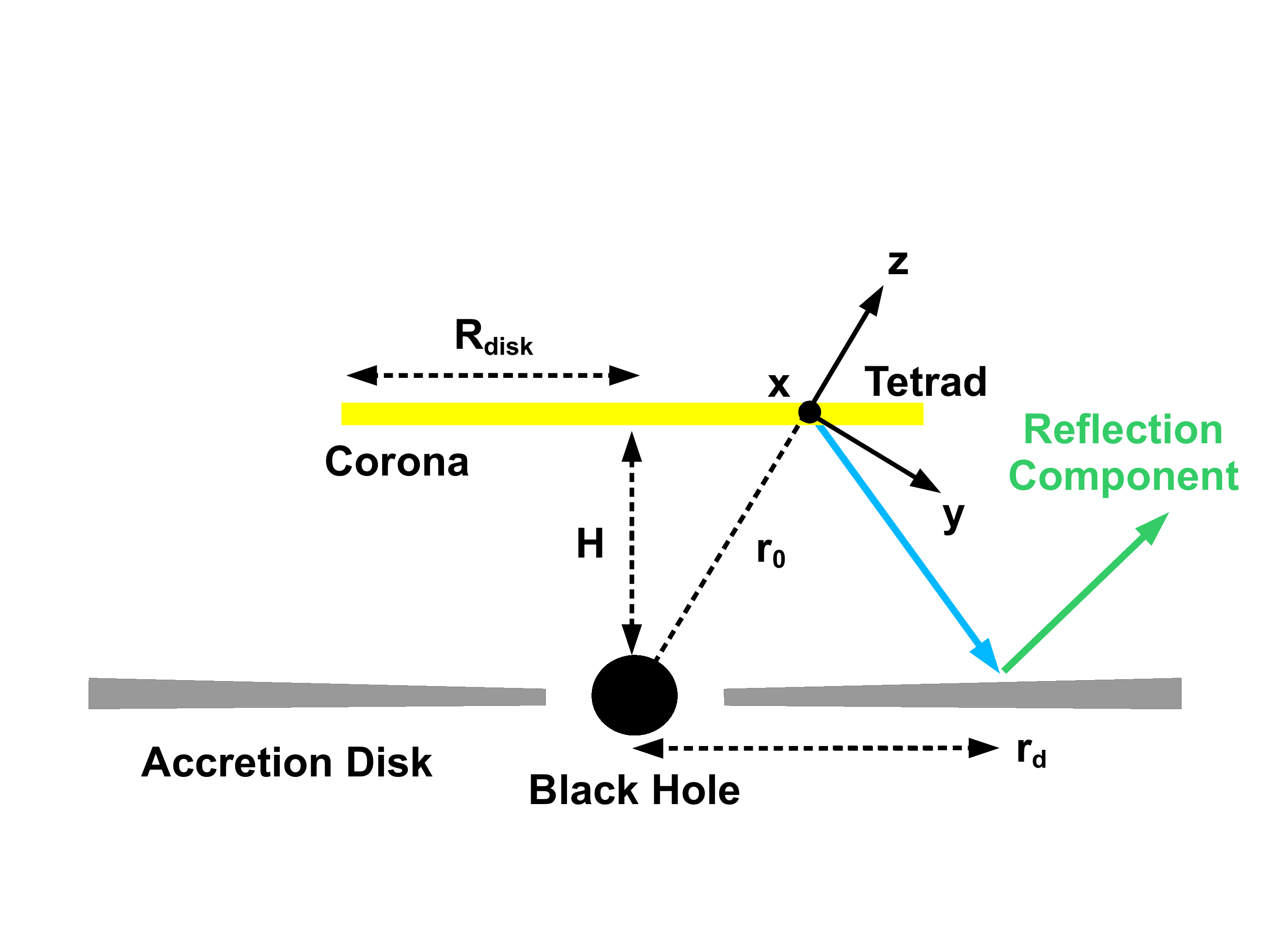}
\end{center}
\vspace{-1.0cm}
\caption{Cartoon of the astrophysical system. The corona is described by an infinitesimal thin disk of radius $R_{\rm disk}$ at an height $H$ above the equatorial plane. The tetrad of orthogonal basis vectors represents the locally Minkowskian reference frame of the emitter in the corona and is introduced to write the photon initial conditions. \label{corona-sketch}}
\vspace{1.2cm}
\end{figure*}

\section{Disk-like coronal geometry}\label{s-corona}

{Disk-like and ring-like coronae have previously been  studied in ~\citet{Miniutti:2003yd}, \citet{Suebsuwong:2006mq}, and \citet{2012MNRAS.424.1284W}, where the authors computed the theoretical emissivity profiles of the accretion disk due to the irradiation of these coronal geometries assuming the Kerr spacetime. In this section, we follow the same strategy as described in~\citet{2012MNRAS.424.1284W} to construct disk-like and ring-like coronae and compute their emissivity profiles in deformed Kerr spacetimes. The simplest and most popular coronal geometry is the so-called lamppost model: an isotropic, stationary, and point-like source residing along the rotational axis of the black hole \citep[see, e.g.,][]{Dauser:2013xv}. However, in reality the corona is more likely extended over a finite region rather than being a point-like source. Disk-like and ring-like coronae are simple extensions of the lamppost model to give the corona a finite size~\citep{Miniutti:2003yd,Suebsuwong:2006mq,2012MNRAS.424.1284W}.}

Fig.~\ref{corona-sketch} shows the astrophysical system that we want to consider in this work. The corona is an infinitesimally thin disk of radius $R_{\rm disk}$. The plane of the corona is parallel to the plane of the accretion disk and their distance is $H$\footnote{Technically, $H$ reduces to the distance of every emitting point in the corona from the accretion disk only in the Newtonian limit. Here it is a parameter linked to the coordinates of the emitting point through Eq.~(\ref{eq-x0}). Such a difference is not a problem in the model, because $H$ can be determined while we fit the data.}. We also assume that the system is perfectly axisymmetric, so the central axis of the corona coincides with that of the accretion disk as well as with the rotational axis of the black hole. The corona can either be static (vanishing angular velocity, $\Omega = 0$) or corotate with the accretion disk ($\Omega = \Omega_{\rm K}$, where $\Omega_{\rm K} = \Omega_{\rm K}(r)$ is the angular velocity of our Keplerian accretion disk).

\begin{figure*}[t]
\begin{center}
\includegraphics[width=0.45\textwidth,trim={0cm 0cm 0cm 0cm},clip]{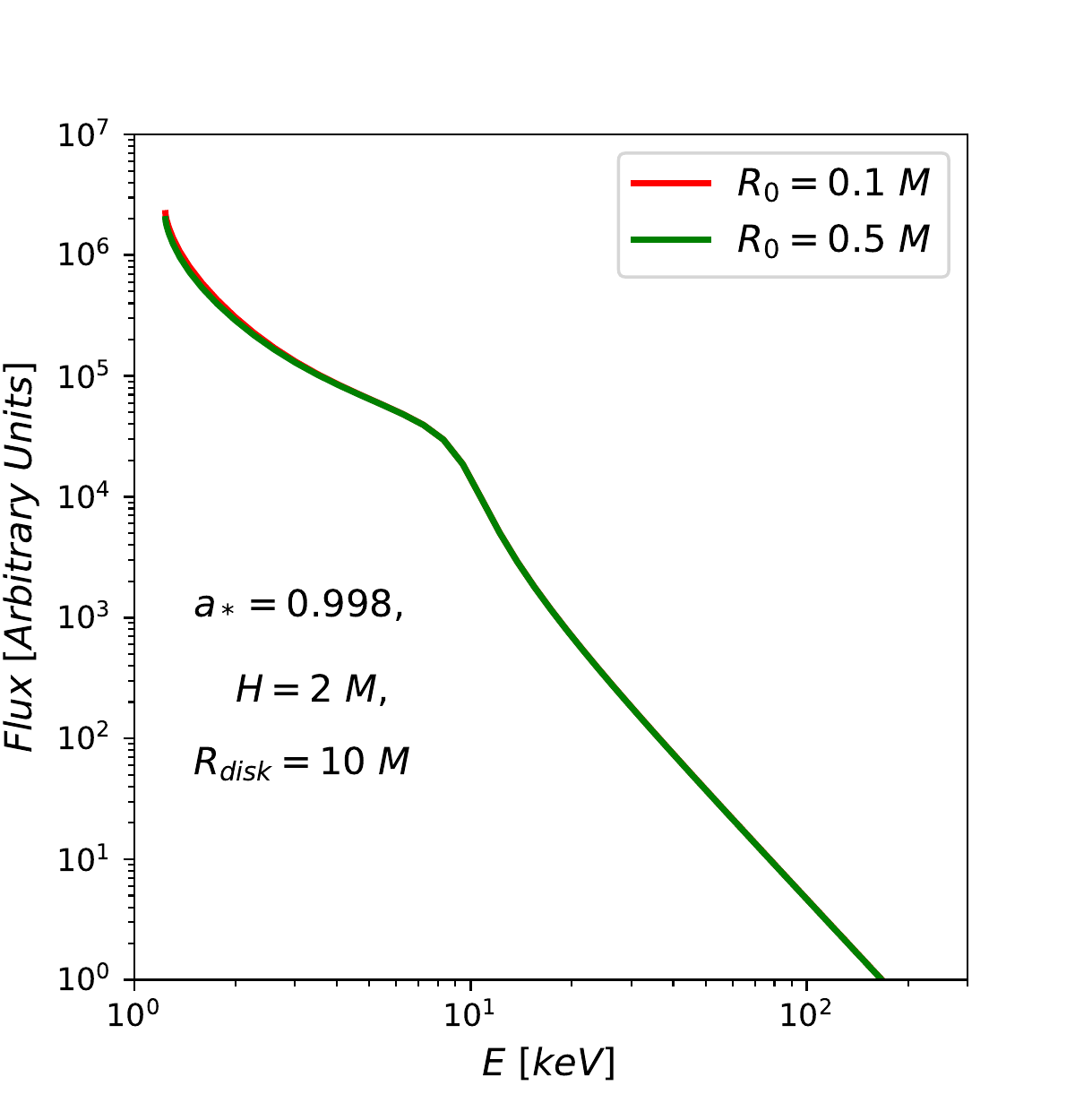}
\end{center}
\vspace{-0.4cm}
\caption{Static disk-like coronae.  Emissivity profiles in Kerr spacetime with $a_* = 0.998$.  The height and the radius of the corona are $H = 2~M$ and $R_{\rm disk} = 10$~$M$, respectively.  Red and green curves represent the emissivity profile of the corona with inner radius $R_{0} = 0.1~M$ and 0.5~$M$, respectively. See the text for more details.   \label{em01_vs_em05}} 
\vspace{0.8cm}
\end{figure*}

In this work, we assume that the background metric is described by the Johannsen metric~\citep{johannsen2013} with the deformation parameter $\alpha_{13}$ while all other deformation parameters vanish. The expression of the metric is reported in Appendix~\ref{app:metric}. The Kerr solution is recovered when $\alpha_{13} = 0$ while deviations from the Kerr spacetime are present in the presence of a non-vanishing $\alpha_{13}$. The reason to use a non-Kerr metric is motivated by the fact we want to study the impact of the coronal geometry on the possibility of testing the Kerr nature of the source and this will be done in Section~\ref{s-sim}.

Thanks to the axial symmetry of the system, we can limit our calculations along a certain radial direction. We consider a set of point-like sources, each of them at a certain radial coordinate $R$. In our calculations, we set the first source at the radial coordinate $R_0 = 0.5$~$M$ and the last source at the radial coordinate $R_{\rm disk}$, which is the coronal radius.  
The central gap in the coronal disk is due to two reasons: 1) the rays starting from the point-like source at $R_{0} < 0.5~M$ take much more computational time to reach the accretion disk, and 2) most of the photons starting from $R_{0}<0.5~M$ fall into the black hole and produce a negligible impact on the emissivity profile. A comparison between the emissivity profiles for the case $R_{0} = 0.1~M$ (red curve) and $R_{0} = 0.5~M$ (green curve) for a Kerr black hole with spin parameter $a_* = 0.998$ is shown in Fig.~\ref{em01_vs_em05}. The height and radius of the coronal disk are set to be $H = 2~M$ and $R_{\rm disk} = 10~M$, respectively. The maximum difference between the two emissivity profiles is less than $0.1 \%$ within the inner few gravitational radii.

The spacing of the point-like sources can be used to regulate the intensity profile of the disk-like corona.  For example, if we assume that the surface of the corona has constant luminosity\footnote{These considerations are in the Newtonian limit. The generalization to a consistent general relativity framework would be quite straightforward, but a more natural extension of our set-up would be to model the intensity profile of the corona with a power-law and then determine the power-law index from the fit.}, we need a point-like source for every equal-area annulus. The area of the annulus of radius $R$ is $A = 2 \pi R \Delta R$ and therefore we would need $\Delta R \propto 1/R$. In the rest of the paper, we employ $\Delta R = {\rm const.}$ independent of $R$, which is equivalent to assume that the intensity profile of the corona scales as $1/R$. We also assume that every point-like source emits isotropically, i.e.~equal power is emitted into equal solid angles in the rest-frame of the source.  The trajectory of each photon is traced numerically by taking all the relativistic effects into account until it hits the surface of the disk or falls into the black hole.  The photons that land on the surface of the disk are radially binned over the accretion disk.  The same procedure is applied for each point-like source in the coronal disk until $R_{\rm disk}$ is reached.  The separation between two consecutive point-like sources is fixed at 0.1~$M$.  If we decrease the separation further, the computational time to calculate the emissivity profile increases, and the model's accuracy does not improve much.

In the locally Minkowskian reference frame of the source, the initial 4-momentum of the photon with energy $E$ is
\be\label{4-momnetum}
k^{(\alpha)}_0 = \left( E, E \sin\chi \cos\psi, E \sin\chi \sin\psi, E \cos\chi \right) \, ,
\ee 
where $\chi$ and $\psi$ are the polar angles of the solid angle element, i.e.~$d\tilde{\Omega} = d(\cos\chi)d\psi$ with $\cos\chi \in [-1,1]$ and $\psi \in [0,2\pi]$. From every point-like source, we shoot photons at equal intervals of the solid angle with $\Delta(\cos\chi) = 0.00008$ and $\Delta\psi = 0.5$, which leads to a grid 25,000$\times$12. The high number of grid points for $\cos\chi$ is required to properly calculate the emissivity profile, especially near the inner edge of the disk, while there is not such a problem for $\psi$. We tested various grid sizes and eventually we chose a grid such that the run-time was as short as possible without a noticeable reduction in the accuracy.

In Boyer-Lindquist-like coordinates, the initial conditions for the photon position are
\be\label{eq-x0}
t_0 &=& 0 \, , \nonumber\\
r_0 &=& \sqrt{H^2 + R^2} \, , \nonumber\\
\theta_0 &=& \arctan \left( \frac{R}{H} \right) \, , \nonumber\\
\phi_0 &=& 0 \, ,
\ee
and the initial conditions for the photon 4-momentum are
\be\label{eq-k0}
k^t_0 &=& k^{(t)}_0 E^{t}_{(t)} + k^{(x)}_0 E_{(x)}^{t} \, , \nonumber\\
k^r_0 &=& k^{(z)}_0 E^r_{(z)} \, , \nonumber\\
k^\theta_0 &=& k^{(y)}_0 E^{\theta}_{(y)} \, , \nonumber\\
k^\phi_0 &=& k^{(t)}_0 E^{\phi}_{(t)} + k^{(x)}_0 E^{\phi}_{(x)} \, ,
\ee
where $k^{(\alpha)}_0$ is in Eq.~(\ref{4-momnetum}) and $\{ E^\mu_{(\alpha)} \}$ is the tetrad of orthogonal basis vectors associated to the locally Minkowskian reference frame of the source. The derivation and expressions of $\{ E^\mu_{(\alpha)} \}$ are reported in Appendix~\ref{app:photon}.

The photon trajectories are calculated by solving the geodesic equations~\citep{2012ApJ...745....1P} using a modified version of the ray-tracing code described in~\citet{Abdikamalov:2019yrr},~\citet{Ayzenberg:2018jip}, and~\citet{Gott:2018ocn}. The photon trajectory starts with the initial conditions in Eq.~(\ref{eq-x0}) and Eq.~(\ref{eq-k0}). The calculations stop when the photon reaches the equatorial plane $\theta = \pi/2$. If the photon hits the accretion disk, namely its radial coordinate on the equatorial plane, say $r_{\rm d}$, is between the inner edge of the accretion disk (set at the innermost stable circular orbit of the spacetime, ISCO, $r_{\rm in} = r_{\rm ISCO}$) and the outer edge ($r_{\rm out} = 1000$~$M$ in our code), it is collected into a radial bin, $N(r, \Delta r)$. By repeating the same scheme for each photon, we get the count of rays in each radial bin. 
We note that similar calculations were presented in \citet{2012MNRAS.424.1284W} and \citet{Dauser:2013xv}, where the reader can also find all the relevant formulas.

The photon redshift factor between the corona and the accretion disk is 
\begin{equation}
g = \frac{k_\mu u^\mu}{k_\nu U^\nu} \, .
\label{red-shift-eq}
\end{equation}             
$k^\mu$ is the photon 4-momentum in the Boyer-Lindquist coordinate system. $k^\mu$ is evaluated at the incident point on the accretion disk at the numerator (which we know in numerical form at the end of every ray-tracing calculation) and at the emission point in the corona at the denominator (i.e. $k^\mu = k^\mu_0$). $u^\mu = u^t \left( 1 , 0 , 0, \Omega_{\rm K} \right)$ is the 4-velocity of the particles in the accretion disk, where
\be
u^t = \frac{1}{\sqrt{ - g_{tt} - 2 \Omega_{\rm K} g_{t\phi} - \Omega_{\rm K}^2 g_{\phi\phi}}} \Big|_{r=r_{\rm d} , \theta=\pi/2} \, , 
\ee
and $\Omega_{\rm K}$ is the Keplerian angular velocity of the particles in the accretion disk
\be\label{eq-omegak}
\Omega_{\rm K} \left( r=r_{\rm d} , \theta=\pi/2 \right) 
= \frac{- \left(\partial_r g_{t\phi}\right) \pm \sqrt{\left( \partial_r g_{t\phi} \right)^2 
- \left(\partial_r g_{tt}\right)\left(\partial_r g_{\phi \phi}\right) }  }{\left(\partial_r g_{\phi \phi}\right)} \, .
\ee
$U^\mu$ is the 4-velocity of the emitting point in the corona. In the case of a static corona, we have $U^\mu_{\rm stat} = \left( 1/\sqrt{-g_{tt}} , 0 , 0, 0 \right)$. In the case of a corona corotating with the accretion disk, we have $U^\mu_{\rm corot} = U^t \left( 1 , 0 , 0, \Omega_{\rm K} \right)$, where
\be
U^t = \frac{1}{\sqrt{ - g_{tt} - 2 \Omega_{\rm K} g_{t\phi} - \Omega_{\rm K}^2 g_{\phi\phi}}} \Big|_{r=r_0, \theta=\theta_0} \, ,
\ee
$r_0$ and $\theta_0$ are the photon initial conditions in Eq.~(\ref{eq-x0}), and $\Omega_{\rm K}$ is still the Keplerian angular velocity in Eq.~(\ref{eq-omegak}) calculated on the equatorial plane and at the radial coordinate $r_{\rm d} = r_0 \sin\theta_0$.

We consider 100~radii on the disk ($i = 0$, 1, ..., 99), where $r_0 = r_{\rm in}$, $r_{99} = r_{\rm out}$, and $r_i$ with $1 \le i \le 98$ are calculated by the algorithm already used to tabulate the transfer function for {\tt relxill\_nk}~\citep[see][]{Abdikamalov:2019yrr}. The area of the annulus $i$ is
\begin{equation}
A(r_i, \Delta r_i) = 2\pi \sqrt{g_{rr}g_{\phi \phi}}\Delta r_i \, , 
\label{proper-area} 
\end{equation}
where $g_{rr}$ and $g_{\phi \phi}$ are evaluated at the radial coordinate $r_i$ and $\Delta r_i$ is
\begin{equation}
\Delta r_i =  \left[ \left( \frac{r_{i+1}}{99} \right)^4 - \left( \frac{r_{i}}{99} \right)^4 \right] \left( r_{\rm out} - r_{\rm in} \right)
\end{equation}
The area of the radial bin in the reference frame of the particles in the accretion disk is obtained by multiplying Eq.~(\ref{proper-area}) with the Lorentz factor of the particles in the accretion disk, $\gamma$. We have~\citep{1972ApJ...178..347B}
\begin{equation}
\gamma = \left[ 1 + \frac{(\Omega_{\rm K} g_{\phi\phi} - g_{t \phi})^2}{g_{tt}g_{\phi \phi} - g_{t \phi}^2} \right]^{-1/2} \, .
\label{Lorentz-factor}
\end{equation}

After firing all photons from the point-like source at the coronal radius $R$, we have the ray number per radial bin in the disk from that source, $\mathcal{N}_R(r_i, \Delta r_i)$. With the area of the annuli of the accretion disk for the distant observer, $A(r_i, \Delta r_i)$, and the Lorentz factor of the particles in the accretion disk, $\gamma$, we can write the ray number density per radial bin generated by the annulus $R$ of the corona 
\begin{equation}
n_R(r_i) = \frac{\mathcal{N}_R(r_i, \Delta r_i)}{A(r_i, \Delta r_i) \gamma} \, .
\end{equation} 
The photon number flux at the emission point in the corona can be approximated by a power-law
\be
\frac{dN_{\rm c}}{dt_{\rm c} dE_{\rm c}} = K E^{-\Gamma}_{\rm c} \, ,
\ee
where $K$ is a constant, $\Gamma$ is the photon index, and the subindex c refers to the fact these quantities are evaluated at the emission point in the corona. The photon number is conserved, so $N = K E^{-\Gamma} \, \Delta t \, \Delta E$ is a constant along the photon path and can be associated to the number of photons for every ray. The energy density illuminating the disk per radial bin generated by the annulus $R$ of the corona is thus
\be
\mathcal{E}_R (r_i) = E_{\rm d} \, N \, n_R(r_i) 
= E_{\rm d} \left( K E^{-\Gamma}_{\rm c} \, \Delta t_{\rm c} \, \Delta E_{\rm c} \right) n_R(r_i)
= g^\Gamma \, K E^{-\Gamma+1}_{\rm d} \, \Delta t_{\rm d} \, \Delta E_{\rm d} \, n_R(r_i) \, ,
\ee 
where $g = \Delta t_{\rm c}/\Delta t_{\rm d} = E_{\rm d}/E_{\rm c}$ is the redshift factor between the emission point in the corona and the incident point in the disk calculated in Eq.~(\ref{red-shift-eq}) and the subindex d is used for the quantities on the disk. The emissivity profile can be written as
\be
\varepsilon(r_i) \propto \sum_R \frac{\mathcal{E}_R (r_i)}{\Delta t_{\rm d} \, \Delta E_{\rm d}}
= \sum_R g^\Gamma \, K E^{-\Gamma+1}_{\rm d} \, n_R(r_i) \, ,
\ee
where we have to sum over all annuli of the disk-like corona, ranging from $R_0$ to $R_{\rm disk}$.

The coronal spectrum is normally described by a power-law with an exponential high energy cut-off and we have thus two parameters, namely the photon index $\Gamma$ and high energy cut-off $E_{\rm cut}$. The value of the photon index does not change from the emission to the detection point, but the high energy cut-off scales with the redshift factor. In {\tt relxilllp}, the lamppost model in the {\tt relxill} package~\citep{Dauser:2013xv, garcia2013x}, the model parameter $E_{\rm cut}$ refers to the {\it high energy cut-off at the detection point}, but since the source is point-like and we know its location, it is straightforward to infer $E_{\rm cut}$ at the emission point of the corona. Moreover, we can calculate the high energy cut-off at every incident point on the accretion disk, and thus calculate the reflection spectrum produced by the correct $E_{\rm cut}$ of the radiation illuminating the accretion disk. In models with broken power-law emissivity profiles for arbitrary coronal geometries, like the normal {\tt relxill}, we do not know the the location of the corona and therefore the redshift factor between the emission point and the detection point and at the incident points on the accretion disk. In such a case, the parameter $E_{\rm cut}$ of the model still refers to the high energy cut-off at the detection point but the same value is also used for the spectrum illuminating the disk. Such a simplification might have some (weak) impact on the predicted reflection spectrum at the emission points of the accretion disk even at low energies \citep[see, e.g.,][]{2015ApJ...808L..37G}.

In this work, in the presence of an extended corona, we employ the same simplification as in the models with a broken power-law emissivity profile: our parameter $E_{\rm cut}$ referring to the high energy cut-off at the detection point is also used for the radiation illuminating the disk. If we did not do so and we assumed that all emission points in the corona have the same value of the high energy cut-off $E_{\rm cut}$, we would find that every point on the accretion disk and the distant observer receive a spectrum resulting from the combination of power-law spectra with different $E_{\rm cut}$, as every point of the disk-like corona has a different redshift factor. In other words, our coronal geometry is only used to calculate the corresponding emissivity profile on the disk, and we neglect other minor relativistic effects entering the calculation of the reflection spectrum of an accretion disk.

\begin{figure*}[t]
\begin{center}
\includegraphics[width=0.95\textwidth,trim={0cm 0.5cm 0cm 0cm},clip]{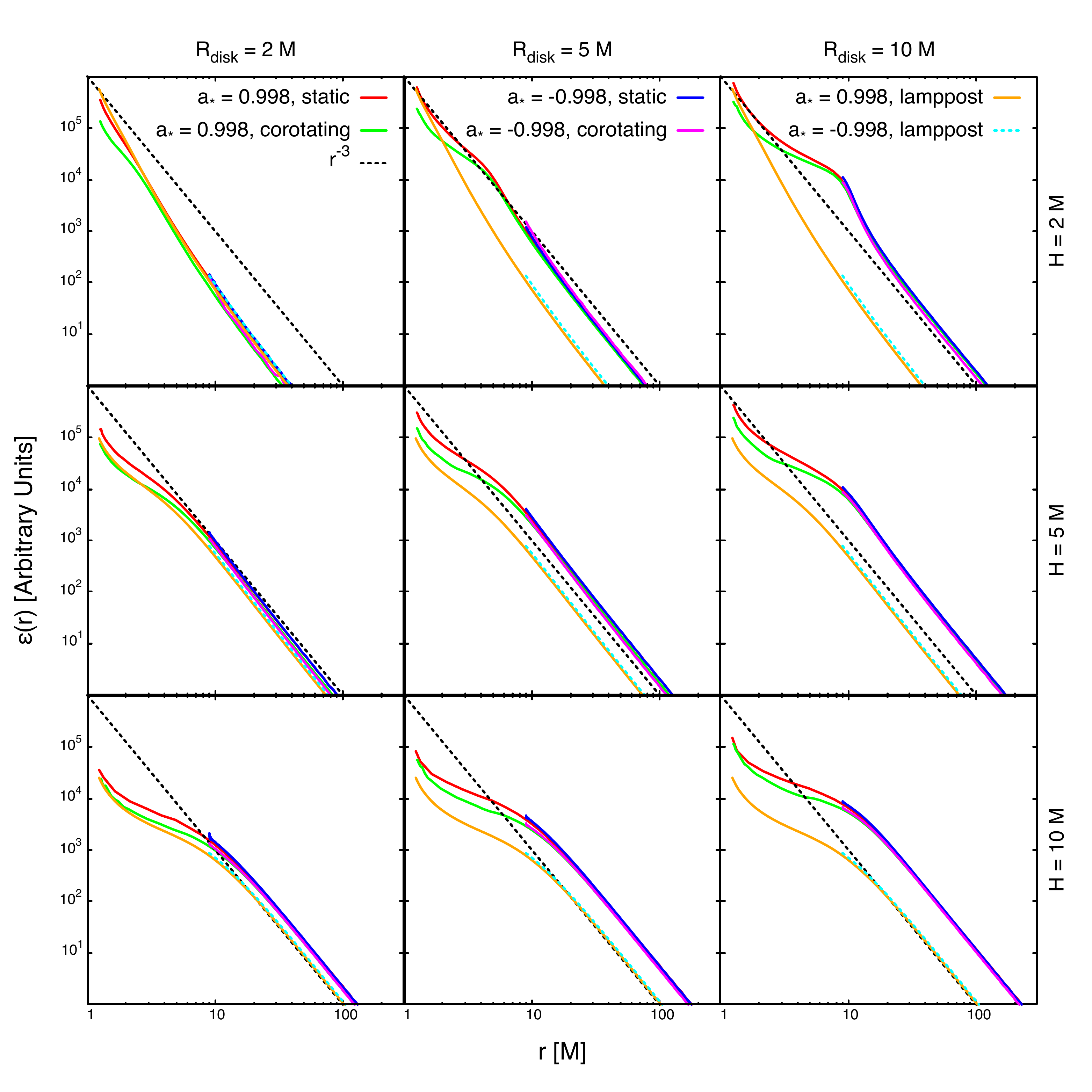}
\end{center}
\vspace{-0.5cm}
\caption{Emissivity profiles in the Kerr spacetime for different values of the corona radius $R_{\rm disk}$ and of the corona height $H$. Red curves are for static disk-like coronae and a black hole spin parameter $a_* = 0.998$. Green curves are for corotating disk-like coronae and a black hole spin parameter $a_* = 0.998$. Blue curves are for static disk-like coronae and a black hole spin parameter $a_* = -0.998$. Magenta curves are for corotating disk-like coronae and a black hole spin parameter $a_* = -0.998$. Black dashed lines are for the canonical emissivity profile $\varepsilon \propto r^{-3}$. Orange curves and cyan dashed curves are for the lamppost model with coronal height $h = H$ and a black hole spin parameter, respectively, $a_* = 0.998$ and $-0.998$. 
\label{f-emissivity}}
\end{figure*}

Fig.~\ref{f-emissivity} shows some emissivity profiles for different values of the coronal radius $R_{\rm disk}$ ($R_{\rm disk} = 2$~$M$, 5~$M$, and 10~$M$ for, respectively, left, central, and right panels) and coronal height $H$ ($H = 2$~$M$, 5~$M$, and 10~$M$ for, respectively, top, central, and bottom panels). In every panel, we show both the static and the corotating coronae, and we can see that the difference between the two models is marginal and only evident at small radii ($r < 10$~$M$). In every panel, we assume the Kerr metric ($\alpha_{13} = 0$) and we show two values of the black hole spin parameter: $a_* = 0.998$ and $a_* = -0.998$. The black hole spin determines the inner edge of the accretion disk, as here we assume it is at the ISCO, but for radii larger than the ISCO radius of the case $a_* = -0.998$ the two curves almost overlap: the impact of the black hole spin is very weak on the photon trajectories and the value of $a_*$ mainly manifests on the inner edge of the disk. A similar conclusion would hold for a non-vanishing $\alpha_{13}$: this deformation parameter has a very weak impact on the emissivity profile and mainly determines the ISCO radius. For comparison, every panel also reports the canonical emissivity profile $\varepsilon \propto r^{-3}$ and the emissivity profile of the lamppost model with coronal height $h = H$ (still for the Kerr metric with $a_* = 0.998$ and $a_* = -0.998$).

When the coronal radius is small ($R_{\rm disk} = 2$~$M$), there is not much difference between the emissivity profiles of the disk-like corona and of the lamppost corona. This is understandable because in the limit $R_{\rm disk} \rar 0$ the disk-like corona reduces to the lamppost one. The impact of the extended corona is maximum when the coronal height is low and the coronal radius is large, see the top right panel in Fig.~\ref{f-emissivity}. At large radii, all emissivity profiles match well (modulo their normalization) with the canonical emissivity profile $\varepsilon \propto r^{-3}$.

\begin{figure*}[t]
\begin{center}
\includegraphics[width=0.90\textwidth,trim={0cm 0.5cm 0cm 0cm},clip]{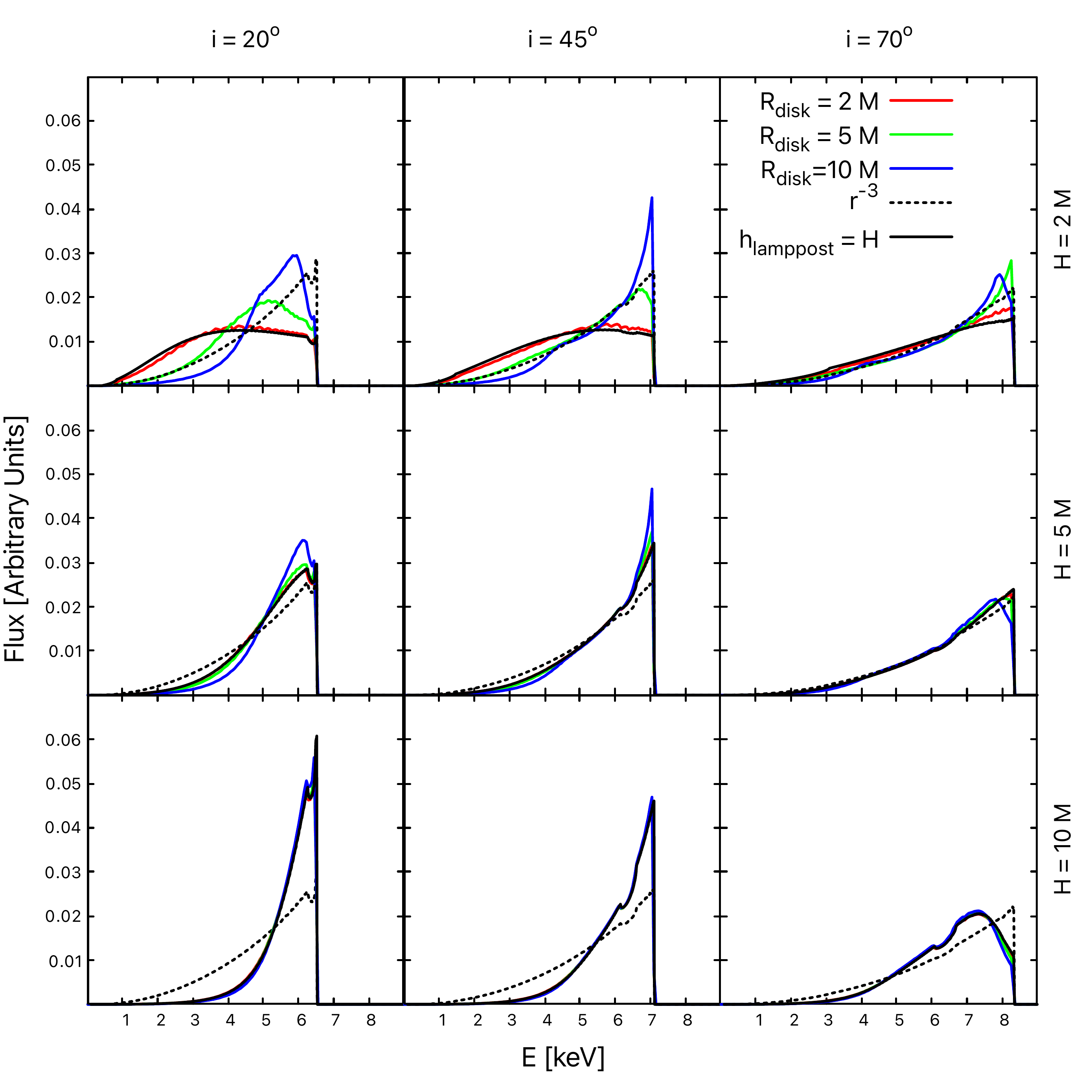}
\end{center}
\vspace{-0.4cm}
\caption{Static disk-like coronae. Iron line profiles in Kerr spacetime with $a_* = 0.998$. The radius of the corona is $R_{\rm disk} = 2$~$M$ (red profiles), 5~$M$ (green profiles), and 10~$M$ (blue profiles). In every panel we also show an iron line for a power-law emissivity profile with emissivity index $q = 3$ (black dotted profiles) and for a lamppost corona (black solid profiles). \label{f-linesstat}}
\end{figure*}

\begin{figure*}[t]
\begin{center}
\includegraphics[width=0.90\textwidth,trim={0cm 0.5cm 0cm 0cm},clip]{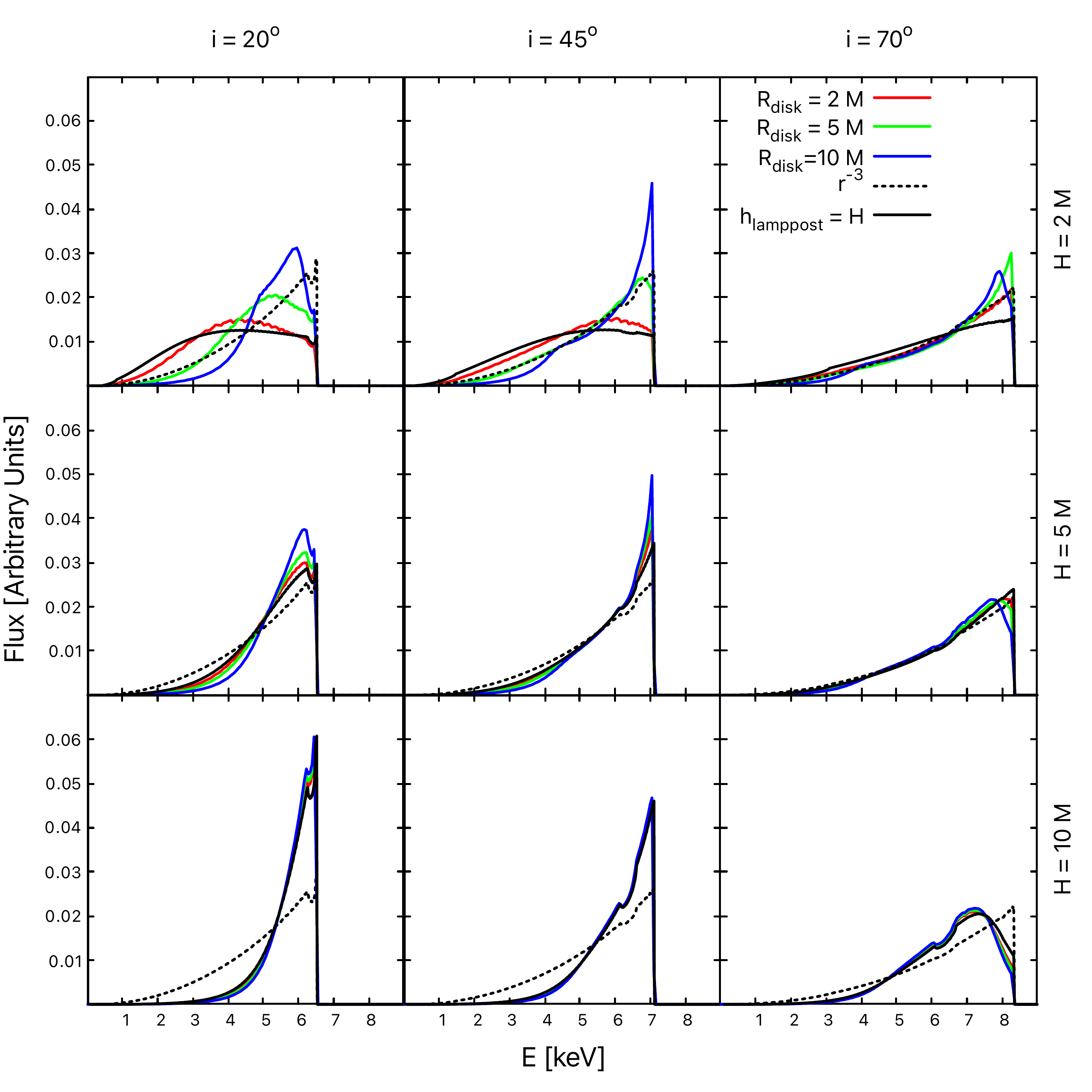}
\end{center}
\vspace{-0.4cm}
\caption{Corotating disk-like coronae. Iron line profiles in Kerr spacetime with $a_* = 0.998$. The radius of the corona is $R_{\rm disk} = 2$~$M$ (red profiles), 5~$M$ (green profiles), and 10~$M$ (blue profiles). In every panel we also show an iron line for a power-law emissivity profile with emissivity index $q = 3$ (black dotted profiles) and for a lamppost corona (black solid profiles). \label{f-linescorot}}
\end{figure*}

\begin{figure*}[t]
\begin{center}
\includegraphics[width=0.85\textwidth,trim={0cm 0.5cm 0cm 7.0cm},clip]{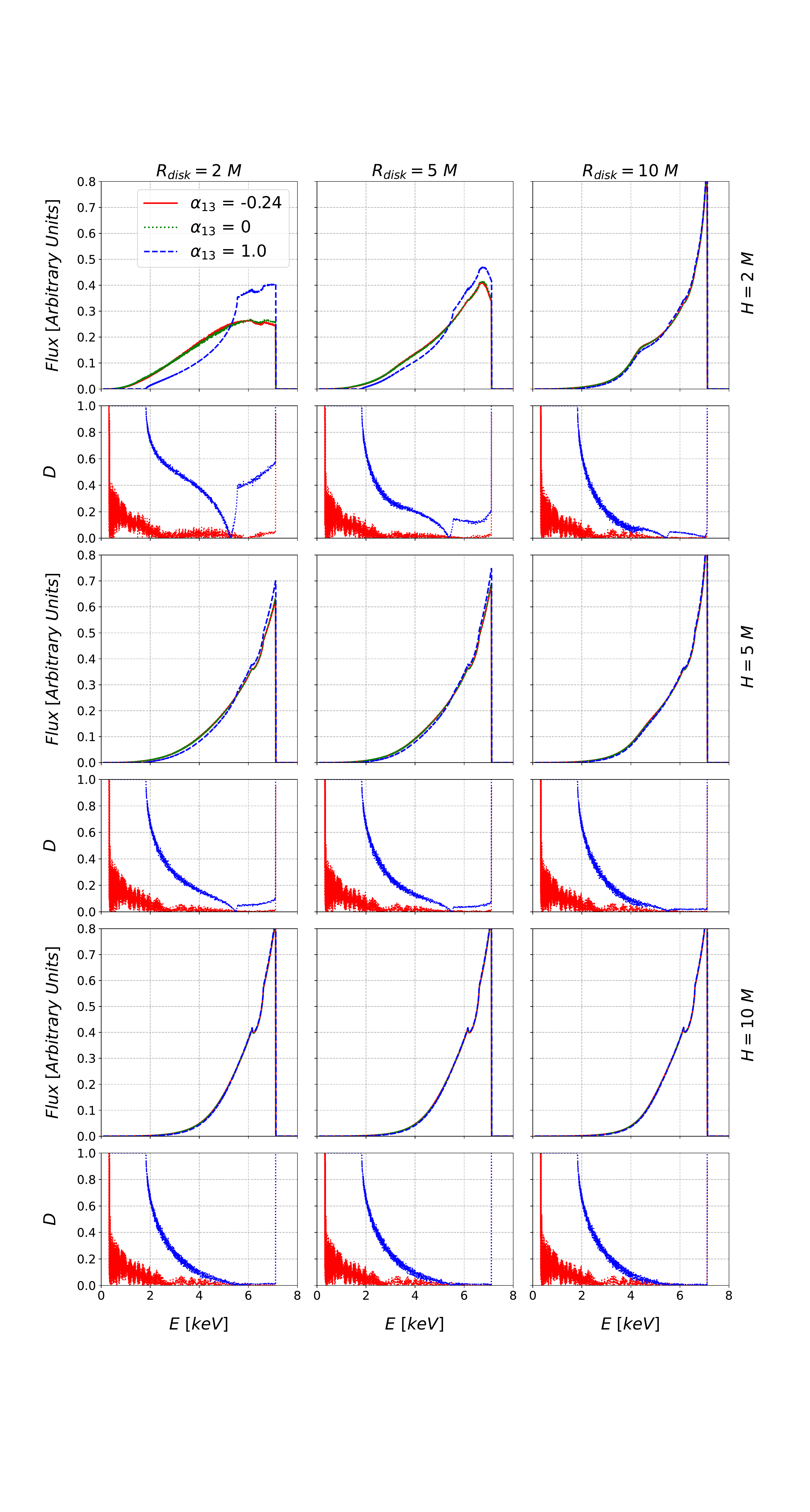}
\end{center}
\vspace{-3.4cm}
\caption{{Static disk-like coronae.  Iron line profiles for $a_* = 0.998$ and deformation parameter $\alpha_{13} = -0.24$ (red solid curves),  $\alpha_{13} = 0$ (green dotted curves), and $\alpha_{13} = 1.0$ (blue dashed curves). The small quadrants show the relative difference between the non-Kerr and the Kerr iron lines: $D = |F_{\rm NK} - F_{\rm K}|/F_{\rm K}$, where $F_{\rm K}$ and $F_{\rm NK}$ are the fluxes in the Kerr and non-Kerr models, respectively. The observer's viewing angle is set to be $45^{\circ}$. } \label{f-lines_i45_nk}}
\vspace{0.5cm}
\end{figure*}


\section{Iron line profiles of accretion disks illuminated by disk-like coronae}\label{s-refl}

In this section, we assume that the spectrum at every emission point on the accretion disk is a narrow line at 6.4~keV and we calculate the spectrum at the detection point far from the source employing the emissivity profiles calculated in the previous section. The advantage of the narrow line with respect to the full reflection spectrum is that it can better illustrate the impact of the emissivity profile on the reflection features of the disk.

The calculation of line profiles from geometrically thin and optically thick accretion disks have been extensively discussed in the literature~\citep[see, e.g.,][]{bambi2012code,Bambi:2017khi}. The accuracy of the output of our code was compared with the {\tt xspec} model {\tt relline} in Fig.~1 in \citet{Riaz:2019bkv}. The photons trajectories are calculated backward in time, from the image plane of the distant observer to the accretion disk. The integration stops when the photon hits the equatorial plane, which is the surface of our infinitesimally thin accretion disk. The redshift factor is computed on the disk surface as 
\begin{equation}
g  =  \frac{\sqrt{-g_{tt} - 2g_{t\phi}\Omega_{\rm K} -g_{\phi \phi}\Omega^{2}_{\rm K} }}{1 + \lambda \Omega_{\rm K}} \, , 
\end{equation}          
where $\Omega_{\rm K}$ is the (Keplerian) velocity of the fluid element in the accretion disk, $\lambda = k_{t} / k_{\phi}$, and $k^{t}$ and $k^{\phi}$ are the $t$ and the $\phi$ components of the photon 4-momentum. Since $k_{t}$ and $k_{\phi}$ are constants of motion, $\lambda$ can be computed from the photon initial conditions. The iron line profile detected by the distant observer is computed by integrating over the disk image
\begin{equation}
N(E_{\rm o}) = \frac{1}{E_{\rm o}} \int g^3 I_{\rm e} ( E_{\rm e}) \frac{dX dY}{D^2} \, , 
\label{iron-line} 
\end{equation}
where $N(E_{\rm o})$ is the photon number flux with energy $E_{\rm o}$ measured by the distant observer, $E_{\rm e}$ is the photon energy at the emission point, $g = E_{\rm o}/E_{\rm e}$ is the redshift factor, $I_{\rm e}$ is the specific intensity of the radiation at the emission point, $D$ is the distance between the observer and the source, and $X$ and $Y$ are the Cartesian coordinates on the image plane of the observer. $I_{\rm e}$ is a narrow line with the normalization determined by the emissivity profile induced by the disk-like coronae calculated in the previous section.

Fig.~\ref{f-linesstat} and Fig.~\ref{f-linescorot} show iron line profiles of accretion disks illuminated by disk-like coronae.  The spacetime metric is described by the Kerr solution with $a_* = 0.998$ and the inclination angle of the disk with respect to the line of sight of the observer is $i = 20^\circ$, 45$^\circ$, and 70$^\circ$ (left, central, and right panels, respectively). For every inclination angle, we show the case of a coronal height $H = 2$~$M$, 5~$M$, and 10~$M$ (top, central, and bottom panels, respectively). In every panel, we show the iron line profiles of accretion disks illuminated by disk-like coronae with radius $R_{\rm disk} = 2$~$M$, 5~$M$, and 10~$M$ (red, green, and blue curves, respectively), by a lamppost corona (black solid curves), and by a disk with canonical emissivity profile $\varepsilon \propto r^{-3}$ (black dashed curve). Fig.~\ref{f-linesstat} is for the case of static coronae ($\Omega = 0$) and Fig.~\ref{f-linescorot} is for corotating coronae ($\Omega = \Omega_{\rm K}$). From the comparison of Fig.~\ref{f-linesstat} and Fig.~\ref{f-linescorot} we see that the rotation of the corona has quite a weak impact on the iron line profile while the two key quantities are the coronal radius and the coronal height. As the coronal radius $R_{\rm disk}$ decreases, the iron line profile approaches that of the lamppost model. 

Fig.~\ref{f-lines_i45_nk} shows the iron line shapes of an accretion disk illuminated by a static disk-like corona for different values of the Johannsen deformation parameter $\alpha_{13}$. The black hole spin and the observer's viewing angle are set to be $a_* = 0.998$ and $i = 45^{\circ}$, respectively. The height of the disk-like corona is taken to be $H = 2~M$,  $5~M$, and $10~M$ (top, central, and bottom panels, respectively). The radius of the disk-like corona is set to be $R_{\rm disk} = 2~M$, $5~M$, and $10~M$ (left, middle, and right panels, respectively). 
{In every large quadrant, we show the iron line profiles for deformation parameter $\alpha_{13} = -0.24$, 0, and 1.0 (red, green, and blue curves, respectively). A smaller quadrant at the bottom of every large quadrant shows the relative difference between the non-Kerr and the Kerr iron lines. The impact of the value of the deformation parameter on the iron line is stronger for a lower height and a smaller coronal radius $R_{\rm disk}$.  As we increase the coronal height and/or $R_{\rm disk}$, the iron line profiles become less sensitive to the exact value of the deformation parameter $\alpha_{13}$. This is due to the fact that a compact corona closer to the black hole illuminates better the inner part of the accretion disk, where the gravitational field is stronger and the spectrum is more affected by relativistic effects. As the corona moves away from the black hole and/or increases in size, it illuminates better the disk at larger radii and the relative weight of the spectrum from the region very close to the black hole on the total spectrum decreases. For any given height and size of the corona, the most significant difference between the Kerr and the non-Kerr iron lines comes from the low-energy part of the profiles. Again, this is because the photons in the low-energy tail of the iron lines come from the inner part of the accretion disk, where the gravitational field is stronger.}



\begin{figure*}[t]
\begin{center}
\includegraphics[width=0.45\textwidth,trim={0cm 0cm 0cm 0cm},clip]{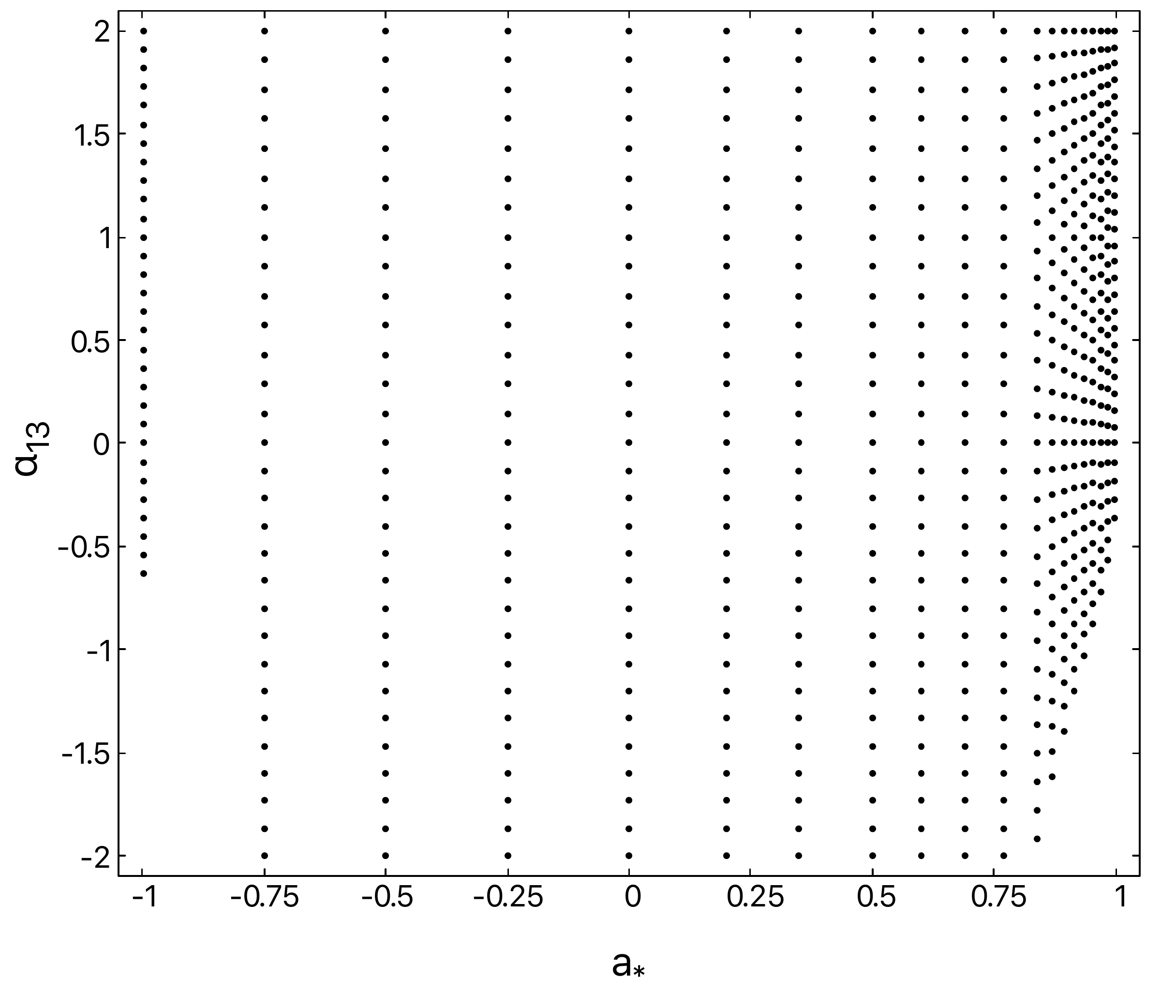}
\hspace{0.5cm}
\includegraphics[width=0.45\textwidth,trim={0cm 0cm 0cm 0cm},clip]{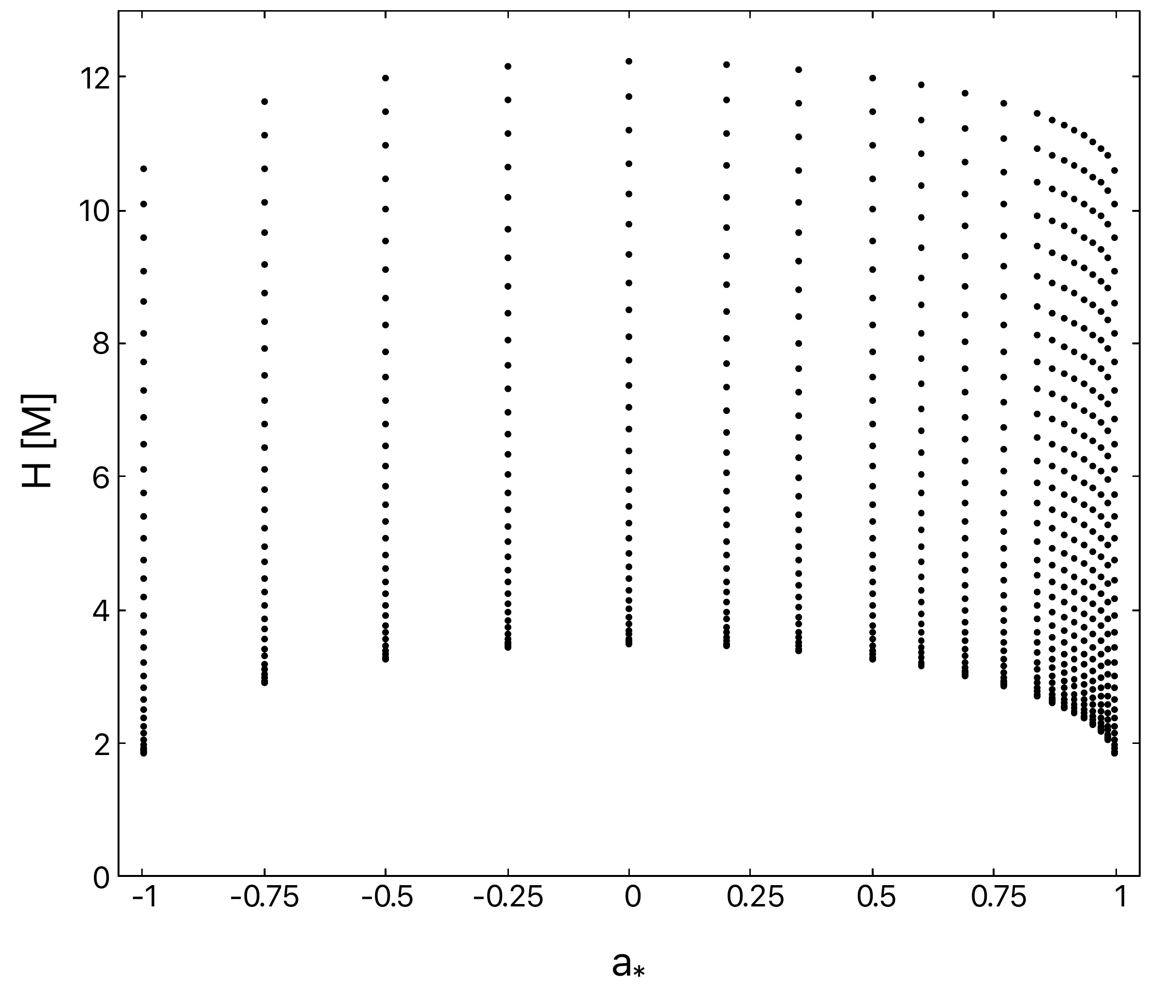}
\end{center}
\vspace{-0.3cm}
\caption{Grids $a_*$ vs $\alpha_{13}$ and $a_*$ vs $H$ of the FITS file of the emissivity profiles for our model {\tt relxilldisk\_nk}. \label{f-grid}}
\vspace{0.5cm}
\end{figure*}

\section{Simulations}\label{s-sim}

In the previous section, we have calculated the iron line profiles of accretion disks illuminated by disk-like coronae for different values of the viewing angle $i$, the coronal height $H$, and the coronal radius $R_{\rm disk}$ in either the Kerr or non-Kerr spacetime. In this section, we want to explore the impact of disk-like coronae on the measurement of the properties of a source. In particular, we want to figure out whether we are able to recover the correct parameter values in the case the corona is an infinitesimally thin disk and we fit the data with theoretical models that assume either a broken power-law emissivity profile or a lamppost coronal geometry. Since the emissivity profiles generated by static and corotating coronae are quite similar, here we consider the static case only.

First, we incorporate the emissivity profile for disk-like coronae in our reflection model {\tt relxill\_nk}.  This is done by adding a new flavor called {\tt relxilldisk\_nk} which reads an external FITS file where we have tabulated the emissivity profiles generated by ring-like coronae (see Appendix~\ref{app:ring}) for a grid 245$\times$34$\times$20$\times$30 of, respectively, coronal radii $R_{\rm disk}$ from 0.5 to 25~$M$, heights $H$ from $\sim$2 to $\sim$10~$M$ (the exact range depends on the value of the black hole spin $a_*$, see the right panel in Fig.~\ref{f-grid}), dimensionless black hole spins $a_*$ from $-0.998$ to 0.9982, and Johannsen deformation parameters $\alpha_{13}$ from $-2$ to 2 with the constraint in Eq.~(\ref{eq-constraints}).  The grids $a_*$ vs $\alpha_{13}$ and $a_*$ vs $H$ are shown in Fig.~\ref{f-grid}. The procedure was already described in \citet{Abdikamalov:2019yrr} in the case of the lamppost model. The model {\tt relxilldisk\_nk} reads the FITS file and sums up the contribution of the necessary rings to calculate the emissivity of the disk-like corona. We note that we have the angle-resolved calculation of the reflection spectrum of the disk of the {\tt relxill} package~\citep[for a discussion on these calculations and the angle-averaged ones, see, e.g.,][]{2020arXiv201104792B}.

We simulate some observation of a black hole binary with \textsl{NuSTAR}~\citep{Harrison:2013md}. \textsl{NuSTAR} is currently the most suitable X-ray mission for X-ray reflection spectroscopy of black hole binaries because of its broad energy band covering both the iron line and the Compton hump and the fact that it can observe bright sources like black hole binaries without incurring the pile-up issue. For simplicity, we consider a spectrum described by a power-law component (the direct radiation from the corona) and the relativistic reflection spectrum of the disk. In {\tt xspec} language~\citep{xspec}, the model is

\vspace{0.2cm}
{\tt tbabs $\times$ (cutoffpl + relxilldisk\_nk)}             
\vspace{0.2cm}

\noindent where {\tt tbabs} describes the Galactic absorption due to the interstellar medium~\citep{wilms2000absorption}, {\tt cutoffpl} is a power-law component describing the direct radiation from the corona, and {\tt relxilldisk\_nk} is the reflection spectrum of the new model for a disk-like corona emissivity profile. The reflection fraction in {\tt relxilldisk\_nk} is frozen to $-1$ because we already have {\tt cutoffpl} to describe the direct radiation from the corona.  In the simulations, we assume both cases,  the Kerr and non-Kerr metric.

We assume the observation of a bright black hole binary and we require the energy flux $\Phi = 4 \cdot 10^{-9}$~erg~cm$^{-2}$~s$^{-1}$ in the 1-10~keV energy range. We adjust the normalization parameters of {\tt cutoffpl} and {\tt relxilldisk\_nk} so that 1/3 of this energy flux comes from {\tt cutoffpl} and 2/3 from {\tt relxilldisk\_nk}.  We use the {\tt xspec} command {\tt fakeit} to simulate a 100~ks observation with FPMA/\textsl{NuSTAR} (essentially equivalent to a 50~ks observation employing both FPMA and FPMB),  which gives us about 2.9~million counts in the 3-79~keV energy range.

The simulated data are fitted with the model

\vspace{0.2cm}
{\tt tbabs $\times$ (cutoffpl + relxill(lp)\_nk/relxilldisk\_nk)}       
\vspace{0.2cm}

\noindent where {\tt relxill\_nk} is our relativistic reflection model~\citep{bambi2017testing, Abdikamalov:2019yrr}\footnote{The public version of the model is available at \url{http://www.physics.fudan.edu.cn/tps/people/bambi/Site/RELXILL_NK.html}.}. {\tt relxill\_nk} is an extension of the {\tt relxill} package~\citep{Dauser:2013xv, garcia2013x} to non-Kerr spacetimes. Here we use the version in which the background metric is described by the Johannsen metric~\citep{johannsen2013} with only one possible non-vanishing deformation parameter, $\alpha_{13}$.

\vspace{0cm}
\begin{table}
\centering
\caption{\rm {List of the simulated observations and the input values of their key-parameters along with the fitting models. ``Yes'' and ``No'' indicate whether the simulated observation is fitted or not with the fitting models {\tt relxill\_nk}, {\tt relxilllp\_nk}, or {\tt relxilldisk\_nk}. \label{simulations-table} }}
\begin{center}
{
{\renewcommand{\arraystretch}{1.4}
\begin{tabular}{lcccccccccc}
\hline
& \hspace{0.1cm} $a_*$ \hspace{0.1cm} & \hspace{0.1cm} $\alpha_{13}$ \hspace{0.1cm} & \hspace{0.1cm} $i$ [deg] \hspace{0.1cm} &  $H/M$ \hspace{0.1cm} & \hspace{0.1cm} $R_{\rm disk}/M$  && \multicolumn{3}{c}{fitting model}   \\ &&&&&& \hspace{0.1cm} {\tt relxill\_nk} && {\tt relxilllp\_nk} && {\tt relxilldisk\_nk} \\
\hline
Simulation~A & 0.99 & 0 & 20 & 2 & 2 & Yes && Yes && No \\
Simulation~B & 0.99 & 0 & 20 & 2 & 6 & Yes && Yes && Yes  \\
Simulation~C & 0.99 & 0 & 70 & 2 & 2 & Yes && Yes  && No\\
Simulation~D & 0.99 & 0 & 70 & 2 & 6 & Yes  && Yes && Yes\\
Simulation~E & 0.99 & $-0.24$ & 20 & 2 & 2 & Yes && Yes && No\\
Simulation~F & 0.99 & $-0.24$ & 20 & 2 & 6 & Yes && Yes && Yes\\
Simulation~G & 0.99 & $-0.24$ & 70 & 2 & 2 & Yes && Yes && No\\
Simulation~H & 0.99 & $-0.24$ & 70 & 2 & 6 & Yes && Yes && Yes\\
Simulation~I & 0.99 & 1.0 & 20 & 2 & 2 & Yes && Yes && No\\
Simulation~J & 0.99 & 1.0 & 20 & 2 & 6 & Yes && Yes && Yes\\
Simulation~K & 0.99 & 1.0 & 70 & 2 & 2 & Yes && Yes && No\\
Simulation~L & 0.99 & 1.0 & 70 & 2 & 6 & Yes && Yes && Yes\\
\hline
\end{tabular}}
}
\end{center}
\end{table}

\begin{figure*}[t]
	\begin{center}
		\includegraphics[width=0.45\textwidth,trim={0cm 0cm 0cm 0cm},clip]{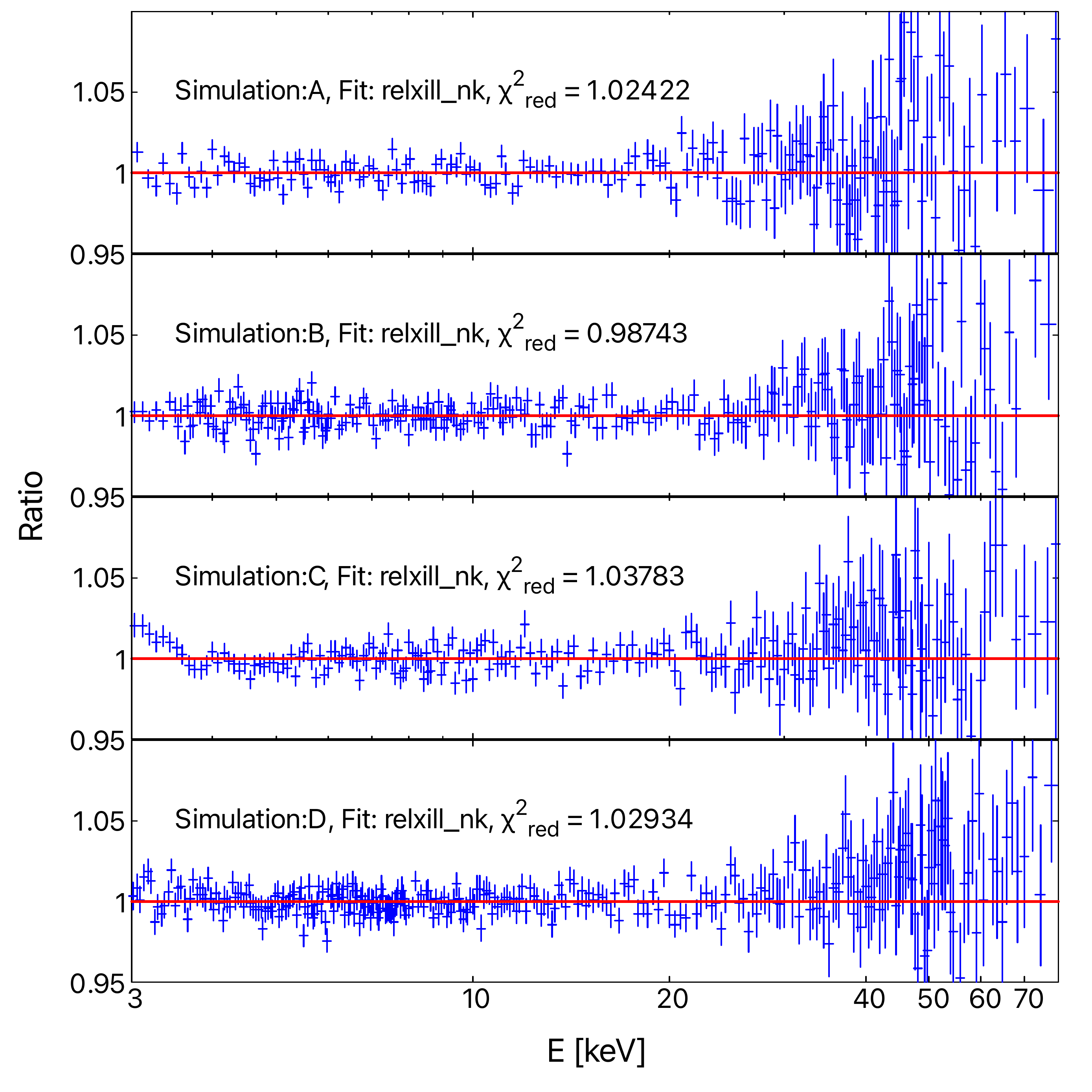}
		\hspace{0.4cm}
		\includegraphics[width=0.45\textwidth,trim={0cm 0cm 0cm 0cm},clip]{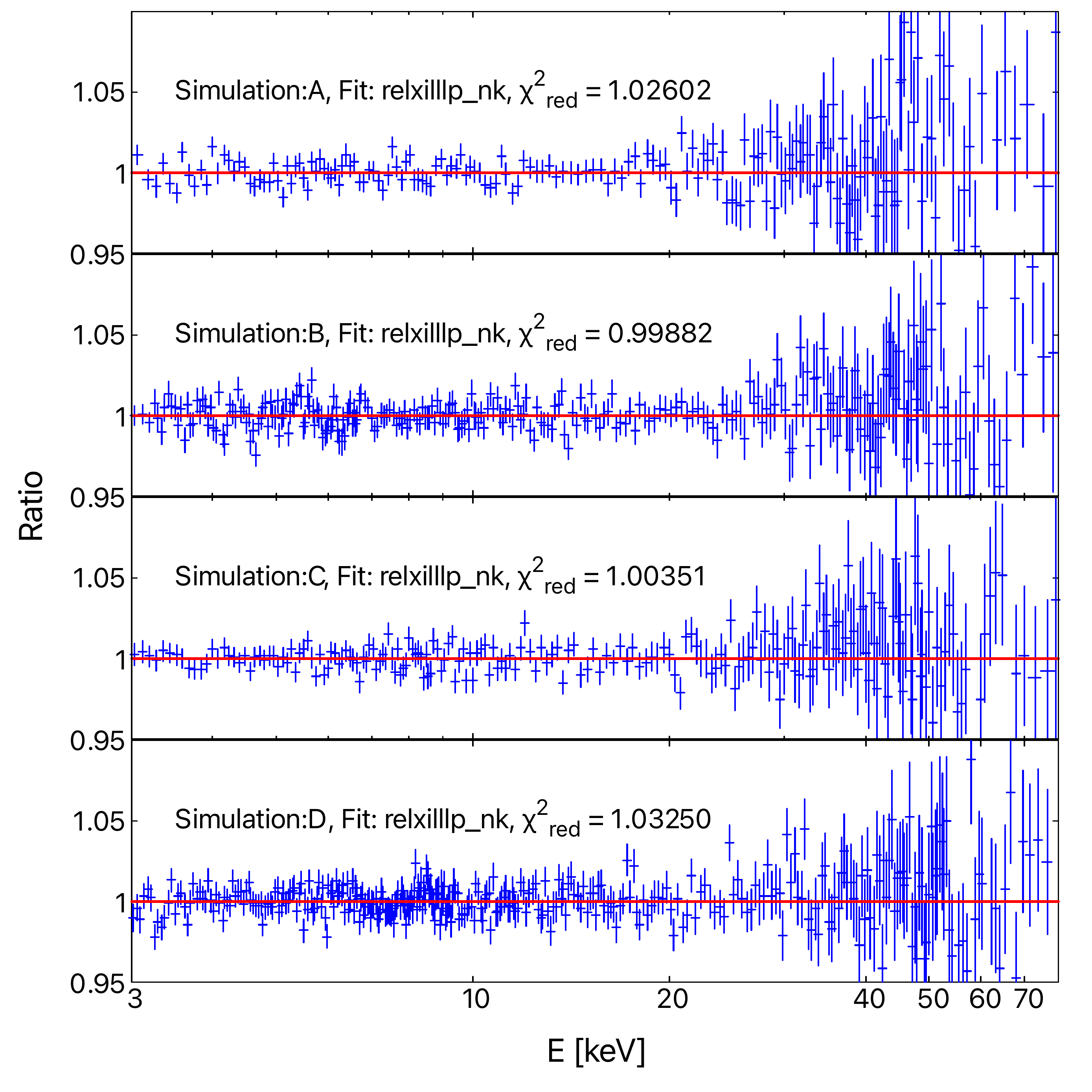}
	\end{center}
	\vspace{-0.3cm}
	{\caption{Data to best-fit model ratios for the fits in Tab.~\ref{t-fit1} with {\tt relxill\_nk} (left panel) and Tab.~\ref{t-fit2} with {\tt relxilllp\_nk} (right panel). See the text for more details.  \label{f-ratio-a13_0}}}
	\vspace{0.4cm}
\end{figure*}

\begin{figure*}[t]
	\begin{center}
		\includegraphics[width=0.45\textwidth,trim={0cm 0cm 0cm 0cm},clip]{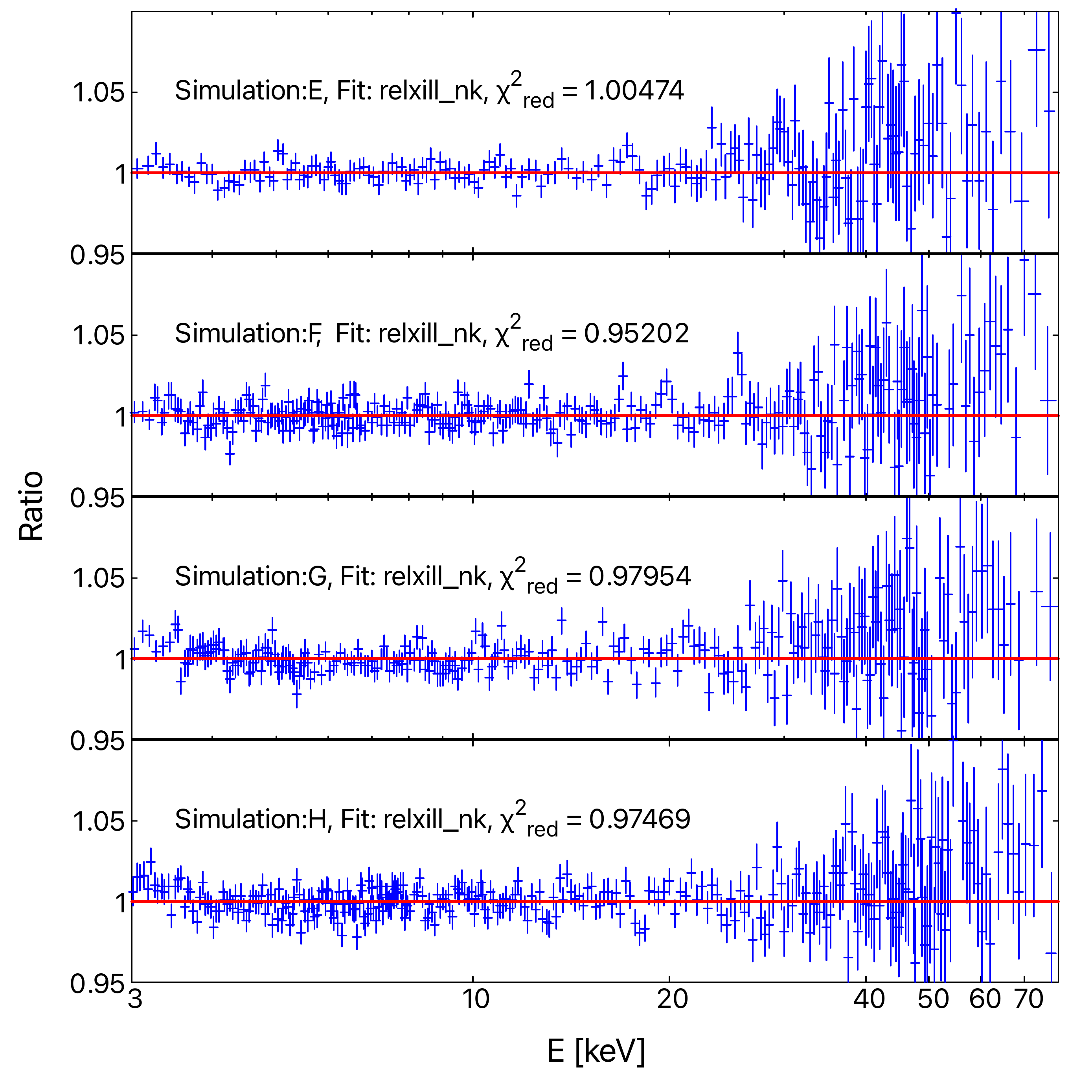}
		\hspace{0.4cm}
		\includegraphics[width=0.45\textwidth,trim={0cm 0cm 0cm 0cm},clip]{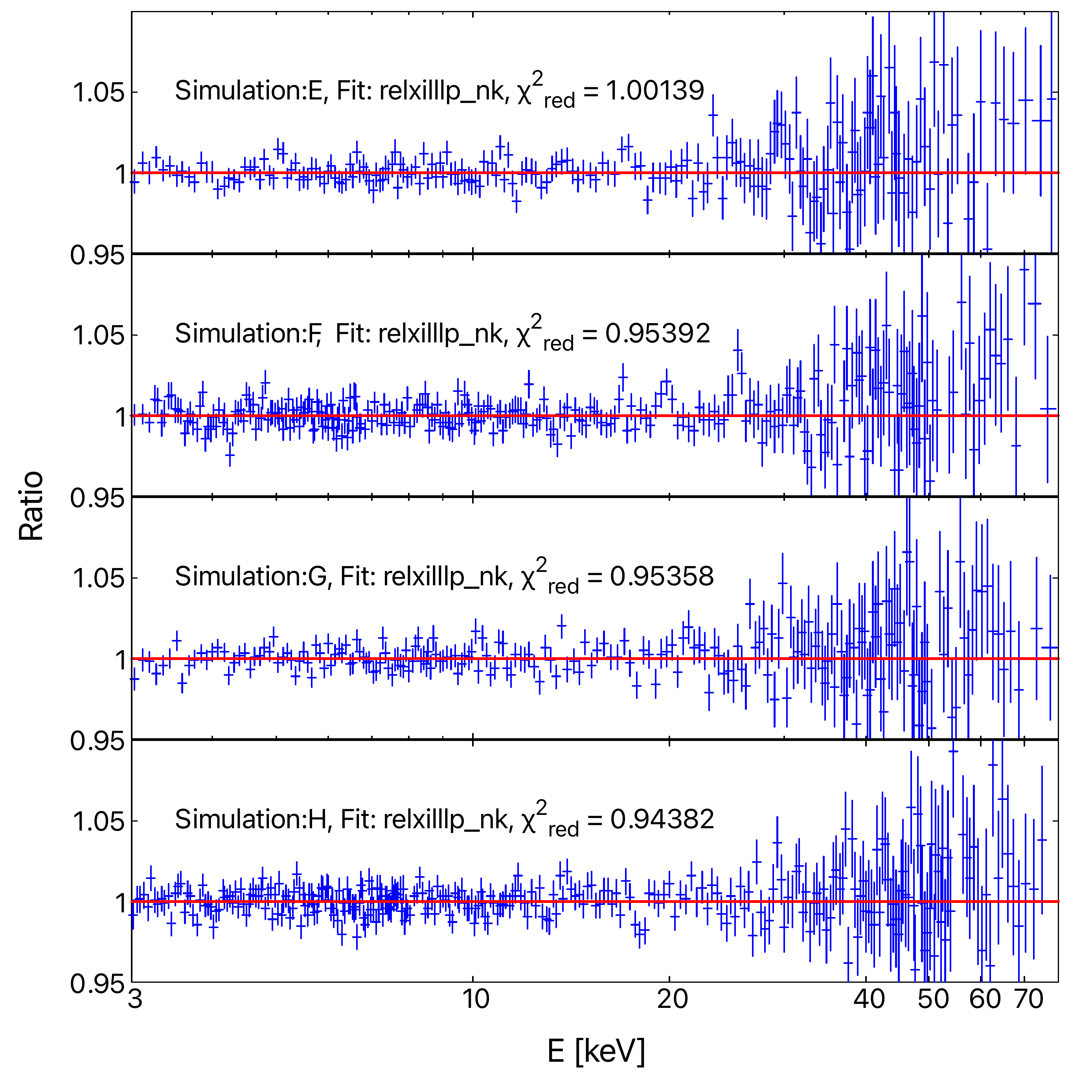}
	\end{center}
	\vspace{-0.3cm}
	{\caption{Data to best-fit model ratios for the fits in Tab.~\ref{t-fit3} with {\tt relxill\_nk} (left panel) and Tab.~\ref{t-fit4} with {\tt relxilllp\_nk} (right panel). See the text for more details.  \label{f-ratio-a13_-0.5}}}
	\vspace{0.4cm}
\end{figure*}

\begin{figure*}[t]
	\begin{center}
		\includegraphics[width=0.45\textwidth,trim={0cm 0cm 0cm 0cm},clip]{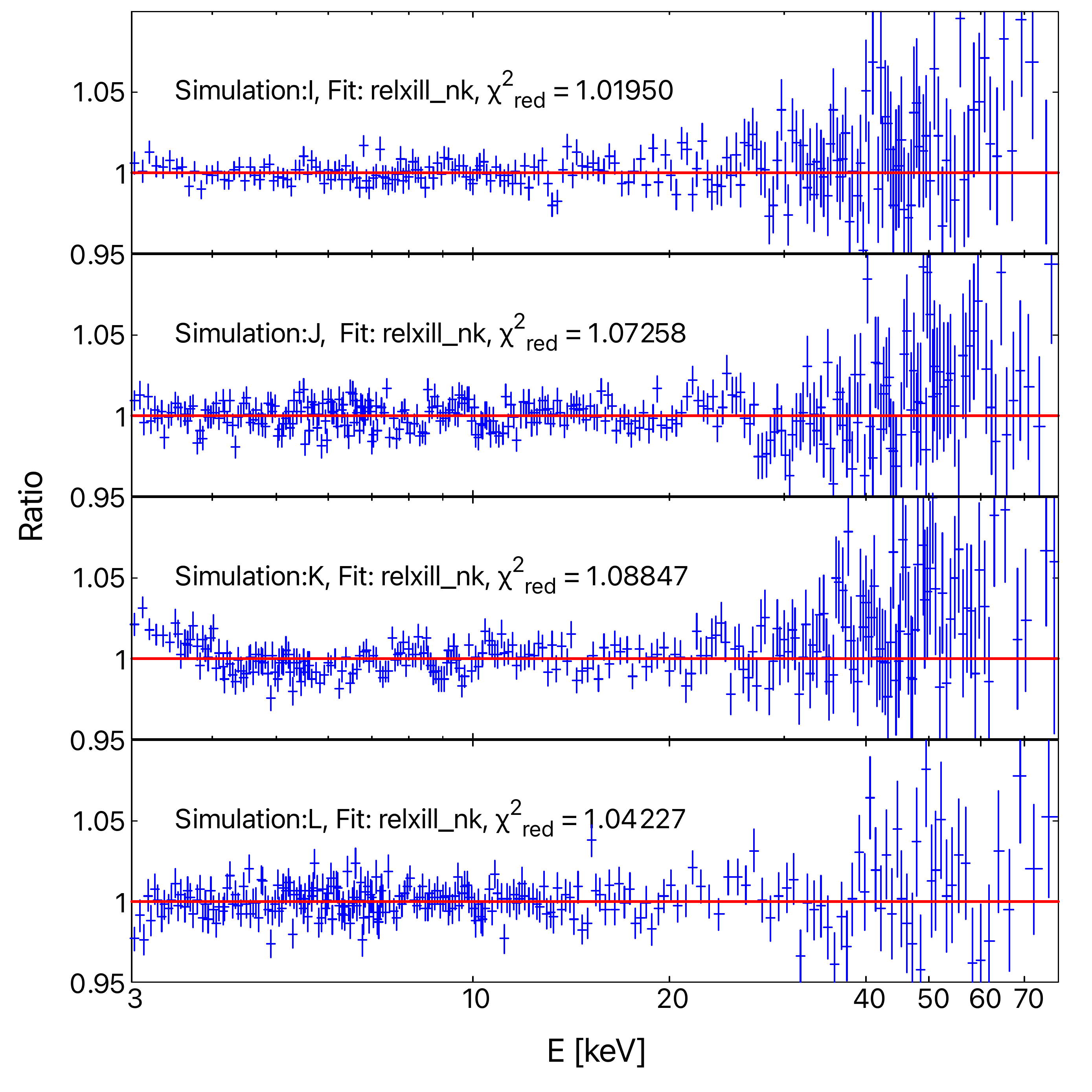}
		\hspace{0.4cm}
		\includegraphics[width=0.45\textwidth,trim={0cm 0cm 0cm 0cm},clip]{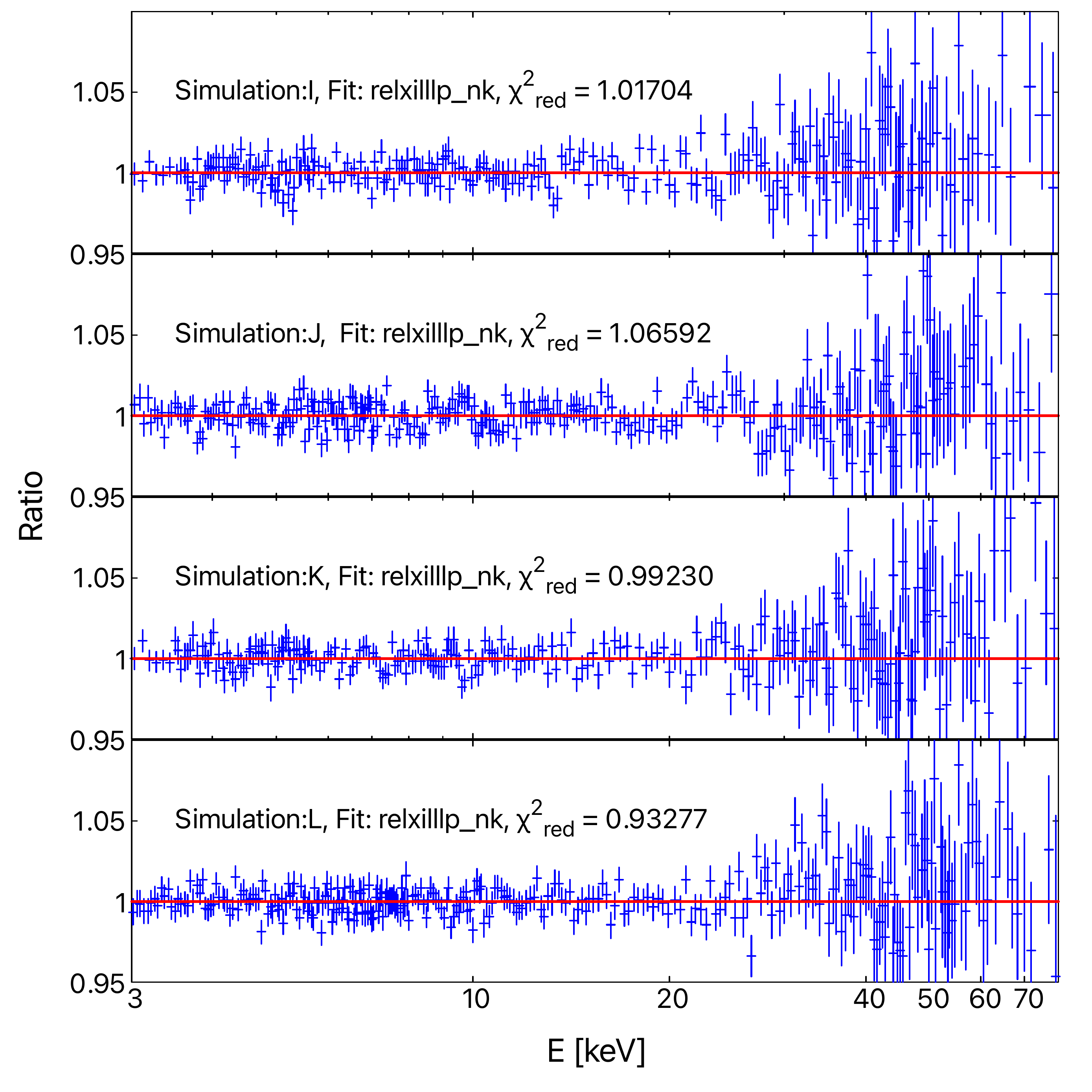}
	\end{center}
	\vspace{-0.3cm}
	{\caption{Data to best-fit model ratios for the fits in Tab.~\ref{t-fit5} with {\tt relxill\_nk} (left panel) and Tab.~\ref{t-fit6} with {\tt relxilllp\_nk} (right panel). See the text for more details.  \label{f-ratio-a13_0.5}}}
	\vspace{0.4cm}
\end{figure*}

{We simulate 12 configurations of the astrophysical system (4 in the Kerr metric and 8 in the non-Kerr metric), which we call Simulations A,  B,  C,  D,  E,  F,  G,  H,  I,  J,  K, and L (see Tab.\ref{simulations-table}).} In all simulations, the ionization of the disk is $\log\xi  = 3.1$ ($\xi$ in units erg~cm~s$^{-1}$), the iron abundance of the disk is $A_{\rm Fe} = 1$ (i.e. the Solar value), the photon index of the coronal spectrum is $\Gamma = 1.7$, and the high energy cut-off of the coronal spectrum  is $E_{\rm cut} = 300$~keV.  The data are then fitted with {\tt relxill\_nk} assuming a broken power-law emissivity profile (inner emissivity profile $q_{\rm in}$, outer emissivity profile $q_{\rm out}$, and breaking radius $r_{\rm br}$ free).  We also leave the deformation parameter $\alpha_{13}$ free in the fit, as we are interested to see whether we can test the Kerr metric.  The results of our fits are summarized in Tabs.~\ref{t-fit1},~\ref{t-fit3}, and~\ref{t-fit5}, and the data to best-fit model ratios are shown in the left panel in Figs.~\ref{f-ratio-a13_0}-\ref{f-ratio-a13_0.5}.

To check whether the lamppost model can describe the emissivity profile generated by a disk-like corona better than the broken power-law model, we repeat our fits with {\tt relxilllp\_nk}, namely the {\tt relxill\_nk} version with lamppost emissivity profile~\citep{Abdikamalov:2019yrr}. The summary of the second set of fits is reported in Tabs.~\ref{t-fit2},~\ref{t-fit4},  and~\ref{t-fit6}, and the data to best-fit model ratios are in the right panel in Figs.~\ref{f-ratio-a13_0}-\ref{f-ratio-a13_0.5}.

\begin{figure*}[t]
	\begin{center}
		\includegraphics[width=0.45\textwidth,trim={0cm 0cm 0cm 0cm},clip]{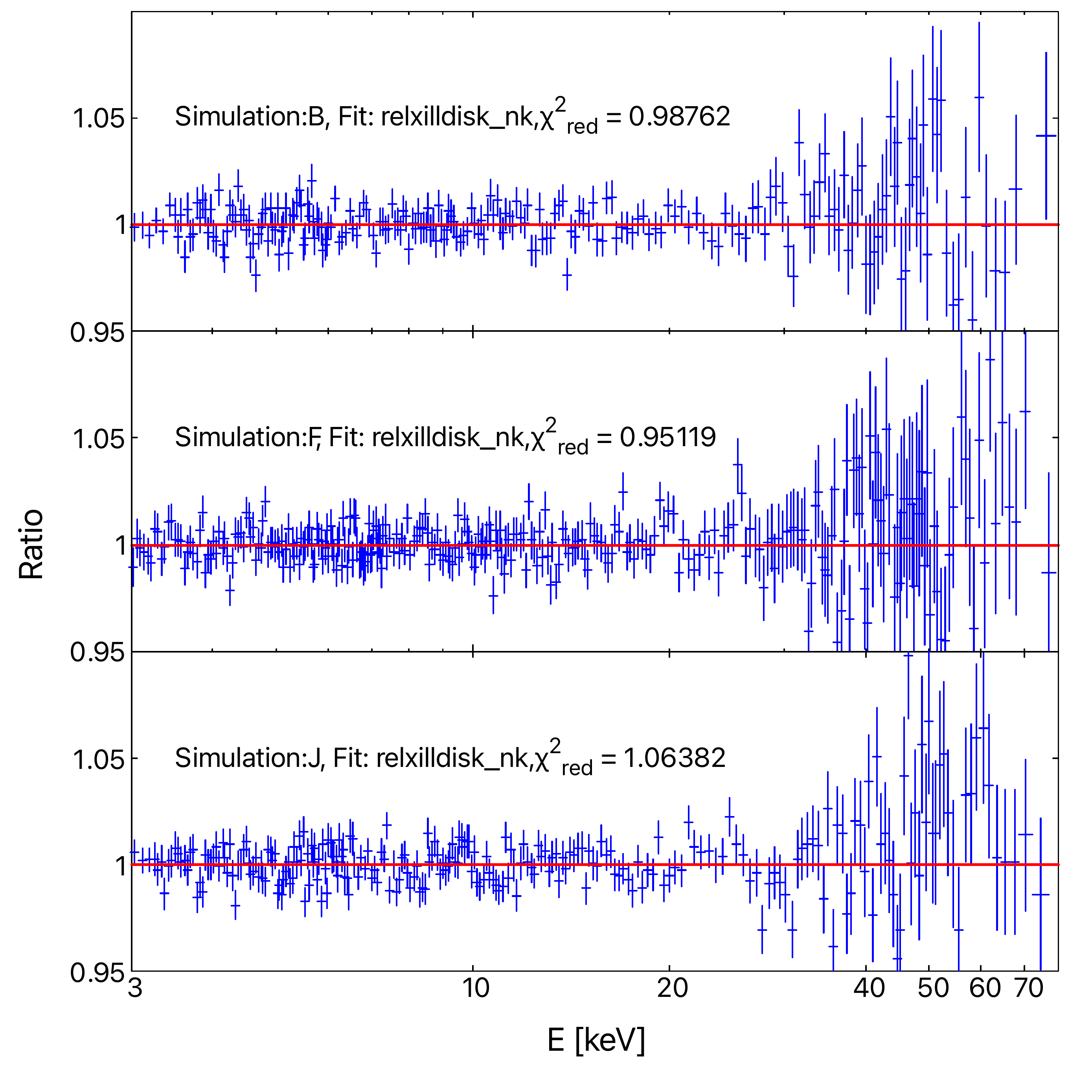}
		\hspace{0.4cm}
		\includegraphics[width=0.45\textwidth,trim={0cm 0cm 0cm 0cm},clip]{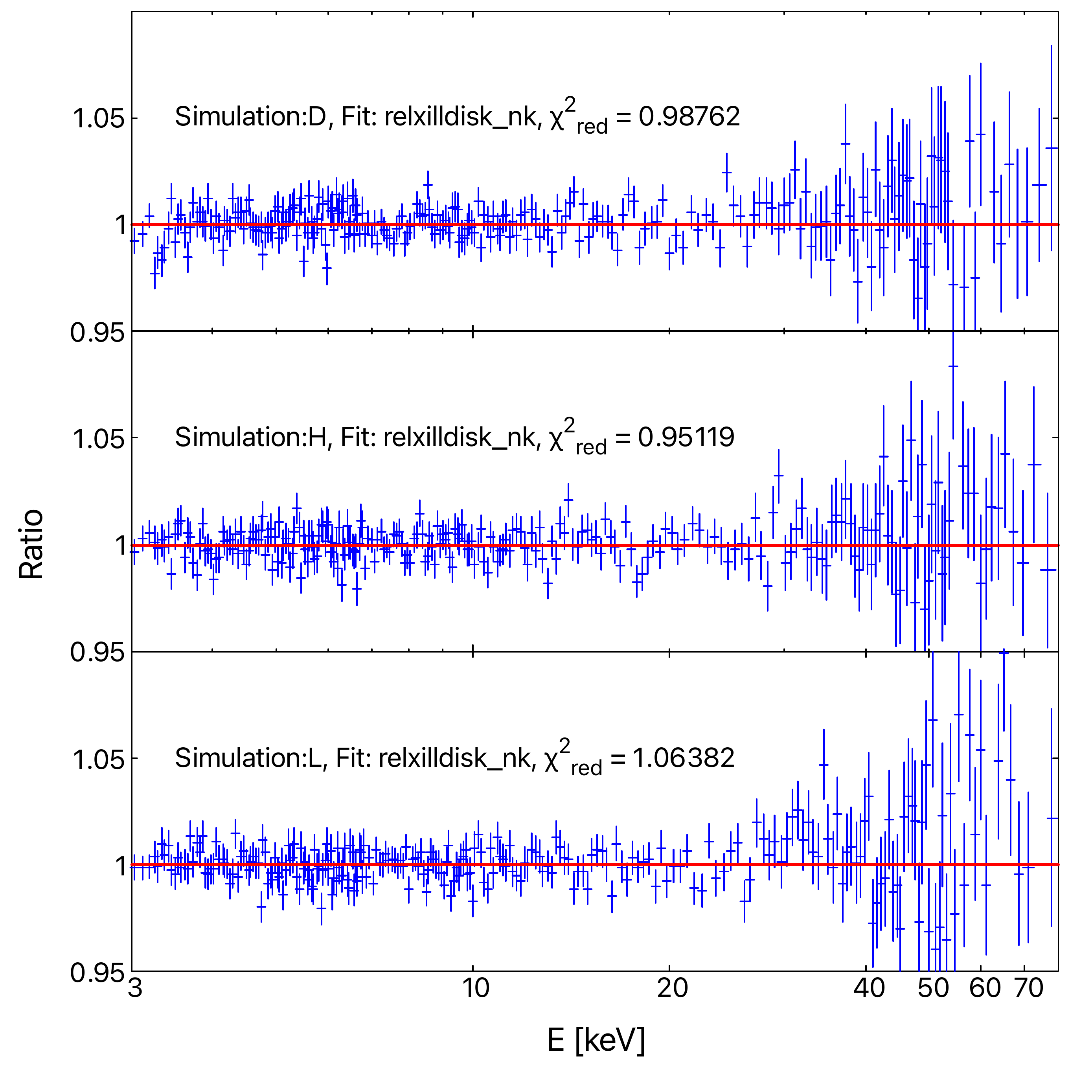}
	\end{center}
	\vspace{-0.3cm}
	{\caption{Data to best-fit model ratios for the fits in Tab.~\ref{t-fit7} and Tab.~\ref{t-fit8} with {\tt relxilldisk\_nk}.  See the text for more details.  \label{f-ratio-fitdisk}}}
	\vspace{0.4cm}
\end{figure*}

We also fit some of simulations with the correct model, i.e. {\tt relxilldisk\_nk}. This would allow us to determine how reliably the extent of a disk-like corona can be measured by the model and to determine whether the illumination of the disk-like corona intrinsically limits the ability to estimate the deviation from the Kerr metric. The best-fit values are shown in Tabs.~\ref{t-fit7} and~\ref{t-fit8}. The data to best-fit model ratios are in Fig.~\ref{f-ratio-fitdisk}.

The discussion of all fits is postponed to the next section.


\begin{table*}
	\centering
	\caption{\rm Best-fit values for simulations~A-D when we model the disk's emissivity profile with a broken power-law. $\xi$ in units of erg~cm~s$^{-1}$. The reported uncertainties correspond to a 90\% confidence level for one relevant parameter ($\Delta\chi^2 = 2.71$). $^\star$ indicates that the parameter is frozen in the fit. (P) indicates that the parameter boundary is within the 90\% confidence level. When there is no upper/lower uncertainty, it means that the parameter is stuck at the upper/lower boundary of the range in which it is allowed to vary. See the text for more details. \label{t-fit1}}
	{\renewcommand{\arraystretch}{1.4}
		\begin{tabular}{lcccccccc}
			\hline\hline
			& \multicolumn{2}{c}{Simulation~A} & \multicolumn{2}{c}{ Simulation~B} & \multicolumn{2}{c}{Simulation~C} & \multicolumn{2}{c}{Simulation~D} \\
			& Input & Fit & Input & Fit & Input & Fit & Input & Fit \\
			\hline
			{\tt tbabs} &&&&&&&& \\
			$N_{\rm H}$ [$10^{20}$~cm$^{-2}$] & $6.74$ & $6.74^\star$ & $6.74$ & $6.74^\star$ & $6.74$ & $6.74^\star$ & $6.74$ & $6.74^\star$ \\
			\hline
			{\tt relxill\_nk} &&&&&&&& \\
			$H$ [$M$] & $2$ & -- & $2$ & -- & $2$ & -- & $2$ & -- \\
			$R_{\rm disk}$ [$M$] & $2$ & -- & $6$ & -- & $2$ & -- & $6$ & -- \\
			$q_{\rm in}$ & -- & $4.0^{+1.5}_{-0.8}$ & -- & $3.0^{+0.4}_{-0.4}$ & -- & $10.0^{\rm +(P)}_{-0.5}$ & -- & $3.2^{+0.3}_{-0.5}$ \\
			$q_{\rm out}$ & -- & $2.97^{+0.42}_{-0.24}$ & -- & $10^{+\rm (P)}_{-5.4}$ & -- & $3.65^{+0.21}_{-0.50}$ & -- & $10^{}_{-5}$ \\
			$r_{\rm br}$ [$M$] & -- & $8^{+12}_{-5}$ & -- & $19^{+8}_{-3}$ & -- & $1.66^{+0.05}_{-0.08}$ & -- & $7.7^{+3.5}_{-2.3}$ \\
			$i$ [deg] & $20$ & $17^{+8}_{-8}$ & $20$ & $24.9^{+1.6}_{-1.8}$ & $70$ & $68.1^{+1.0}_{-0.4}$ & $70$ & $70.5^{+1.0}_{-0.8}$ \\
			$a_*$ & $0.99$ & $0.95^{\rm +(P)}_{-0.43}$ & $0.99$ & $0.85^{\rm +(P)}_{-0.11}$ & $0.99$ & $0.9980^{}_{-0.0004}$ & $0.99$ & $0.998^{}_{-0.004}$ \\
			$A_{\rm Fe}$ & $1$ & $1.00^{+0.39}_{-0.13}$ & $1$ & $1.20^{+0.19}_{-0.16}$ & $1$ & $2.19^{+0.11}_{-0.13}$ & $1$ & $1.49^{+0.07}_{-0.10}$ \\
			$\Gamma$ & $1.7$ & $1.687^{+0.021}_{-0.026}$ & $1.7$ & $1.688^{+0.012}_{-0.016}$ & $1.7$ & $1.628^{+0.009}_{-0.008}$ & $1.7$ & $1.648^{+0.008}_{-0.005}$ \\
			$\log\xi$ & $3.1$ & $3.10^{+0.04}_{-0.04}$ & $3.1$ & $3.099^{+0.026}_{-0.019}$ & $3.1$ & $3.188^{+0.020}_{-0.020}$ & $3.1$ & $3.176^{+0.016}_{-0.016}$ \\
			$E_{\rm cut}$ [keV] & $300$ & $300^{\star}$ & $300$ & $300^{\star}$ & $300$ & $300^{\star}$ & $300$ & $300{^\star}$ \\
			$\alpha_{13}$ & $0$ & $-0.12^{+1.56}_{-\rm (P)}$ & $0$ & $-0.9^{+0.4}_{-0.9}$ & $0$ & $0.000^{+0.010}_{-0.091}$ & $0$ & {$0.00^{+0.04}_{-0.21}$} \\
			\hline
			$\chi^2/\nu$ && $\quad 1324.32/1293 \quad$ && $\quad 1367.60/1385 \quad$ && $\quad 1421.82/1370\quad$ && $\quad 1501.81/1459 \quad$ \\
			&& =1.02422 && =0.98743 && =1.03783 && =1.02934 \\
			\hline\hline
	\end{tabular}}
	
\vspace{0.8cm}
	\centering
	{\caption{\rm Best-fit values for simulations~A-D when we employ the disk's emissivity profile of a lamppost corona. $\xi$ in units of erg~cm~s$^{-1}$. The reported uncertainties correspond to a 90\% confidence level for one relevant parameter ($\Delta\chi^2 = 2.71$). $^\star$ indicates that the parameter is frozen in the fit. (P) indicates that the parameter boundary is within the 90\% confidence level. When there is no upper/lower uncertainty, it means that the parameter is stuck at the upper/lower boundary of the range in which it is allowed to vary. See the text for more details. \label{t-fit2}}}
	{\renewcommand{\arraystretch}{1.4}
		\begin{tabular}{lccccccccccc}
			\hline\hline
			& \multicolumn{2}{c}{Simulation~A} & \multicolumn{2}{c}{Simulation~B} & \multicolumn{2}{c}{Simulation~C} & \multicolumn{2}{c}{Simulation~D}\\
			& Input & Fit & Input & Fit & Input & Fit & Input & Fit \\
			\hline
			{\tt tbabs} &&&&&&&& \\
			$N_{\rm H}$ [$10^{20}$~cm$^{-2}$] & $6.74$ & $6.74^\star$ & $6.74$ & $6.74^\star$ & $6.74$ & $6.74^\star$ & $6.74$ & $6.74^\star$ \\
			\hline
			{\tt relxilllp\_nk} &&&&&&&& \\
			$H$ [$M$] & $2$ & -- & $2$ & -- & $2$ & -- & $2$ & -- \\
			$R_{\rm disk}$ [$M$] & $2$ & -- & $6$ & -- & $2$ & -- & $6$ & -- \\
			$h$ [$M$] & -- & $2.0^{+0.3}_{-\rm (P)}$ & -- & $3.00^{+0.19}_{- \rm (P)}$ & -- & $2.1^{+1.1}_{-\rm (P)}$ & -- & $4.84^{+0.04}_{-0.04}$\\
			$i$ [deg] & $20$ & $23.2^{+2.0}_{-1.8}$ & $20$ & $15.4^{+0.8}_{-1.3}$ & $70$ & $70.4^{+0.9}_{-1.4}$ & $70$ & $75.0^{+2.5}_{-3.0}$ \\
			$a_*$ & $0.99$ & $0.988^{\rm +(P)}_{-0.047}$ & $0.99$ & $0.71^{+0.08}_{-0.04}$ & $0.99$ & $0.991^{+0.003}_{-0.006}$ & $0.99$ & $0.71^{+0.21}_{-0.50}$ \\
			$A_{\rm Fe}$ & $1$ & $1.00^{+0.21}_{-0.07}$ & $1$ & $0.929^{+0.020}_{-0.008}$ & $1$ & $0.95^{+0.03}_{-0.04}$ & $1$ & $0.926^{+0.023}_{-0.022}$ \\
			$\Gamma$ & $1.7$ & $1.706^{+0.026}_{-0.013}$ & $1.7$ & $1.705^{+0.006}_{-0.011}$ & $1.7$ & $1.714^{+0.013}_{-0.015}$ & $1.7$ & $1.691^{+0.005}_{-0.005}$ \\
			$\log\xi$ & $3.1$ & $3.094^{+0.027}_{-0.023}$ & $3.1$ & $3.10^{+0.03}_{-0.04}$ & $3.1$ & $3.087^{+0.027}_{-0.022}$ & $3.1$ & $3.116^{+0.012}_{-0.009}$ \\
			$E_{\rm cut}$ [keV] & $300$ & $300^{\star}$ & $300$ & $300^{\star}$ & $300$ & $300^{\star}$ & $300$ & $300{^\star}$ \\
			$\alpha_{13}$ & $0$ & $-0.09^{+0.53}_{-0.07}$ & $0$ & $0.0^{+0.5}_{-0.3}$ & $0$ & $-0.20^{+0.55}_{-0.13}$ & $0$ & $-1.8^{+3.0}_{-\rm (P)}$ \\
			\hline
			$\chi^2/\nu$ && $\quad 1328.69/1295 \quad$ && $\quad 1385.37/1387 \quad$ && $\quad 1376.81/1372\quad$ && $\quad 1508.48/1461 \quad$ \\
			&& =1.02602 && =0.99882 && =1.00351 && =1.03250 \\
			\hline\hline
	\end{tabular}}
	\vspace{0.5cm}
\end{table*}



\begin{table*}
	\centering
	\caption{\rm Best-fit values for simulations~E-H when we model the disk's emissivity profile with a broken power-law. $\xi$ in units of erg~cm~s$^{-1}$. The reported uncertainties correspond to a 90\% confidence level for one relevant parameter ($\Delta\chi^2 = 2.71$). $^\star$ indicates that the parameter is frozen in the fit. (P) indicates that the parameter boundary is within the 90\% confidence level. When there is no upper/lower uncertainty, it means that the parameter is stuck at the upper/lower boundary of the range in which it is allowed to vary. See the text for more details. \label{t-fit3}} 
	{\renewcommand{\arraystretch}{1.4}
		\begin{tabular}{lcccccccc}
			\hline\hline
			& \multicolumn{2}{c}{Simulation~E} & \multicolumn{2}{c}{ Simulation~F} & \multicolumn{2}{c}{Simulation~G} & \multicolumn{2}{c}{Simulation~H} \\
			& Input & Fit & Input & Fit & Input & Fit & Input & Fit \\
			\hline
			{\tt tbabs} &&&&&&&& \\
			$N_{\rm H}$ [$10^{20}$~cm$^{-2}$] & $6.74$ & $6.74^\star$ & $6.74$ & $6.74^\star$ & $6.74$ & $6.74^\star$ & $6.74$ & $6.74^\star$ \\
			\hline
			{\tt relxill\_nk} &&&&&&&& \\
			$H$ [$M$] & $2$ & -- & $2$ & -- & $2$ & -- & $2$ & -- \\
			$R_{\rm disk}$ [$M$] & $2$ & -- & $6$ & -- & $2$ & -- & $6$ & -- \\
			$q_{\rm in}$ & -- & $3.34^{+0.17}_{-0.03}$ & -- & $0.7^{+1.4}_{-0.3}$ & -- & $3.7^{+0.4}_{-0.3}$ & -- & $10.0^{}_{-0.6}$ \\
			$q_{\rm out}$ & -- & $0.1^{+0.6}_{-\rm (P)}$ & -- & $3.25^{+0.18}_{-0.10}$ & -- & $10^{+\rm (P)}_{-6}$ & -- & $3.23^{+0.17}_{-0.17}$ \\
			$r_{\rm br}$ [$M$] & -- & $107^{+10}_{-6}$ & -- & $3.37^{+2.14}_{-0.24}$ & -- & $8^{+50}_{-4}$ & -- & $1.59^{+0.05}_{-0.05}$ \\
			$i$ [deg] & $20$ & $4^{+19}_{-\rm (P)}$ & $20$ & $20^{+4}_{-3}$ & $70$ & $68.0^{+1.3}_{-0.8}$ & $70$ & $70.19^{+0.38}_{-0.28}$ \\
			$a_*$ & $0.99$ & $0.93^{+0.03}_{-0.10}$ & $0.99$ & $0.91^{\rm +(P)}_{-0.21}$ & $0.99$ & $0.998^{}_{-0.004}$ & $0.99$ & $0.998^{}_{-0.062}$ \\
			$A_{\rm Fe}$ & $1$ & $0.930^{+0.015}_{-0.095}$ & $1$ & $1.018^{+0.115}_{-0.040}$ & $1$ & $1.79^{+0.10}_{-0.12}$ & $1$ & $1.74^{+0.11}_{-0.11}$ \\
			$\Gamma$ & $1.7$ & $1.690^{+0.007}_{-0.007}$ & $1.7$ & $1.692^{+0.012}_{-0.016}$ & $1.7$ & $1.626^{+0.005}_{-0.008}$ & $1.7$ & $1.649^{+0.006}_{-0.005}$ \\
			$\log\xi$ & $3.1$ & $3.107^{+0.040}_{-0.021}$ & $3.1$ & $3.103^{+0.020}_{-0.020}$ & $3.1$ & $3.215^{+0.024}_{-0.024}$ & $3.1$ & $3.164^{+0.012}_{-0.011}$ \\
			$E_{\rm cut}$ [keV] & $300$ & $300^{\star}$ & $300$ & $300^{\star}$ & $300$ & $300^{\star}$ & $300$ & $300{^\star}$ \\
			$\alpha_{13}$ & $-0.24$ & $-0.9^{+1.3}_{-0.6}$ & $-0.24$ & $-1.1^{+1.6}_{-0.7}$ & $-0.24$ & $0.01^{+0.04}_{-0.18}$ & $-0.24$ & $0.00^{+0.01}_{-0.05}$ \\
			\hline
			$\chi^2/\nu$ && $\quad 1290.08/1284 \quad$ && $\quad 1308.08/1374 \quad$ && $\quad 1346.88/1375\quad$ && $\quad 1429.87/1467 \quad$ \\
			&& =1.00474 && =0.95202 && =0.97954 && =0.97469 \\
			\hline\hline
	\end{tabular}}
	
\vspace{0.8cm}
	\centering
	\caption{\rm {Best-fit values for simulations~E-H when we employ the disk's emissivity profile of a lamppost corona. $\xi$ in units of erg~cm~s$^{-1}$. The reported uncertainties correspond to a 90\% confidence level for one relevant parameter ($\Delta\chi^2 = 2.71$). $^\star$ indicates that the parameter is frozen in the fit. (P) indicates that the parameter boundary is within the 90\% confidence level. When there is no upper/lower uncertainty, it means that the parameter is stuck at the upper/lower boundary of the range in which it is allowed to vary. See the text for more details.} \label{t-fit4}}
	{\renewcommand{\arraystretch}{1.4}
		\begin{tabular}{lccccccccccc}
			\hline\hline
			& \multicolumn{2}{c}{Simulation~E} & \multicolumn{2}{c}{Simulation~F} & \multicolumn{2}{c}{Simulation~G} & \multicolumn{2}{c}{Simulation~H}\\
			& Input & Fit & Input & Fit & Input & Fit & Input & Fit \\
			\hline
			{\tt tbabs} &&&&&&&& \\
			$N_{\rm H}$ [$10^{20}$~cm$^{-2}$] & $6.74$ & $6.74^\star$ & $6.74$ & $6.74^\star$ & $6.74$ & $6.74^\star$ & $6.74$ & $6.74^\star$ \\
			\hline
			{\tt relxilllp\_nk} &&&&&&&& \\
			$H$ [$M$] & $2$ & -- & $2$ & -- & $2$ & -- & $2$ & -- \\
			$R_{\rm disk}$ [$M$] & $2$ & -- & $6$ & -- & $2$ & -- & $6$ & -- \\
			$h$ [$M$] & -- & $2.06^{+0.23}_{-\rm (P)}$ & -- & $2.28^{+0.15}_{-0.03}$ & -- & $2.6^{+0.5}_{-\rm (P)}$ & -- & $3.4^{+0.8}_{-\rm (P)}$\\
			$i$ [deg] & $20$ & $23.1^{+1.6}_{-1.8}$ & $20$ & $17.6^{+0.9}_{-1.1}$ & $70$ & $69.1^{+0.6}_{-1.3}$ & $70$ & $70.0^{+0.5}_{-1.7}$ \\
			$a_*$ & $0.99$ & $0.997^{\rm +(P)}_{-0.059}$ & $0.99$ & $0.996^{+\rm (P)}_{-0.305}$ & $0.99$ & $0.988^{+0.005}_{-0.004}$ & $0.99$ & $0.35^{+0.23}_{-0.11}$ \\
			$A_{\rm Fe}$ & $1$ & $0.94^{+0.03}_{-0.05}$ & $1$ & $0.960^{+0.017}_{-0.016}$ & $1$ & $0.98^{+0.05}_{-0.03}$ & $1$ & $0.957^{+0.014}_{-0.017}$ \\
			$\Gamma$ & $1.7$ & $1.706^{+0.019}_{-0.013}$ & $1.7$ & $1.698^{+0.008}_{-0.011}$ & $1.7$ & $1.709^{+0.010}_{-0.007}$ & $1.7$ & $1.692^{+0.004}_{-0.004}$ \\
			$\log\xi$ & $3.1$ & $3.11^{+0.03}_{-0.03}$ & $3.1$ & $3.109^{+0.019}_{-0.013}$ & $3.1$ & $3.094^{+0.013}_{-0.013}$ & $3.1$ & $3.118^{+0.011}_{-0.007}$ \\
			$E_{\rm cut}$ [keV] & $300$ & $300^{\star}$ & $300$ & $300^{\star}$ & $300$ & $300^{\star}$ & $300$ & $300{^\star}$ \\
			$\alpha_{13}$ & $-0.24$ & $0.21^{+0.17}_{-0.29}$ & $-0.24$ & $1.52^{+0.17}_{\rm - (P)}$ & $-0.24$ & $-0.03^{+0.15}_{-0.17}$ & $-0.24$ & $-1.8^{+2.3}_{- \rm (P)}$ \\
			\hline
			$\chi^2/\nu$ && $\quad 1287.78/1286 \quad$ && $\quad 1312.60/1376 \quad$ && $\quad 1313.09/1377\quad$ && $\quad 1386.48/1469 \quad$ \\
			&& =1.00139 && = 0.95392 && = 0.95358 && =0.94382 \\
			\hline\hline
	\end{tabular}}
	\vspace{0.5cm}
\end{table*}



\begin{table*}
	\centering
	\caption{\rm Best-fit values for simulations~I-L when we model the disk's emissivity profile with a broken power-law. $\xi$ in units of erg~cm~s$^{-1}$. The reported uncertainties correspond to a 90\% confidence level for one relevant parameter ($\Delta\chi^2 = 2.71$). $^\star$ indicates that the parameter is frozen in the fit. (P) indicates that the parameter boundary is within the 90\% confidence level. When there is no upper/lower uncertainty, it means that the parameter is stuck at the upper/lower boundary of the range in which it is allowed to vary. See the text for more details. \label{t-fit5}} 
	{\renewcommand{\arraystretch}{1.4}
		\begin{tabular}{lcccccccc}
			\hline\hline
			& \multicolumn{2}{c}{Simulation~I} & \multicolumn{2}{c}{ Simulation~J} & \multicolumn{2}{c}{Simulation~K} & \multicolumn{2}{c}{Simulation~L} \\
			& Input & Fit & Input & Fit & Input & Fit & Input & Fit \\
			\hline
			{\tt tbabs} &&&&&&&& \\
			$N_{\rm H}$ [$10^{20}$~cm$^{-2}$] & $6.74$ & $6.74^\star$ & $6.74$ & $6.74^\star$ & $6.74$ & $6.74^\star$ & $6.74$ & $6.74^\star$ \\
			\hline
			{\tt relxill\_nk} &&&&&&&& \\
			$H$ [$M$] & $2$ & -- & $2$ & -- & $2$ & -- & $2$ & -- \\
			$R_{\rm disk}$ [$M$] & $2$ & -- & $6$ & -- & $2$ & -- & $6$ & -- \\
			$q_{\rm in}$ & -- & $10^{}_{-6}$ & -- & $3.1^{+0.9}_{-2.8}$ & -- & $10.0^{}_{-0.5}$ & -- & $0.0^{+0.4}_{}$ \\
			$q_{\rm out}$ & -- & $3.14^{+0.07}_{-0.10}$ & -- & $3.3^{+1.0}_{-1.4}$ & -- & $3.52^{+0.10}_{-0.25}$ & -- & $10.0^{}_{-1.8}$ \\
			$r_{\rm br}$ [$M$] & -- & $4.2^{+1.2}_{-0.7}$ & -- & $5.6^{+1.7}_{-1.2}$ & -- & $1.527^{+0.016}_{-0.045}$ & -- & $6.3^{+0.5}_{-0.6}$ \\
			$i$ [deg] & $20$ & $19.6^{+2.5}_{-2.7}$ & $20$ & $18^{+5}_{-3}$ & $70$ & $74.4^{+0.5}_{-0.3}$ & $70$ & $66.7^{+0.4}_{-0.4}$ \\
			$a_*$ & $0.99$ & $0.985^{+\rm (P)}_{-0.008}$ & $0.99$ & $0.91^{\rm +(P)}_{-0.32}$ & $0.99$ & $0.998^{}_{-0.046}$ & $0.99$ & $0.998^{}_{-0.046}$ \\
			$A_{\rm Fe}$ & $1$ & $1.43^{+0.10}_{-0.13}$ & $1$ & $1.66^{+0.15}_{-0.11}$ & $1$ & $2.00^{+0.11}_{-0.06}$ & $1$ & $2.03^{+0.06}_{-0.06}$ \\
			$\Gamma$ & $1.7$ & $1.649^{+0.010}_{-0.010}$ & $1.7$ & $1.638^{+0.011}_{-0.013}$ & $1.7$ & $1.620^{+0.007}_{-0.007}$ & $1.7$ & $1.669^{+0.005}_{-0.005}$ \\
			$\log\xi$ & $3.1$ & $3.148^{+0.012}_{-0.021}$ & $3.1$ & $3.159^{+0.024}_{-0.020}$ & $3.1$ & $3.213^{+0.017}_{-0.019}$ & $3.1$ & $3.328^{+0.003}_{-0.003}$ \\
			$E_{\rm cut}$ [keV] & $300$ & $300^{\star}$ & $300$ & $300^{\star}$ & $300$ & $300^{\star}$ & $300$ & $300{^\star}$ \\
			$\alpha_{13}$ & $1.0$ & $1.3^{+0.4}_{-1.5}$ & $1.0$ & $0.1^{+1.5}_{- \rm (P)}$ & $1.0$ & $0^{+0.005}_{-0.034}$ & $1.0$ & $-0.22^{+1.11}_{-\rm (P)}$ \\
			\hline
			$\chi^2/\nu$ && $\quad 1368.17/1342 \quad$ && $\quad 1535.93/1432 \quad$ && $\quad 1524.94/1401\quad$ && $\quad 1542.48/1480 \quad$ \\
			&& =1.01950 && = 1.07258 && = 1.08847 && =1.04222 \\
			\hline\hline
	\end{tabular}}
	
\vspace{0.8cm}
	\centering
	\caption{\rm Best-fit values for simulations~I-L when we employ the disk's emissivity profile of a lamppost corona. $\xi$ in units of erg~cm~s$^{-1}$. The reported uncertainties correspond to a 90\% confidence level for one relevant parameter ($\Delta\chi^2 = 2.71$). $^\star$ indicates that the parameter is frozen in the fit. (P) indicates that the parameter boundary is within the 90\% confidence level. When there is no upper/lower uncertainty, it means that the parameter is stuck at the upper/lower boundary of the range in which it is allowed to vary. See the text for more details. \label{t-fit6}}
	{\renewcommand{\arraystretch}{1.4}
		\begin{tabular}{lccccccccccc}
			\hline\hline
			& \multicolumn{2}{c}{Simulation~I} & \multicolumn{2}{c}{Simulation~J} & \multicolumn{2}{c}{Simulation~K} & \multicolumn{2}{c}{Simulation~L}\\
			& Input & Fit & Input & Fit & Input & Fit & Input & Fit \\
			\hline
			{\tt tbabs} &&&&&&&& \\
			$N_{\rm H}$ [$10^{20}$~cm$^{-2}$] & $6.74$ & $6.74^\star$ & $6.74$ & $6.74^\star$ & $6.74$ & $6.74^\star$ & $6.74$ & $6.74^\star$ \\
			\hline
			{\tt relxilllp\_nk} &&&&&&&& \\
			$H$ [$M$] & $2$ & -- & $2$ & -- & $2$ & -- & $2$ & -- \\
			$R_{\rm disk}$ [$M$] & $2$ & -- & $6$ & -- & $2$ & -- & $6$ & -- \\
			$h$ [$M$] & -- & $2.78^{+0.10}_{-0.03}$ & -- & $3.0^{+1.3}_{-0.2}$ & -- & $2.13^{+0.20}_{-\rm (P)}$ & -- & $2.43^{+0.20}_{-\rm (P)}$\\
			$i$ [deg] & $20$ & $21.4^{+1.2}_{-2.5}$ & $20$ & $17.0^{+1.0}_{-1.6}$ & $70$ & $73.0^{+1.1}_{-0.6}$ & $70$ & $72.7^{+0.9}_{-0.7}$ \\
			$a_*$ & $0.99$ & $0.998^{}_{-0.13}$ & $0.99$ & $0.997^{+\rm (P)}_{-0.009}$ & $0.99$ & $0.991^{+\rm (P)}_{-0.009}$ & $0.99$ & $0.973^{+0.017}_{-0.015}$ \\
			$A_{\rm Fe}$ & $1$ & $1.38^{+0.10}_{-0.07}$ & $1$ & $1.35^{+0.06}_{-0.03}$ & $1$ & $1.00^{+0.07}_{-0.03}$ & $1$ & $1.09^{+0.08}_{-0.06}$ \\
			$\Gamma$ & $1.7$ & $1.662^{+0.008}_{-0.014}$ & $1.7$ & $1.665^{+0.007}_{-0.011}$ & $1.7$ & $1.690^{+0.007}_{-0.008}$ & $1.7$ & $1.688^{+0.006}_{-0.008}$ \\
			$\log\xi$ & $3.1$ & $3.150^{+0.027}_{-0.009}$ & $3.1$ & $3.136^{+0.017}_{-0.013}$ & $3.1$ & $3.124^{+0.017}_{-0.015}$ & $3.1$ & $3.121^{+0.015}_{-0.009}$ \\
			$E_{\rm cut}$ [keV] & $300$ & $300^{\star}$ & $300$ & $300^{\star}$ & $300$ & $300^{\star}$ & $300$ & $300{^\star}$ \\
			$\alpha_{13}$ & $1.0$ & $0.61^{+0.21}_{-0.19}$ & $1.0$ & $1.1^{+0.3}_{-0.4}$ & $1.0$ & $0.247^{+0.040}_{-0.003}$ & $1.0$ & $0.00^{+0.29}_{-0.19}$ \\
			\hline
			$\chi^2/\nu$ && $\quad 1366.90/1344 \quad$ && $\quad 1528.52/1434 \quad$ && $\quad 1392.21/1403\quad$ && $\quad 1382.38/1482 \quad$ \\
			&& =1.01704 && = 1.06592 && = 0.99230 && = 0.93277 \\
			\hline\hline
	\end{tabular}}
	\vspace{0.5cm}
\end{table*}



\begin{table*}
	\centering
	\caption{\rm Best-fit values for simulations~B,  F,  J when we model the disk's emissivity profile with a disk-like corona. $\xi$ in units of erg~cm~s$^{-1}$. The reported uncertainties correspond to a 90\% confidence level for one relevant parameter ($\Delta\chi^2 = 2.71$). $^\star$ indicates that the parameter is frozen in the fit. (P) indicates that the parameter boundary is within the 90\% confidence level. When there is no upper/lower uncertainty, it means that the parameter is stuck at the upper/lower boundary of the range in which it is allowed to vary. See the text for more details. \label{t-fit7}} 
	{\renewcommand{\arraystretch}{1.4}
		\begin{tabular}{lcccccc}
			\hline\hline
			& \multicolumn{2}{c}{Simulation~B} & \multicolumn{2}{c}{ Simulation~F} & \multicolumn{2}{c}{Simulation~J} \\
			& Input & Fit & Input & Fit & Input & Fit \\
			\hline
			{\tt tbabs} &&&&&& \\
			$N_{\rm H}$ [$10^{20}$~cm$^{-2}$] & $6.74$ & $6.74^\star$ & $6.74$ & $6.74^\star$ & $6.74$ & $6.74^\star$  \\
			\hline
			{\tt relxilldisk\_nk} &&&&&& \\
			$H$ [$M$] & $2$ & $2.00^{+0.21}_{}$ & $2$ & $2.31^{+0.29}_{-0.20}$ & $2$ & $2.3^{+0.4}_{-\rm (P)}$   \\
			$R_{\rm disk}$ [$M$] & $6$ & $5.7^{+0.5}_{-0.4}$ & $6$ & $5.28^{+0.19}_{-0.69}$ & $6$ & $5.1^{+0.6}_{-0.9}$  \\
			$i$ [deg] & $20$ & $20.6^{+0.7}_{-0.7}$ & $20$ & $18.6^{+1.1}_{-0.9}$ & $20$ & $19.1^{+2.1}_{-0.9}$  \\
			$a_*$ & $0.99$ & $0.9920^{+0.0019}_{-0.0050}$ & $0.99$ & $0.989^{+\rm (P)}_{-0.054}$ & $0.99$ & $0.981^{+\rm (P)}_{-\rm (P)}$   \\
			$A_{\rm Fe}$ & $1$ & $0.996^{+0.041}_{-0.013}$ & $1$ & $0.992^{+0.018}_{-0.014}$ & $1$ & $1.36^{+0.05}_{-0.03}$  \\
			$\Gamma$ & $1.7$ & $1.710^{+0.006}_{-0.008}$ & $1.7$ & $1.706^{+0.006}_{-0.008}$ & $1.7$ & $1.666^{+0.006}_{-0.010}$  \\
			$\log\xi$ & $3.1$ & $3.088^{+0.012}_{-0.008}$ & $3.1$ & $3.104^{+0.014}_{-0.020}$& $3.1$ & $3.139^{+0.015}_{-0.013}$ \\
			$E_{\rm cut}$ [keV] & $300$ & $300^{\star}$ & $300$ & $300^{\star}$ & $300$ & $300^{\star}$ \\
			$\alpha_{13}$ & $0$ & $-0.0350^{+0.2719}_{-0.0016}$ & $-0.24$ & $-0.178^{+0.003}_{-0.074}$ & $1.0$ & $0.5^{+0.3}_{-\rm (P)}$  \\
			\hline
			$\chi^2/\nu$ && $\quad 1368.85/1386 \quad$ && $\quad 1307.89/1375 \quad$ && $\quad 1524.45/1433\quad$  \\
			&& =0.98762 && = 0.95119 && = 1.06382 \\
			\hline\hline
	\end{tabular}}
	
\vspace{0.8cm}
	\centering
	\caption{\rm Best-fit values for simulations~D,  H,  L when we employ the disk's emissivity profile of a disk-like corona. $\xi$ in units of erg~cm~s$^{-1}$. The reported uncertainties correspond to a 90\% confidence level for one relevant parameter ($\Delta\chi^2 = 2.71$). $^\star$ indicates that the parameter is frozen in the fit. (P) indicates that the parameter boundary is within the 90\% confidence level. When there is no upper/lower uncertainty, it means that the parameter is stuck at the upper/lower boundary of the range in which it is allowed to vary. See the text for more details. \label{t-fit8}}
	{\renewcommand{\arraystretch}{1.4}
		\begin{tabular}{lcccccccc}
			\hline\hline
			& \multicolumn{2}{c}{Simulation~D} & \multicolumn{2}{c}{Simulation~H} & \multicolumn{2}{c}{Simulation~L} \\
			& Input & Fit & Input & Fit & Input & Fit\\
			\hline
			{\tt tbabs} &&&&& \\
			$N_{\rm H}$ [$10^{20}$~cm$^{-2}$] & $6.74$ & $6.74^\star$ & $6.74$ & $6.74^\star$ & $6.74$ & $6.74^\star$  \\
			\hline
			{\tt relxilldisk\_nk} &&&&& \\
			$H$ [$M$] & $2$ & $3.7^{+0.8}_{-0.7}$ & $2$ & $2.5^{+0.7}_{-0.4}$ & $2$ & $2.13^{+0.24}_{-0.07}$  \\
			$R_{\rm disk}$ [$M$] & $6$ & $6.7^{+1.2}_{-2.5}$ & $6$ & $9.0^{+1.6}_{-2.5}$ & $6$ & $4.5^{+0.7}_{-1.2}$   \\
			$i$ [deg] & $70$ & $72.5^{+1.0}_{-1.1}$ & $70$ & $70.2^{+0.3}_{-0.3}$ & $70$ & $72.8^{+0.4}_{-0.4}$  \\
			$a_*$ & $0.99$ & $0.998^{}_{-0.010}$ & $0.99$ & $0.998^{}_{-0.294}$ & $0.99$ & $0.998^{}_{-0.006}$  \\
			$A_{\rm Fe}$ & $1$ & $0.942^{+0.021}_{-0.017}$ & $1$ & $0.969^{+0.017}_{-0.014}$ & $1$ & $1.23^{+0.06}_{-0.04}$  \\
			$\Gamma$ & $1.7$ & $1.701^{+0.005}_{-0.005}$ & $1.7$ & $1.696^{+0.003}_{-0.005}$ & $1.7$ & $1.681^{+0.005}_{-0.006}$  \\
			$\log\xi$ & $3.1$ & $3.102^{+0.007}_{-0.008}$ & $3.1$ & $3.111^{+0.006}_{-0.005}$ & $3.1$ & $3.127^{+0.011}_{-0.006}$  \\
			$E_{\rm cut}$ [keV] & $300$ & $300^{\star}$ & $300$ & $300^{\star}$ & $300$ & $300^{\star}$ \\
			$\alpha_{13}$ & $0$ & $-0.21^{+0.62}_{-0.13}$ & $-0.24$ & $0.0^{+0.3}_{-0.3}$ & $1.0$ & $-0.09^{+1.97}_{-0.13}$  \\
			\hline
			$\chi^2/\nu$ && $\quad 1510.62/1460 \quad$ && $\quad 1385.10/1468 \quad$ && $\quad 1382.82/1481\quad$ \\
			&& =1.03467 && = 0.94352 && = 0.93371  \\
			\hline\hline
	\end{tabular}}
	\vspace{0.5cm}
\end{table*}


\begin{figure*}[t]
	\begin{center}
		\includegraphics[width=0.30\textwidth,trim={0cm 0cm 0cm 0cm},clip, angle=270]{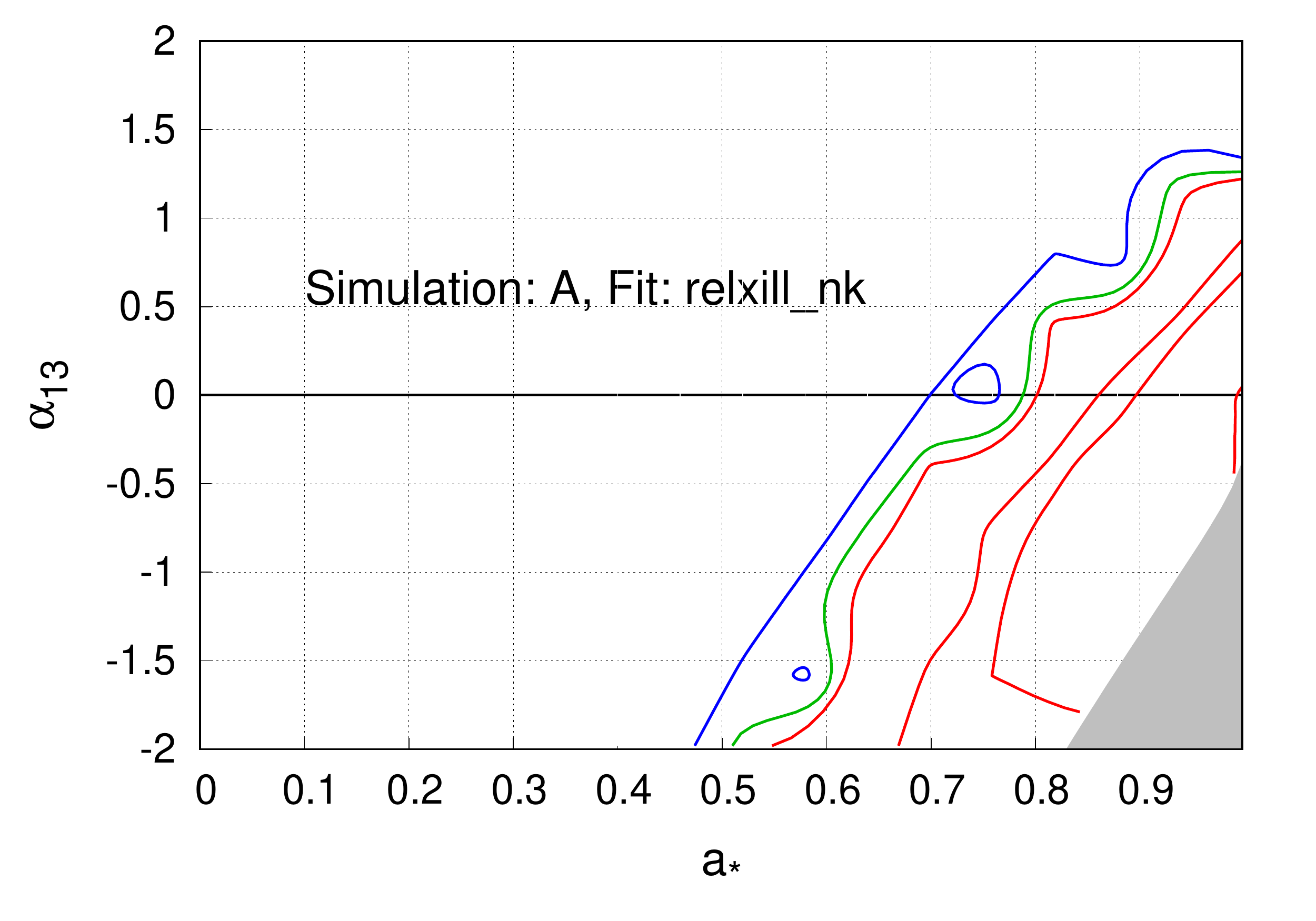}
		\hspace{-0.4cm}
		\includegraphics[width=0.30\textwidth,trim={0cm 0cm 0cm 0cm},clip, angle=270]{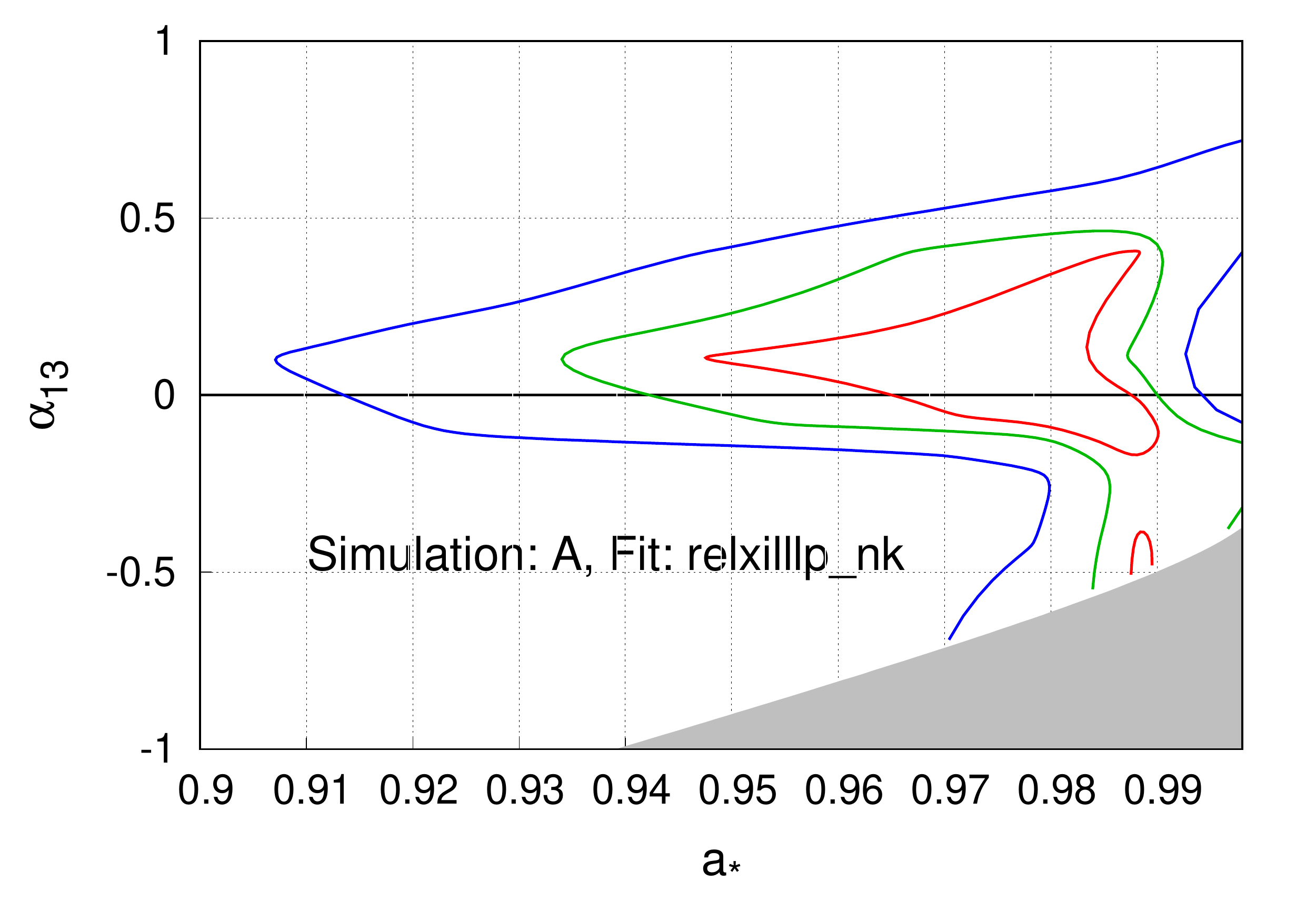}
	\vspace{-0.2cm}
			\includegraphics[width=0.30\textwidth,trim={0cm 0cm 0cm 0cm},clip, angle=270]{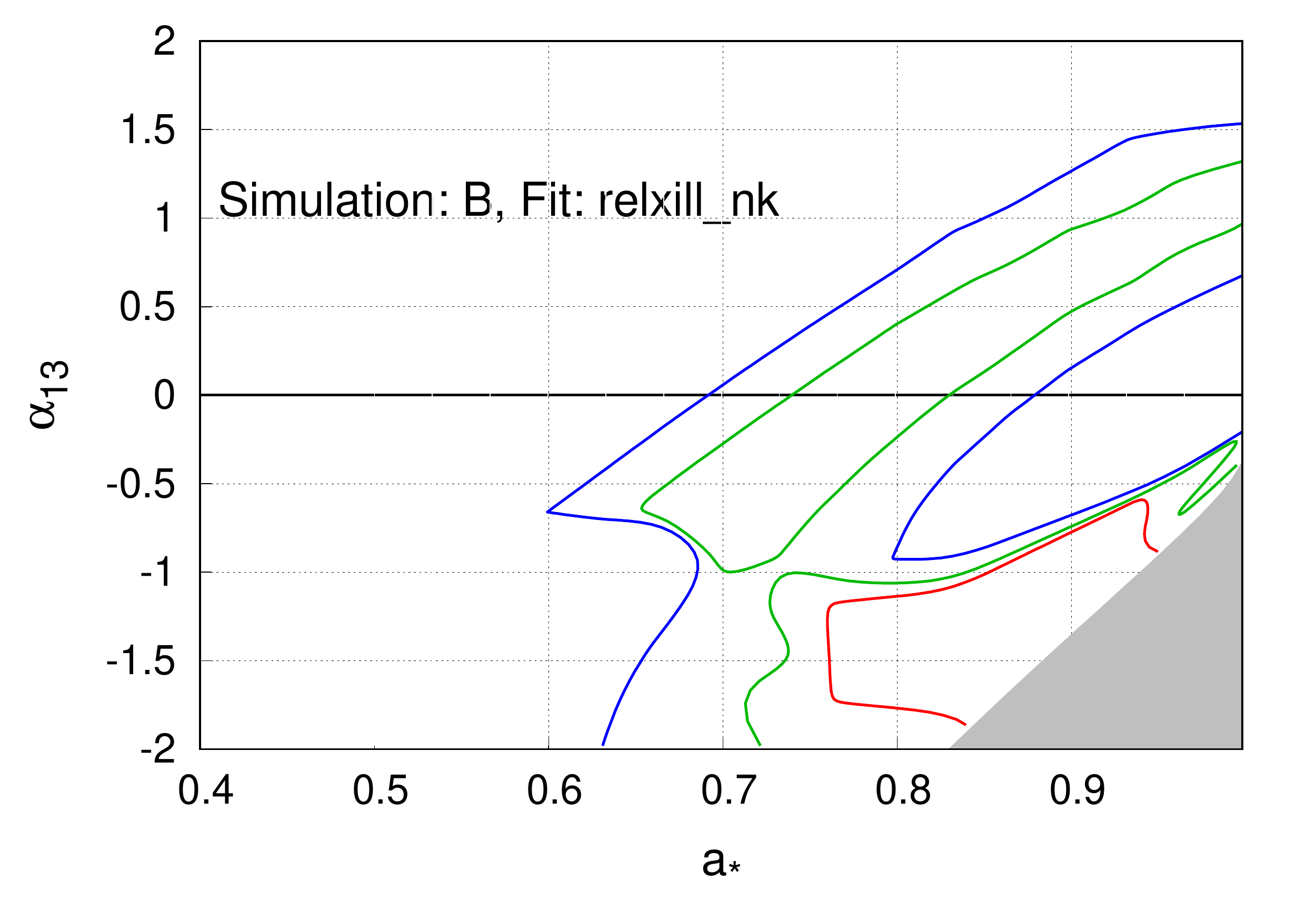}
		\hspace{-0.3cm}
		\includegraphics[width=0.30\textwidth,trim={0cm 0cm 0cm 0cm},clip, angle=270]{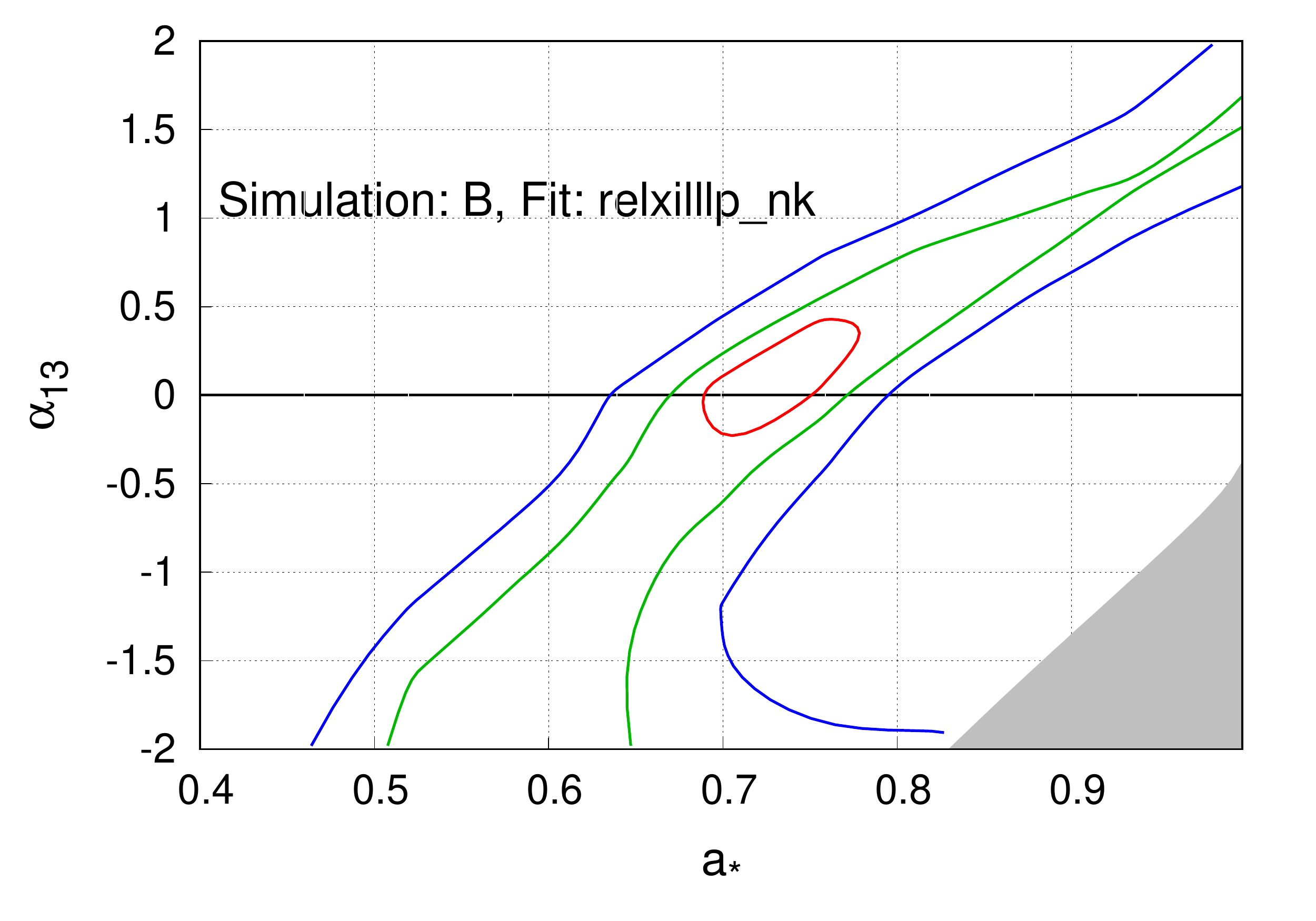}
		\vspace{-0.3cm}
		\includegraphics[width=0.30\textwidth,trim={0cm 0cm 0cm 0cm},clip, angle=270]{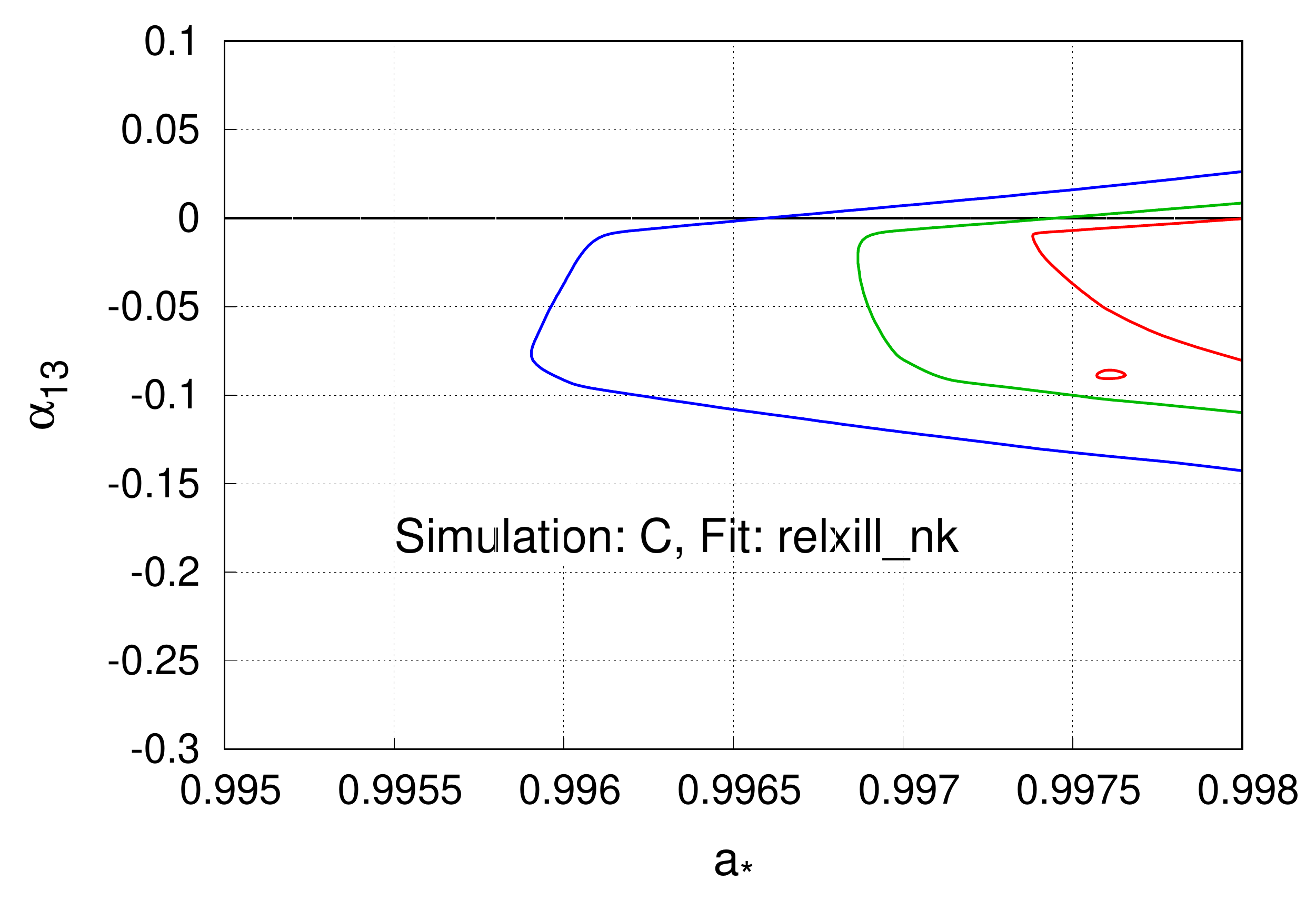}
		\hspace{-0.3cm}
		\includegraphics[width=0.30\textwidth,trim={0cm 0cm 0cm 0cm},clip, angle=270]{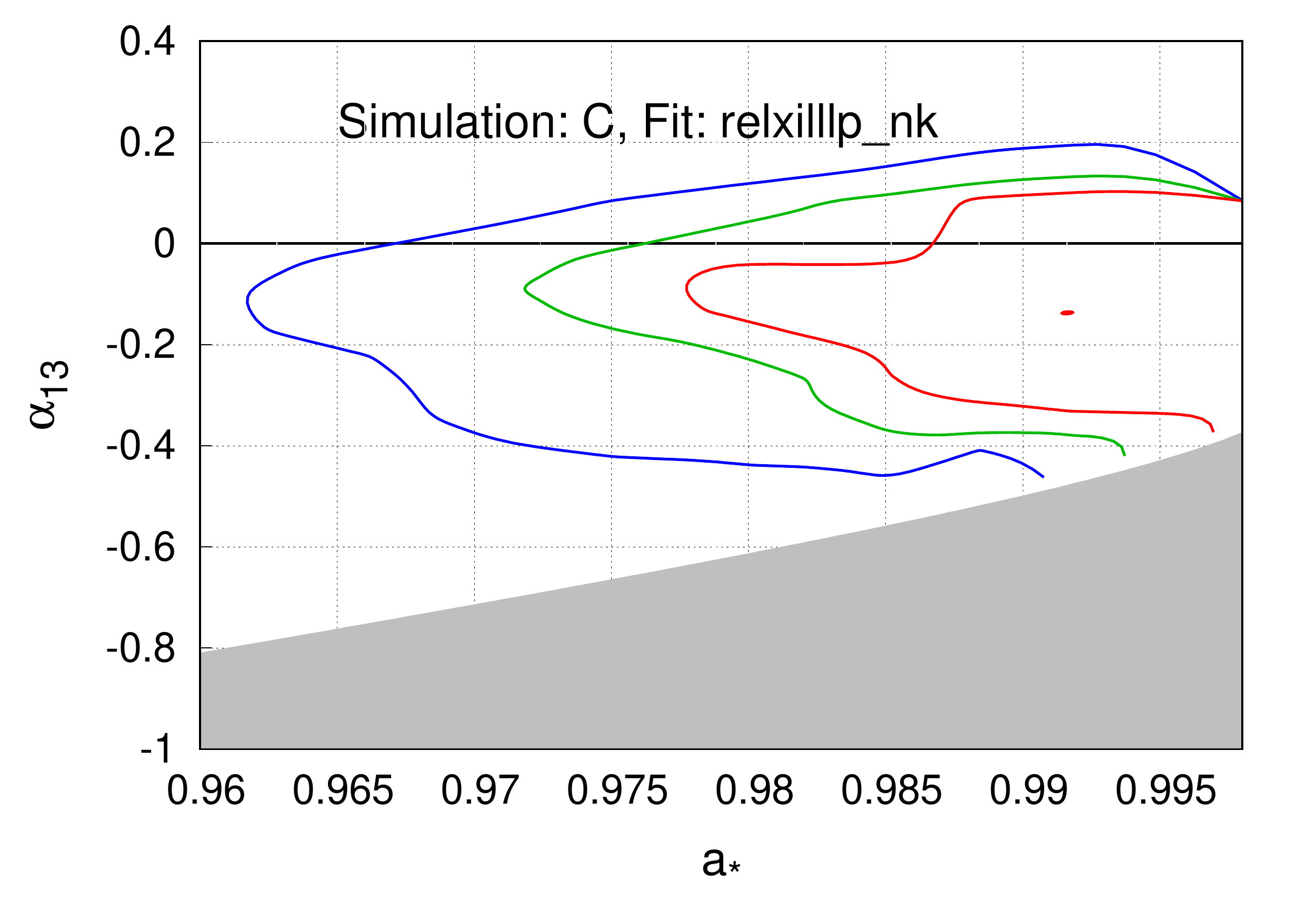}
		\vspace{-0.3cm}
		\includegraphics[width=0.30\textwidth,trim={0cm 0cm 0cm 0cm},clip, angle=270]{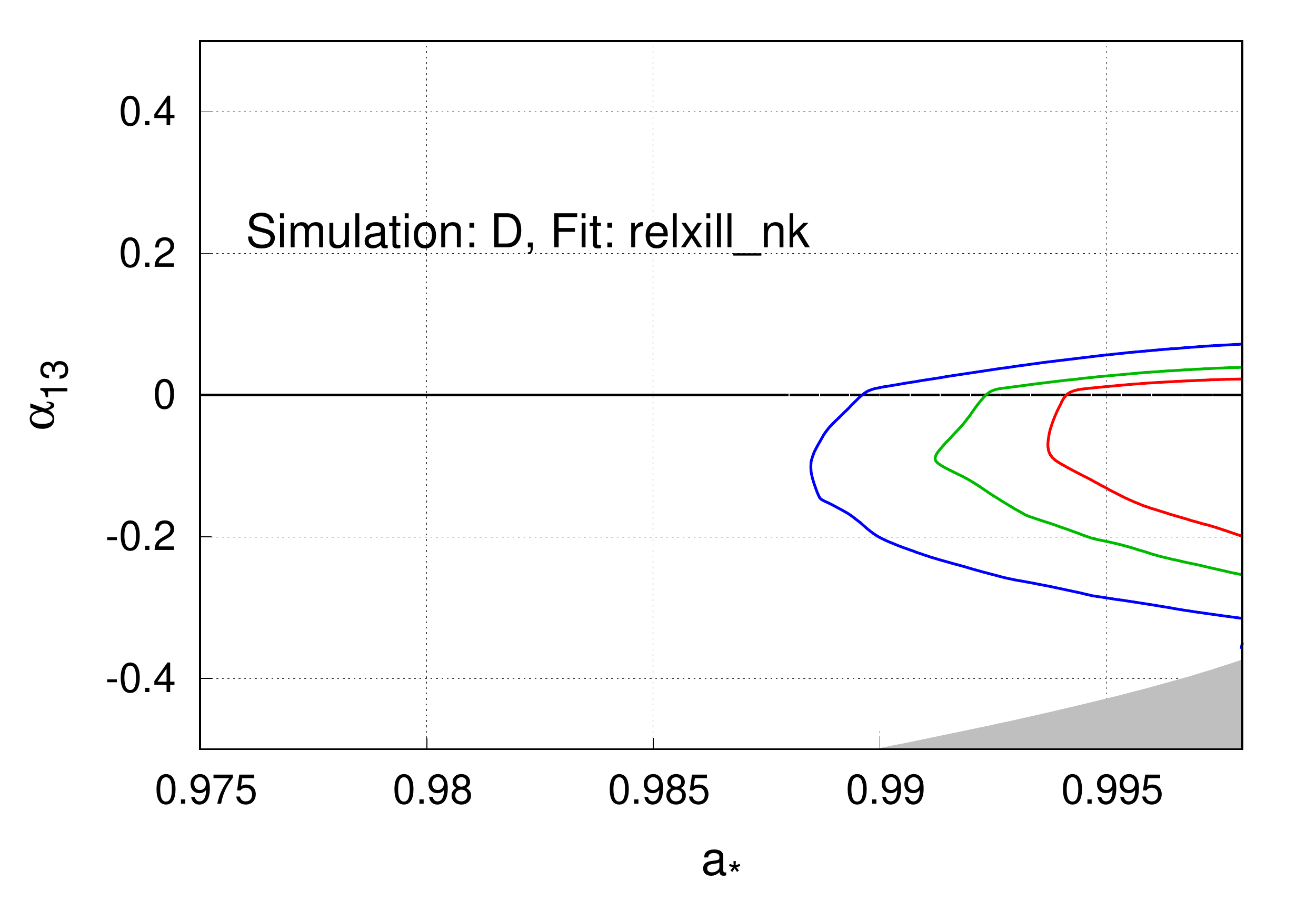}
		\hspace{-0.3cm}
		\includegraphics[width=0.30\textwidth,trim={0cm 0cm 0cm 0cm},clip, angle=270]{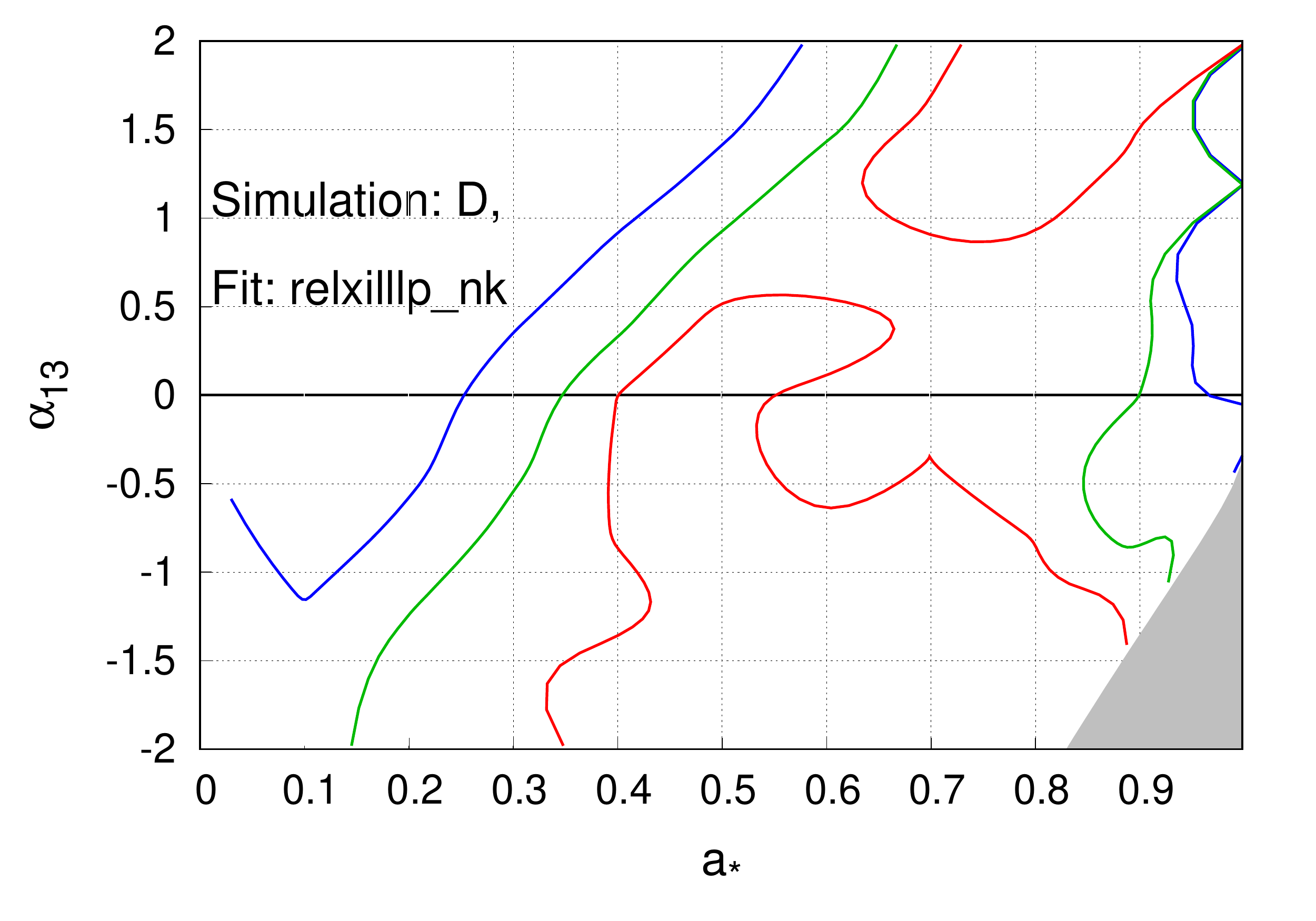}

	\end{center}
	\vspace{-0.3cm}
	{\caption{Constraints on the black hole spin parameter $a_*$ and the deformation parameter $\alpha_{13}$ for simulations~A-D.  The red, green,  and blue curves show,  respectively,   the 68\%, 90\% and 99\% confidence level limits for the two relevant parameters ($\Delta \chi^2$ = 2.30, 4.61, and 9.21).  See the text for more details.  \label{f-contours-A-D}}}
	\vspace{0.4cm}
\end{figure*}


\begin{figure*}[t]
	\begin{center}
		\includegraphics[width=0.28\textwidth,trim={0cm 0cm 0cm 0cm},clip, angle=270]{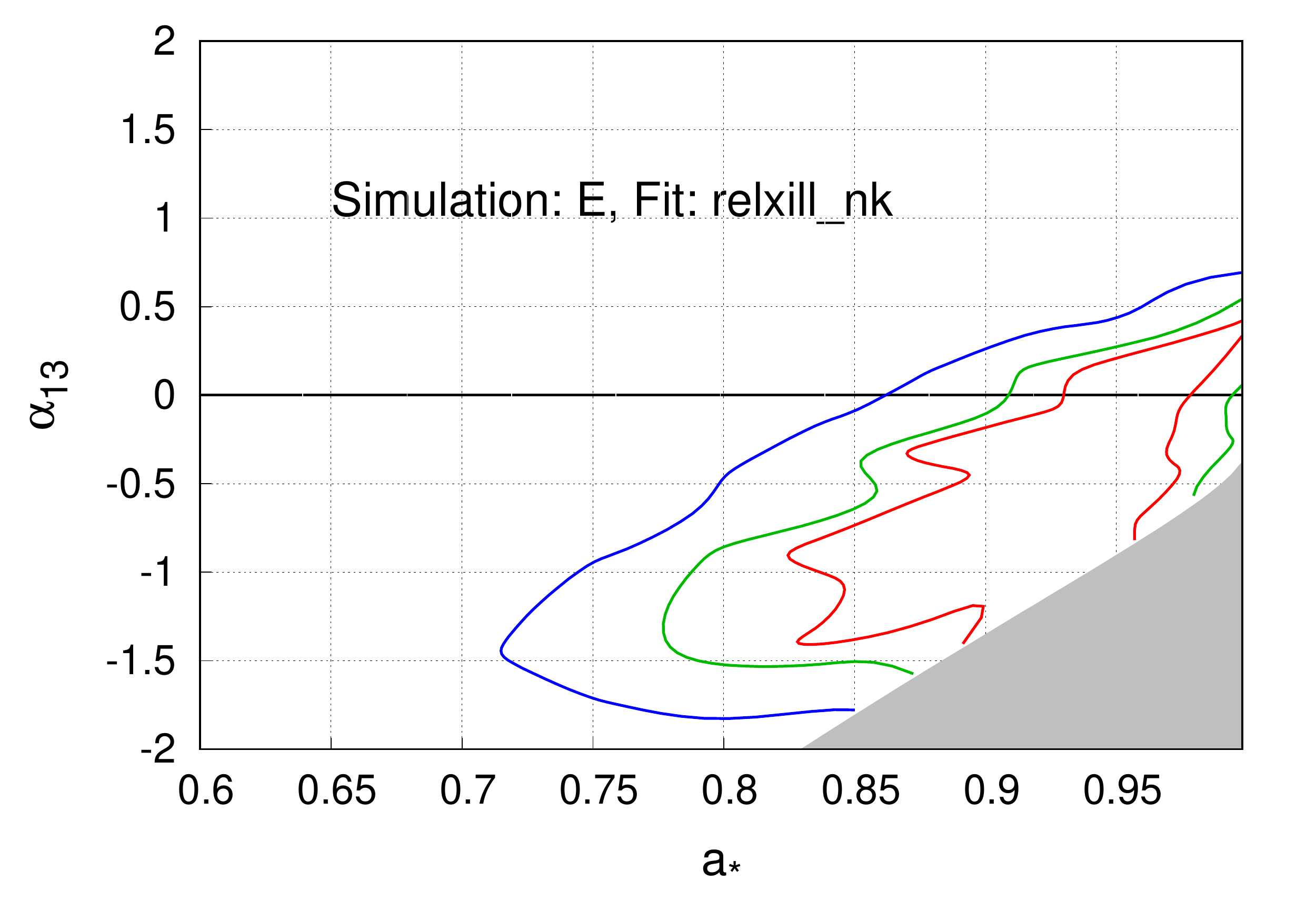}
		\hspace{-0.4cm}
		\includegraphics[width=0.28\textwidth,trim={0cm 0cm 0cm 0cm},clip, angle=270]{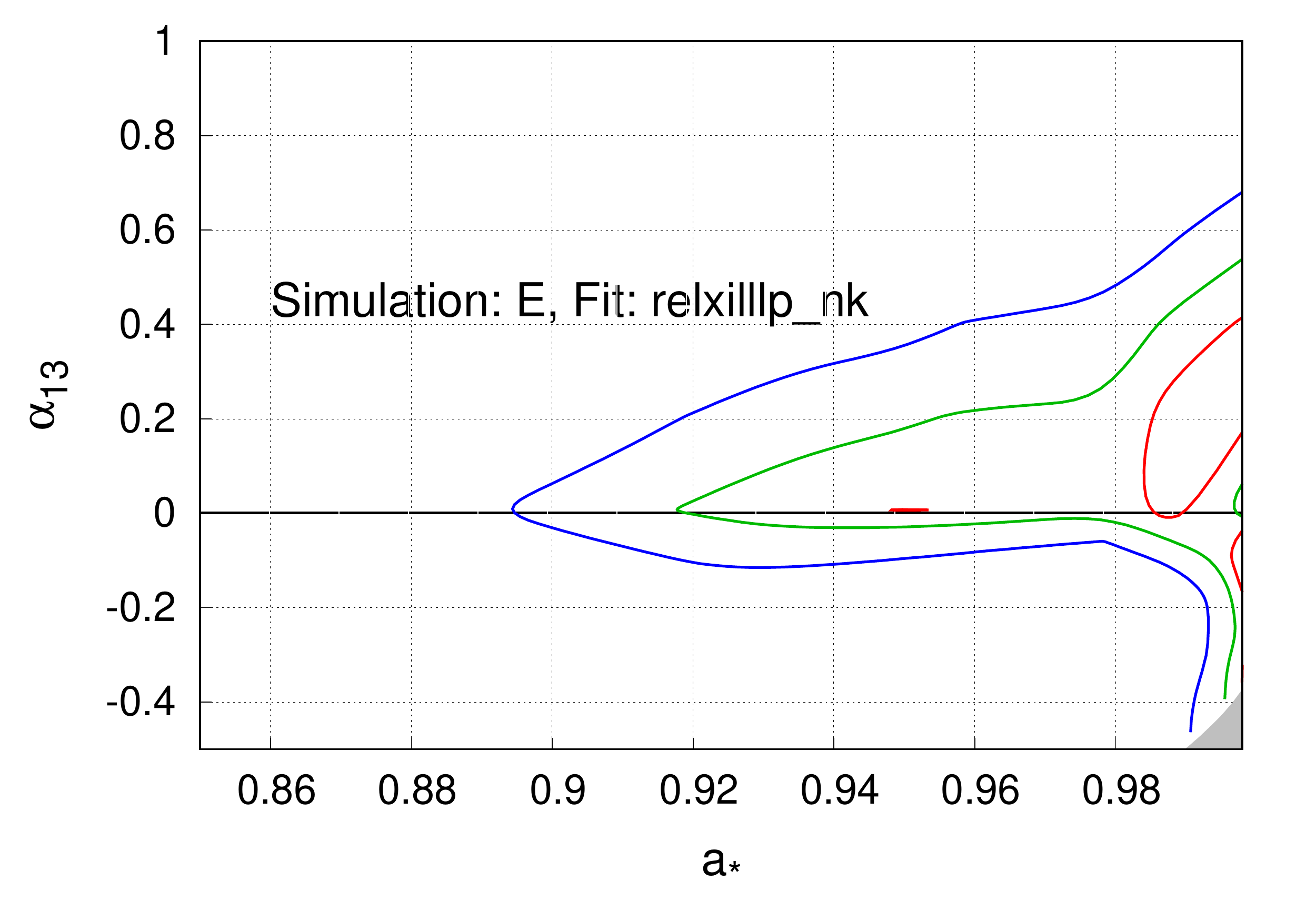}
	\vspace{-0.2cm}
			\includegraphics[width=0.28\textwidth,trim={0cm 0cm 0cm 0cm},clip, angle=270]{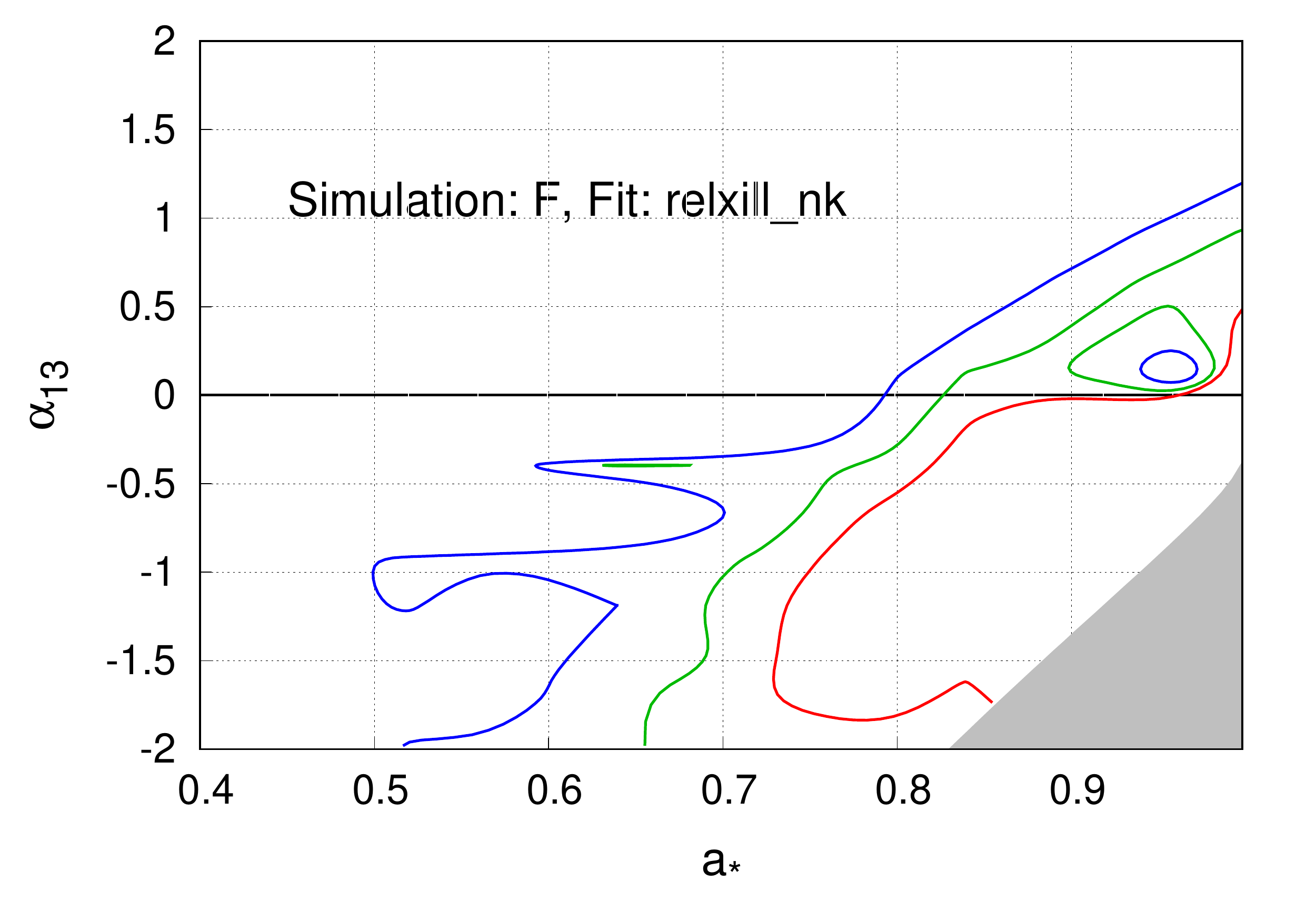}
		\hspace{-0.3cm}
		\includegraphics[width=0.28\textwidth,trim={0cm 0cm 0cm 0cm},clip, angle=270]{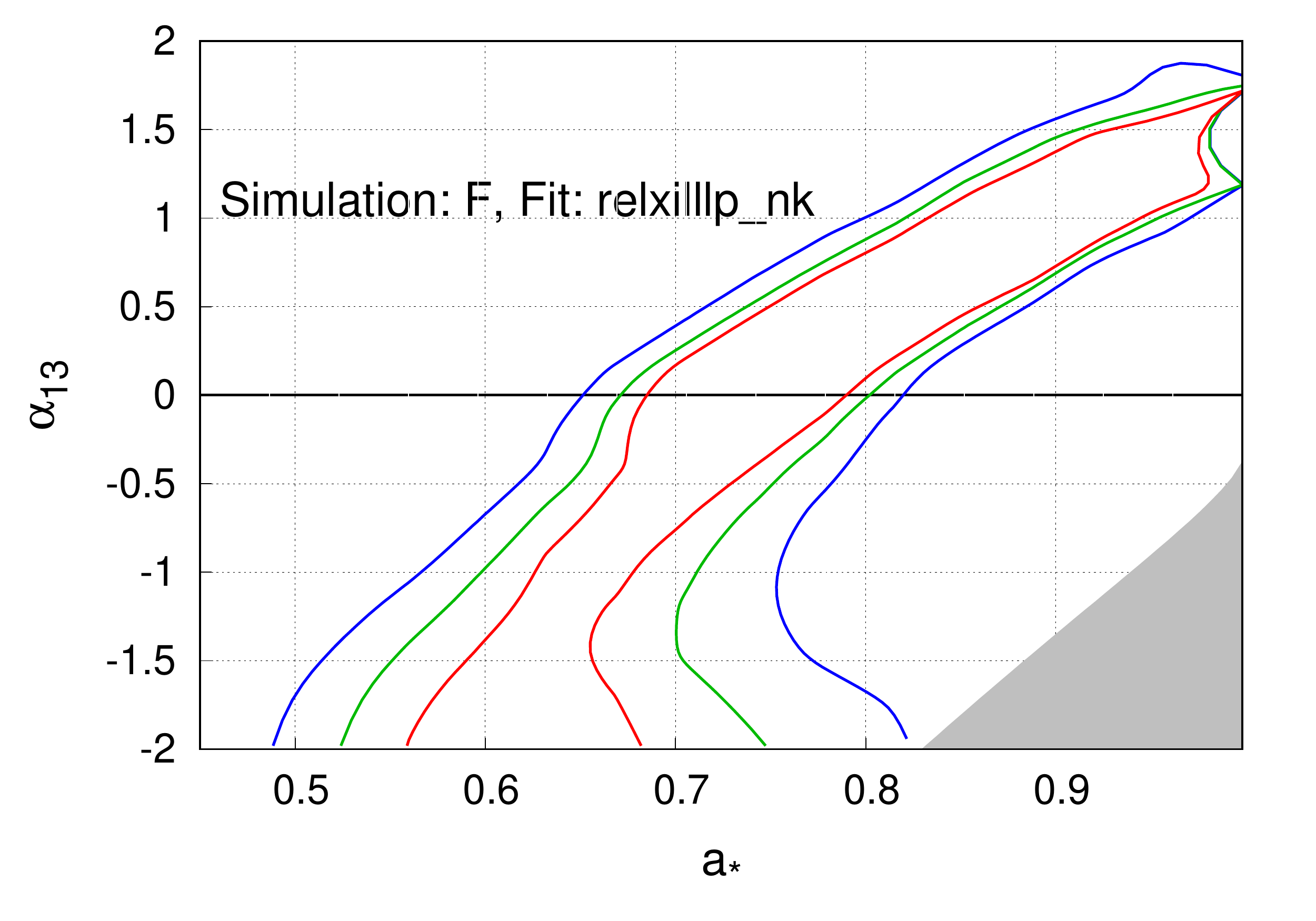}
		\vspace{-0.3cm}
		\includegraphics[width=0.28\textwidth,trim={0cm 0cm 0cm 0cm},clip, angle=270]{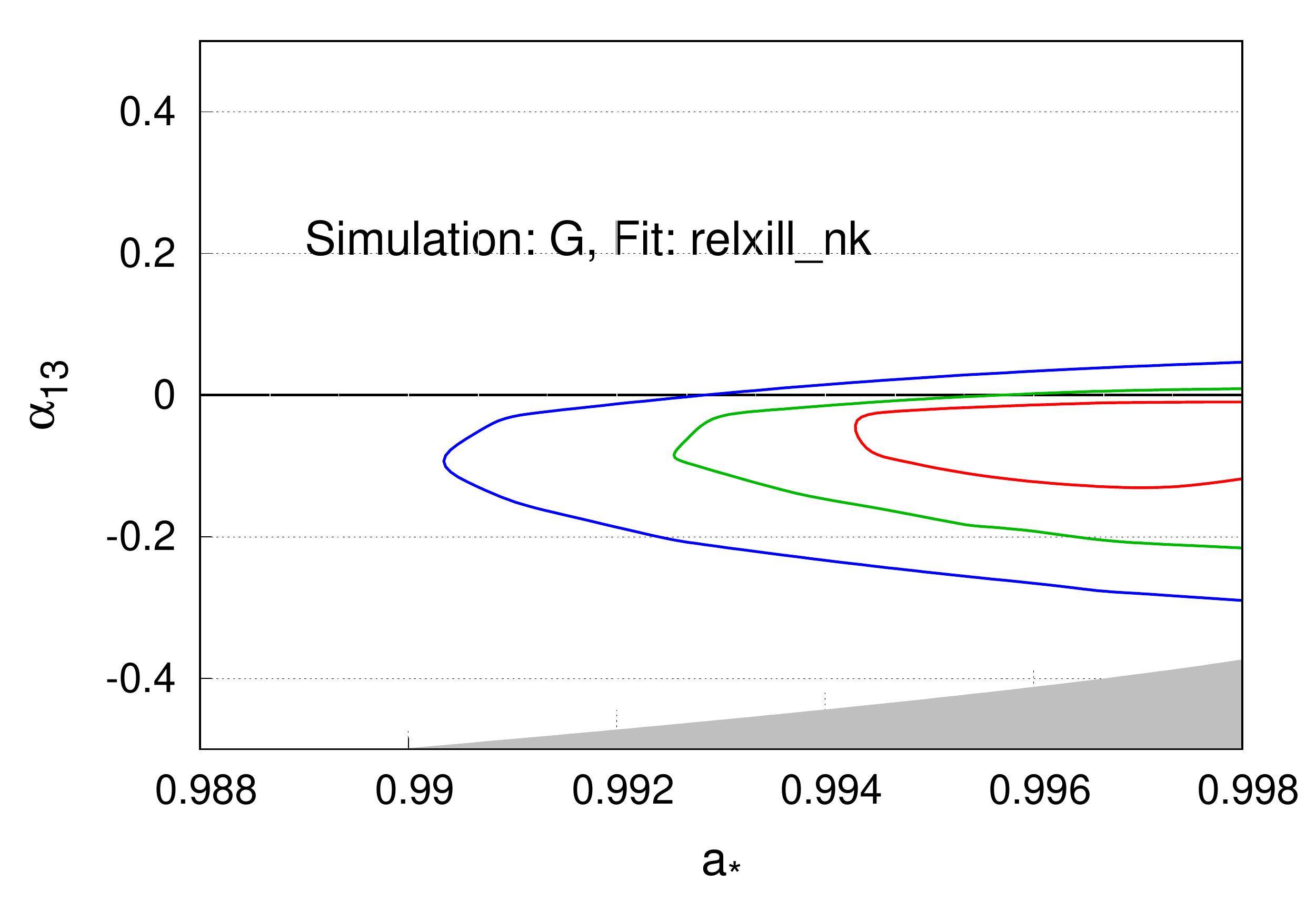}
		\hspace{-0.3cm}
		\includegraphics[width=0.28\textwidth,trim={0cm 0cm 0cm 0cm},clip, angle=270]{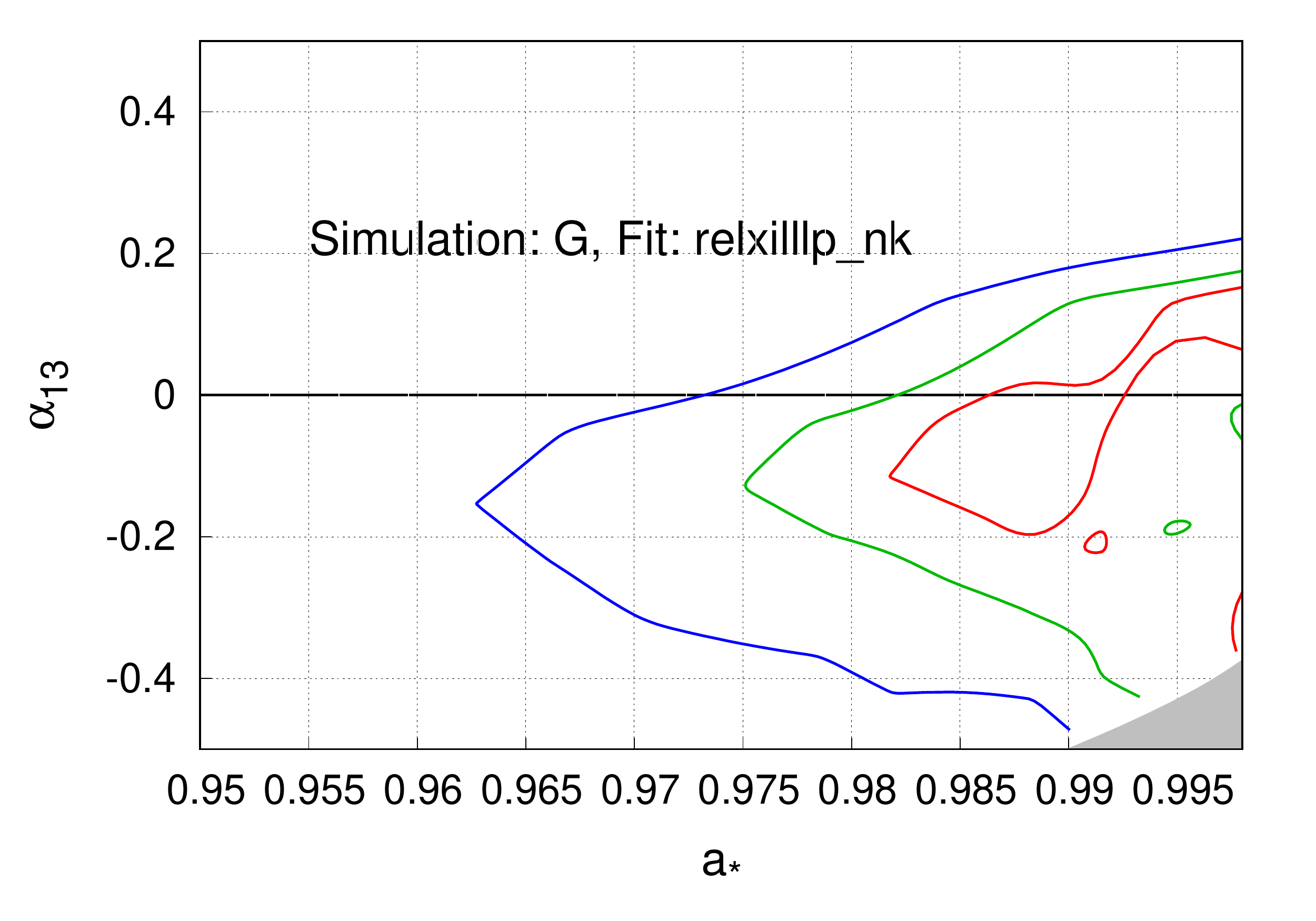}
		\vspace{-0.3cm}
		\includegraphics[width=0.28\textwidth,trim={0cm 0cm 0cm 0cm},clip, angle=270]{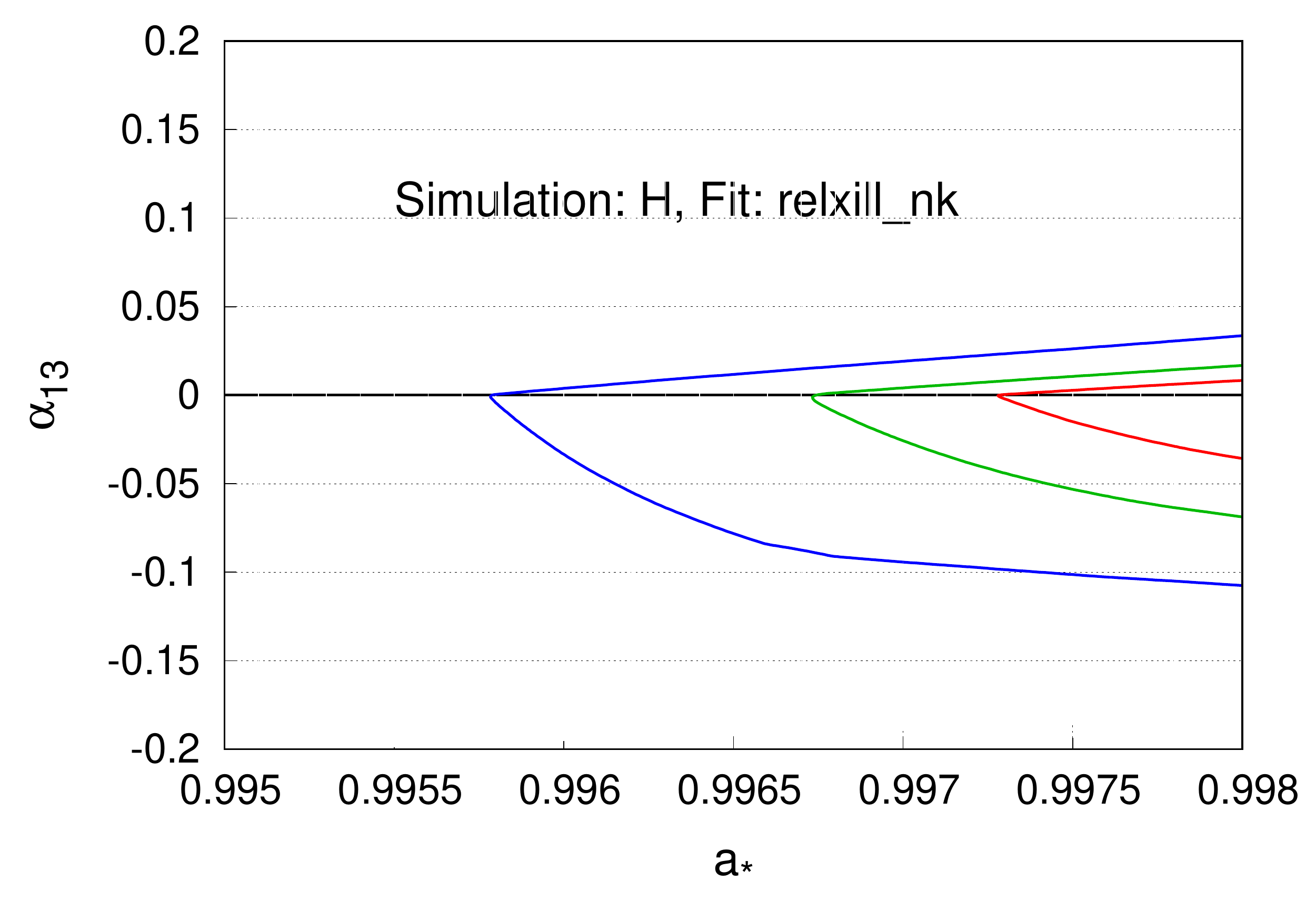}
		\hspace{-0.3cm}
		\includegraphics[width=0.28\textwidth,trim={0cm 0cm 0cm 0cm},clip, angle=270]{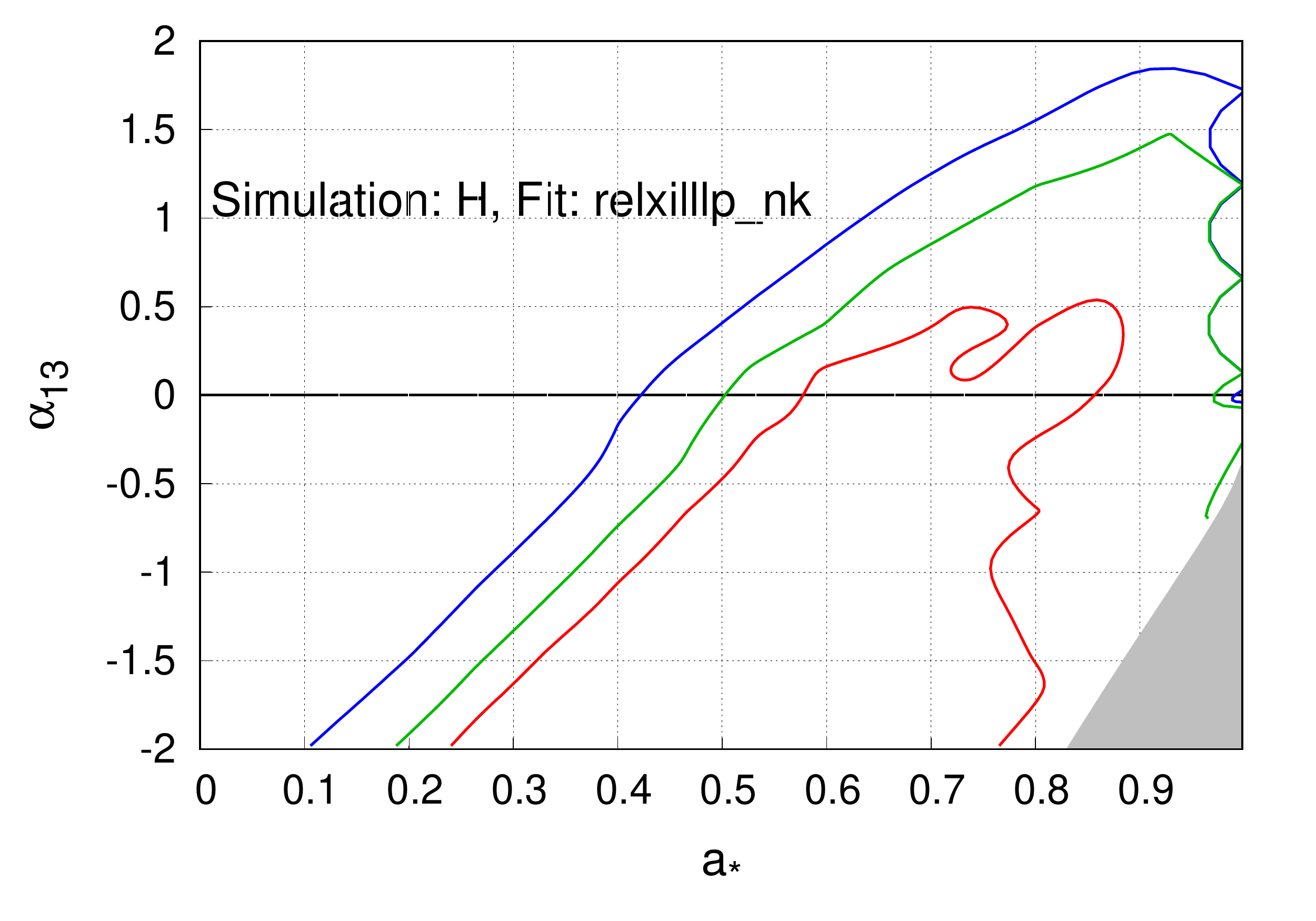}

	\end{center}
	\vspace{-0.3cm}
	{\caption{Constraints on the black hole spin parameter $a_*$ and the deformation parameter $\alpha_{13}$ for simulations~E-H.  The red, green,  and blue curves show,  respectively,   the 68\%, 90\% and 99\% confidence level limits for the two relevant parameters ($\Delta \chi^2$ = 2.30, 4.61, and 9.21).  See the text for more details.  \label{f-contours-E-H}}}
	\vspace{0.4cm}
\end{figure*}

\begin{figure*}[t]
	\begin{center}
		\includegraphics[width=0.90\textwidth,trim={0cm 0.5cm 0cm 0cm},clip]{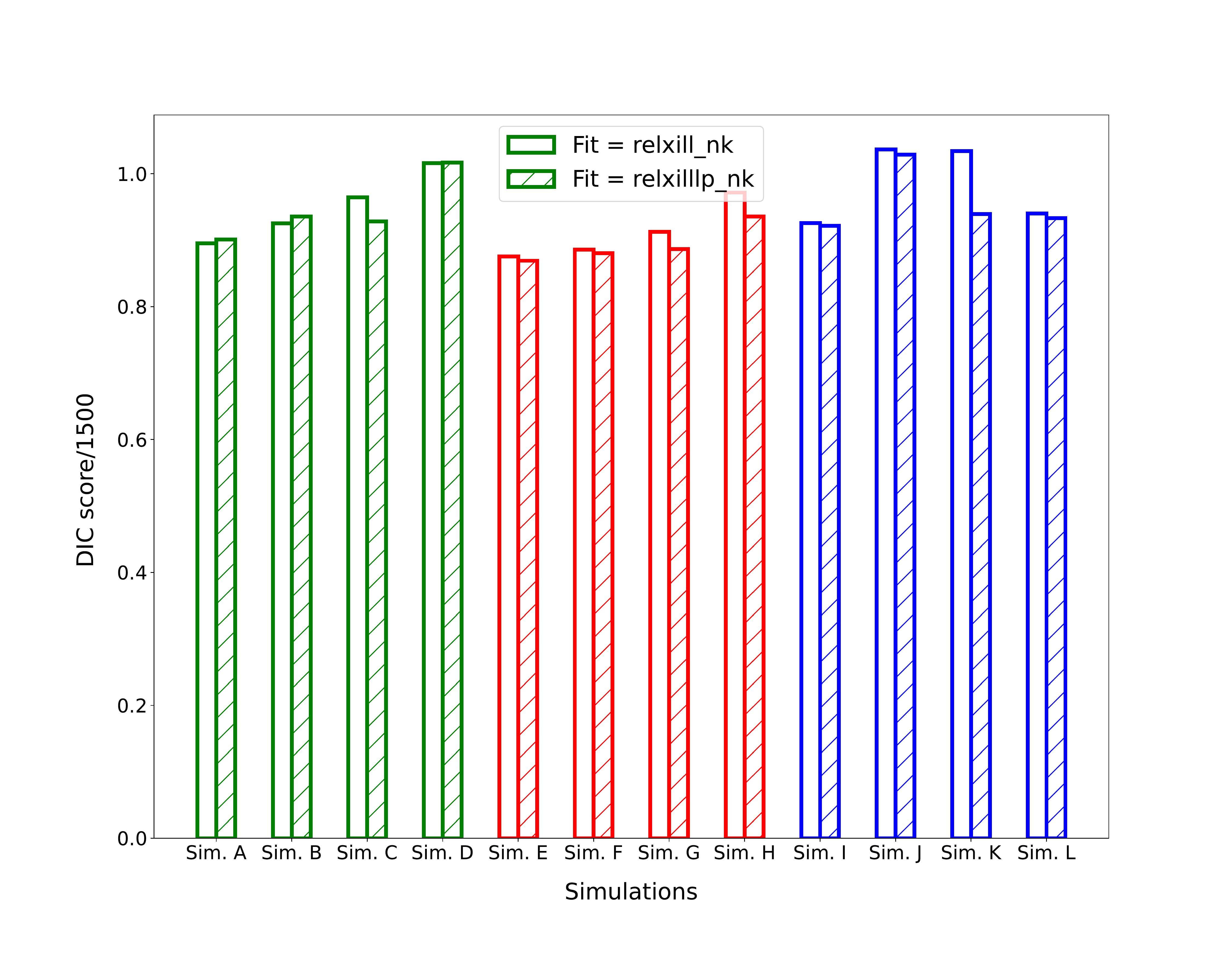}
	\end{center}
	\vspace{-0.5 cm}
	\caption{{Deviance information criteria for model selection for simulations~A-L. The green, red, and blue colors are, respectively, for the simulations with $\alpha_{13}= 0$, $-0.24$, and 1.0. The model with a lower DIC score is preferred.} \label{f-DIC}}
	\vspace{0.4cm}
\end{figure*}

\section{Discussion and conclusions}\label{s-conclusions}

The geometry of the corona determines the emissivity profile of the accretion disk, which is quite an important ingredient when we model the relativistic reflection component in the X-ray spectrum of a black hole. Current data analyses usually model the emissivity profile with a power-law or a broken power-law, or assume a lamppost coronal geometry, namely a point-like source along the black hole spin axis at a height $h$ from the equatorial plane.  In this work, we have considered the possibility that the corona has the shape of infinitesimally thin disk, its central axis the same as the rotational axis of the black hole, and at a height $H$ from the equatorial plane. Within this working hypothesis, we calculated the emissivity profile produced by such a disk-like corona for different values of the coronal radius $R_{\rm disk}$ and the coronal height $H$ in the Johannsen spacetime and we calculated the resulting iron line profiles. While we initially considered that the corona could be either static or corotating with the accretion disk, we then focused our study to the static corona case as the two scenarios lead to quite similar emissivity profiles.

Assuming the static disk-like coronal geometry, we have simulated 12 \textsl{NuSTAR} observations (4 simulations in the Kerr metric and 8 simulations with a non-vanishing deformation parameter $\alpha_{13}$) of a putative bright black hole binary. We have considered two possible viewing angles, $i = 20^\circ$ and $70^\circ$, and two possible values for the coronal radius, $R_{\rm disk} = 2$~$M$ and 6~$M$.  In all simulations, we considered a fast-rotating black hole ($a_* = 0.99$) and a low coronal height ($H = 2$~$M$),  because these are the properties that maximize the relativistic features in the reflection spectrum and, in turn, our ability to test the Kerr metric. The simulated observations are then fitted with a theoretical model that either assumes a broken power-law emissivity profile or employs the emissivity profile of a lamppost corona in order to determine the capability of recovering the correct input parameters in the case that the actual coronal geometry is an infinitesimally thin disk.  Some of the simulations are also fitted with the model that assumes a disk-like corona ({\tt relxilldisk\_nk}).

{First, we discuss the results of the simulations fitted with the correct model, i.e. {\tt relxilldisk\_nk}. The best-fit parameters and the data to best-fit model ratio plots are shown in Tabs.~\ref{t-fit7} and~\ref{t-fit8}, and Fig.~\ref{f-ratio-fitdisk}, respectively.  We choose the simulations with the larger extent of the disk-like corona to fit with the model {\tt relxilldisk\_nk}  because it would allow us to determine how reliably the extent of a corona can be measured and also to determine whether the irradiation of the accretion disk by a disk-like corona intrinsically limits the capability to estimate the deviation from the Kerr metric. The quality of the fit for these simulations is good as the reduced $\chi^2$ is close to 1,  and we do not see unresolved features in the ratio plots except in simulation J; we will return to this simulation shortly. For these simulations, in general,  the model tends to estimate the values of most of the parameters quite close to their input values within the 90\% confidence level. The two crucial parameters of the model, $a_*$ and $\alpha_{13}$, are recovered and constrained well in most of the simulations except in simulations J and L. In simulation L, we find a weaker constraint on $\alpha_{13}$; the positive bound almost reaches the upper limit of the deformation parameter in the grid. Now we return to simulations J; we see some unresolved features in the ratio plot, and the model cannot constrain the height of the corona, the spin of the black hole, and the deformation parameter. This is because, for a given black hole spin, as we increase the value of $\alpha_{13}$, the inner edge of the disk moves to a more considerable distance from the black hole. It would result in a narrower iron line -- high redshifted photons coming from close to the black hole depositing in the low energy tail of the iron line are missing-- which is not suitable for the model to recover and constrain well the geometric properties of the spacetime~\citep{Abdikamalov:2019yrr,Dauser:2013xv}.   Furthermore, for all these simulations, we notice a small deviation from the input value (especially for $\alpha_{13}$), which gives the magnitude of the bias probably associated with the inclusion of Poisson noise and response of the instrument in the synthetic spectra during the {\tt fakeit} procedure in {\tt xspec}.}

{Now we discuss the results of the simulations fitted with a broken power law model {\tt relxill\_nk}.  As we can see from the reduced $\chi^2$ in Tabs.~\ref{t-fit1}, \ref{t-fit3}, and \ref{t-fit5}, as well as from the data to best-fit model ratios (left panels in Figs.~\ref{f-ratio-a13_0}, \ref{f-ratio-a13_-0.5}, and \ref{f-ratio-a13_0.5}),  the theoretical models with broken power-law emissivity profile fit the data well. Overall, in these simulations, the astrophysical properties of the accreting system are recovered well. On the other hand, the geometric properties of the system are sometimes difficult to recover or constrain. In the simulations that are performed assuming the Kerr spacetime with a low inclination angle $i = 20^{\circ}$ (simulations A and B), the spin and the deformation parameter are not constrained well (see Fig.~\ref{f-contours-A-D}). In the case of a high inclination angle $i = 70^{\circ}$ (simulations C and D), the best-fit value of the black hole spin parameter is stuck at the maximum allowed value, and the model can recover and constrain well the deformation parameter. For the simulations with a non-vanishing $\alpha_{13}$, we find a similar trend: it is difficult to constrain the black hole spin and the deformation parameter when the inclination angle is low, while the two parameters can be constrained well for a high inclination angle. The contour plots for the simulations I-L are qualitatively similar to those of A-H and are therefore omitted here.  The inclination angle is recovered quite well for all these fits. The results of these simulations suggest that, with current X-ray missions like \textsl{NuSTAR}, it may be challenging to determine the actual coronal geometry from the analysis of the reflection spectrum. We have simulated a few observations (not shown here) of a similar source with the X-IFU instrument~\citep{2013arXiv1308.6784B}, which is expected to be onboard of \textsl{Athena}~\citep{2013arXiv1306.2307N}, and with \textsl{NICER} \citep{NICER}, and found that the data to best-fit model ratios show clear residuals. Thus, our conclusions with \textsl{NuSTAR} may not apply for other X-ray missions. The residuals found in the case of \textsl{Athena} and \textsl{NICER} are due to the high statistics at low energies, which then tends to drive the fit and so we obtain a poor quality of the fit near the iron line. }

{Now we discuss the results of the simulations fitted with the lamppost corona model.  The quality of the fits are good,  the reduced $\chi^2$ is close to 1 (Tabs.~\ref{t-fit2}, \ref{t-fit4}, and \ref{t-fit6}), the ratio plots do not show clear residuals (right panels in Figs.~\ref{f-ratio-a13_0},~\ref{f-ratio-a13_-0.5}, and~\ref{f-ratio-a13_0.5}).  The fits recover fairly well the input values of most parameters.  However, the geometric properties are hard to recover. Among the simulations assuming the Kerr metric (simulations A-D), simulations A and C provide better constraints on the spin and the deformation parameters (see Fig.~\ref{f-contours-A-D}, right column). The best-fit values of the spin and the deformation parameters are also recovered well within the given confidence level. Note that these are the simulations with a more compact corona. In simulations B and D, the constraints on the spin and the deformation parameter are weak (see Fig.~\ref{f-contours-A-D}, right column). The degeneracy between the spin and the deformation parameter in simulations D has two reasons: 1) a disk-like corona with large size does not irradiate well the inner part of the accretion disk as compared to its compact/lamppost counterpart (see Fig.~\ref{f-emissivity} for instance) and, 2) the interplay between the height of the lamppost, the spin, and the deformation parameter in the fitting model. As a result of 1, highly redshifted photons in the low-energy tail are missing, making the iron line narrower (see Fig.~\ref{f-linesstat}). When such a spectrum is fitted with a lamppost corona model, the model estimates the disk's inner edge at a larger radius, which is compensated by lowering the spin, increasing the lamppost height, and lowering the deformation parameter. This fact highlights the requirement of the compact corona lying close to the black hole to constrain well the geometric properties of the spacetime. The same explanation is also valid for simulation  H, where we also notice the degeneracy between the spin and the deformation parameter of the model. In the simulations assuming non-Kerr spacetime (simulations E-L), the lamppost corona model tends to recover and constrain the deformation parameter when the simulated data has the smaller extent of the disk-like corona, e.g., simulations E, G (see Fig.~\ref{f-contours-E-H}, left column) and J (contour plot not shown here).  For these fits, in general, the measurement of the lamppost height is not too bad within the 90\% confidence level. The black hole spin parameter is recovered quite well within the given uncertainty in most of the cases. The deformation parameter is not recovered within the 90\% confidence level for most of the simulations except in simulations A, B, C, and J, where we do recover the input value of the deformation parameter within the given uncertainty.  Our fits indicate that the value of the deformation parameter $\alpha_{13}$ is the most difficult one to recover and constrain.  We do not see clear differences between the fits employing broken power-law and lamppost emissivity profiles.  We note that -- modeling the emissivity profile with a broken power-law -- past studies have obtained observational constraints from the analysis of real data like $\alpha_{13} = 0.00_{-0.15}^{+0.05}$ for GRS~1915+105 from \textsl{Suzaku} observation~\citep{2020arXiv200309663A} and $\alpha_{13} = 0.00_{-0.20}^{+0.07}$ for MCG--6--30--15 from a set of \textsl{NuSTAR}+\textsl{XMM-Newton} observations~\citep{Tripathi:2018lhx}.  The constraint similar to GRS~1915+105 is found in our simulation C (broken power-law case), indicating that this source probably had a corona $R < 2~M$. Our comparison should be taken with caution because in the simulations, we have explored only a small part of the parameter space, and there are a number of other variables, such as the ionization parameter of the disk, which could influence the constraining power of the model on spin and deformation parameters~\citep{2019_Kammoun, relxill_dgrad, relxill_ion}. When we compare our simulation results with that of the best fit for MCG--6--30--15 found in ~\citet{Tripathi:2018lhx}, we do not find a similar constraint, suggesting that the source had probably a corona with different geometry. All of these considerations are valid if the actual corona has the shape of a disk. In a forthcoming paper, we will apply {\tt relxill\_nk} with the emissivity profiles calculated for disk-like coronae to sources with reflection-dominated spectra to see whether these emissivity profiles can fit the data better than a broken power-law or a lamppost emissivity profile. }

{So far, we have used $\chi^2$ statistics to show whether different models ({\tt relxill\_nk} and {\tt relxilllp\_nk}) can constrain and recover the correct value of the deformation parameter. All $\chi^2_{\rm red}$ values are close to 1, as we can see from Tabs.~\ref{t-fit1}-\ref{t-fit6}, which makes it difficult to draw a conclusion from the best-fits of these models. This problem can be solved in several ways in a Bayesian framework; however, we employ the deviance information criterion (DIC hereafter) for the two competing models, {\tt relxill\_nk} and {\tt relxilllp\_nk}, in order to select the best one; we exclude the {\tt relxilldisk\_nk} from the competing models because this model has been utilized to simulate the observations. The DIC has been proposed by~\citet{spiegelhalter2002bayesian}, and it combines the techniques from both information theory and Bayesian methods~\citep{liddle2007information}. It has already been applied in cosmology and astrophysics to study various problems~\citep{porciani2006luminosity, giles2016xxl,bignone2018metallicity,liang2019self,davari2021mog}. The DIC has exciting properties, such as 1) it accounts for the scenarios in which one or more parameters are poorly constrained by the data and 2) we can calculate it easily from the posterior sample generated by Markov-Chain-Monte-Carlo (MCMC hereafter) simulations~\citep{liddle2007information}. The latter also provides us the advantage of DIC over other model selection criteria in the Bayesian framework because DIC can be calculated easily by running MCMC simulations in {\tt xspec} \footnote{{Other model selection criteria, such as calculation of the Bayes factor, are not currently available natively in {\tt xspec} and require writing some external routines. }} . The DIC is given by
\begin{equation}
	\rm DIC = \langle D \rangle + p_{D}, 
\end{equation}
where,  $\rm \langle D \rangle $ is the mean of the deviance (D = -2ln$\mathcal{L}$, which corresponds with $\chi^2$ in our scenario) calculated over the MCMC chain. $\rm p_{D}$ is the effective number of parameters and penalizes the complexity of the model~\citep{spiegelhalter2002bayesian,porciani2006luminosity,liddle2007information}. A model with a low value of DIC is preferred.}

{We run MCMC simulations for all the simulated observations (A-L) for the employed models {\tt relxill\_nk} and {\tt relxilllp\_nk }. We use the ``chain" command in {\tt xspec} to run MCMC simulations with 100 walkers, 1.0 million steps each, and burn the first 400000 steps. Thus, there are a total of 100 million samples.  Once the MCMC run is over, we use the ``chain dic" command in {\tt xspec} to compute the deviance information criterion. Fig.~\ref{f-DIC} shows the DIC score for all the simulated observations. Based on these DIC scores for the simulations A-D (the Kerr cases), the selection decision between the models is not very clear as the margin of the DIC score is pretty narrow. For simulations A and B, the {\tt relxill\_nk} appears to be more suitable, and for simulation C, the {\tt relxilllp\_nk} is the preferred model. Simulaiton D is inconclusive. For the simulations E-L (the non-Kerr cases), {\tt relxilllp\_nk} is the preferred model; however, the difference of the DIC scores of the two models is quite small.  Our conclusion, based on  the DIC, is that with the current quality of \textsl{NuSTAR} data it may be challenging to choose {\tt relxill\_nk} or {\tt relxilllp\_nk}. However, with high-quality of data of the future X-ray missions such as \textsl{Athena}, we may be able to choose the model by the DIC.                  }

\vspace{0.5cm}


{\bf Acknowledgments --}
This work was supported by the Innovation Program of the Shanghai Municipal Education Commission, Grant No.~2019-01-07-00-07-E00035, the National Natural Science Foundation of China (NSFC), Grant No.~11973019, and Fudan University, Grant No.~JIH1512604.
D.A. is supported through the Teach@T{\"u}bingen Fellowship.
C.B. is a member of the International Team~458 at the International Space Science Institute (ISSI), Bern, Switzerland, and acknowledges support from ISSI during the meetings in Bern.


\appendix

\section{A. Johannsen metric}\label{app:metric}

In this work, we employed the Johannsen metric~\citep{johannsen2013}. In Boyer-Lindquist-like coordinates, the line element is
\be\label{eq-jm}
ds^2 &=&-\frac{\tilde{\Sigma}\left(\Delta-a^2A_2^2\sin^2\theta\right)}{B^2}dt^2
+\frac{\tilde{\Sigma}}{\Delta A_5}dr^2+\tilde{\Sigma} d\theta^2 
-\frac{2a\left[\left(r^2+a^2\right)A_1A_2-\Delta\right]\tilde{\Sigma}\sin^2\theta}{B^2}dtd\phi \nonumber\\
&&+\frac{\left[\left(r^2+a^2\right)^2A_1^2-a^2\Delta\sin^2\theta\right]\tilde{\Sigma}\sin^2\theta}{B^2}d\phi^2
\ee
where $M$ is the black hole mass, $a = J/M$, $J$ is the black hole spin angular momentum, $\tilde{\Sigma} = \Sigma + f$, and
\be
\Sigma = r^2 + a^2 \cos^2\theta \, , \qquad
\Delta = r^2 - 2 M r + a^2 \, , \qquad
B = \left(r^2+a^2\right)A_1-a^2A_2\sin^2\theta \, .
\ee
The functions $A_1$, $A_2$, $A_5$, and $f$ are defined as
\be
A_1 = 1 + \sum^\infty_{n=3} \alpha_{1n} \left(\frac{M}{r}\right)^n \, , \quad
A_2 = 1 + \sum^\infty_{n=2} \alpha_{2n}\left(\frac{M}{r}\right)^n \, , \quad
A_5 = 1 + \sum^\infty_{n=2} \alpha_{5n}\left(\frac{M}{r}\right)^n \, , \quad
f = \sum^\infty_{n=3} \epsilon_n \frac{M^n}{r^{n-2}} \, .
\ee
$\{ \alpha_{1n} \}$, $\{ \alpha_{2n} \}$, $\{ \alpha_{5n} \}$, and $\{ \epsilon_n \}$ are four infinite sets of deformation parameters without constraints from the Newtonian limit and Solar System experiments~\citep{johannsen2013}. If all deformation parameters vanish, the Johannsen metric exactly reduces to the Kerr solution, while deviations from the Kerr geometry require at least a non-vanishing deformation parameter. The leading order deformation parameters are $\alpha_{13}$, $\alpha_{22}$, $\alpha_{52}$, and $\epsilon_3$. In this work, we restricted our attention to the deformation parameter $\alpha_{13}$, which has the strongest impact on the shape of the reflection spectrum~\citep{Bambi:2016sac}. However, our study can be easily extended to any other deformation parameter of the Johannsen spacetime and, more in general, to any stationary, axisymmetric, and asymptotically flat black hole spacetime with one more parameter with respect to the Kerr metric.

Note that the Johannsen spacetime presents some pathological properties for arbitrary values of $a_* = J/M^2$ and $\alpha_{13}$, and therefore we limit our study to the parameter space without these problems. As in the case of the Kerr spacetime, we impose that $| a_* | \le 1$; for $| a_* | > 1$ there is no event horizon and the Johannsen metric describes the spacetime of a naked singularity. We also require that the transfer function~\citep[see][]{Abdikamalov:2019yrr} has only two branches, a condition that is violated at large spin and larger negative values of $\alpha_{13}$ and leads to a stronger condition than previously used~\citep{2018PhRvD..98b3018T}. For $\alpha_{13}$, we impose the following condition
\be
\label{eq-constraints}
\alpha_{13} > 
\begin{cases}
{- \frac{7}{24} a_*^{-0.8} \left( 1 + \sqrt{1 - a^2_*} \right)^4}, & a_* \geq 0
\\
{- \frac{1}{2} \left( 1 + \sqrt{1 - a^2_*} \right)^4}, & a_* < 0
\end{cases} \, .
\ee
The first line for $a_* \geq 0$ is from requiring the transfer function to have two branches. The second line for $a_* < 0$ is from requiring that $B>0$ outside of the event horizon, so that the metric does not diverge in the exterior region.

\section{B. Tetrad for the photon initial conditions}\label{app:photon}

To write the photon initial conditions, it is convenient to choose a locally Minkowskian reference frame for any emission point in the corona. Formally, this is equivalent to a coordinate transformation from the Boyer-Lindquist-like coordinates $\{ x^\mu \}$ to $\{ \tilde{x}^{(\alpha)} \}$ at the emission point
\be
dx^\mu \rar d\tilde{x}^{(\alpha)} = E^{(\alpha)}_\mu dx^\mu \, ,
\ee
such that the new metric tensor is the Minkowski metric
\be
g_{\mu\nu} \rar \eta_{(\alpha)(\beta)} = E_{(\alpha)}^\mu E_{(\beta)}^\nu g_{\mu\nu} \, ,
\ee
where $E_{(\alpha)}^\mu$s are the inverse of $E^{(\alpha)}_\mu$, so $E^{(\alpha)}_\mu E_{(\alpha)}^\nu = \delta_\mu^\nu$ and $E^{(\alpha)}_\mu E_{(\beta)}^\mu = \delta^{(\alpha)}_{(\beta)}$. $\{ E^\mu_{(\alpha)} \}$ is the tetrad of orthogonal basis vectors associated to the locally Minkowskian reference frame of the source. If a vector (dual vector) has components $V^\mu$ ($V_\mu$) in the Boyer-Lindquist-like coordinates $\{ x^\mu \}$, the components of the vector (dual vector) in the locally Minkowskian reference frame are
\be
V^{(\alpha)} = E^{(\alpha)}_\mu V^\mu \, , \quad V_{(\alpha)} = E_{(\alpha)}^\mu V_\mu \, .
\ee
It is straightforward to see that
\be
V^\mu = E_{(\alpha)}^\mu V^{(\alpha)} \, , \quad V_\mu = E^{(\alpha)}_\mu V_{(\alpha)} \, .
\ee

The time-like tetrad basis vector $E^\mu_{(t)}$ is the 4-velocity of the emitter $U^\mu$
\be
E^\mu_{(t)} = U^\mu \, ,
\ee
where $U^\mu = U^\mu_{\rm stat}$ for a static corona, $U^\mu = U^\mu_{\rm corot}$ for a corotating corona, and their expressions are already reported in Section~\ref{s-corona}.

We choose the orientation of the space-like tetrad basis vectors as shown in Fig.~\ref{corona-sketch} with $\hat{r} = \hat{z}$ and $\hat{\theta} = \hat{y}$. Since the Johannsen metric is diagonal in these directions and the $r$ and $\theta$ components of the 4-velocity of the emitter $U^\mu$ vanish, we have
\be
E^\mu_{(z)} &=& \left( 0 , 1/\sqrt{g_{rr}} , 0 , 0 \right) \, , \\
E^\mu_{(y)} &=& \left( 0 , 0 , 1/\sqrt{g_{\theta\theta}} , 0 \right)
\ee

Last, the expression of $E^\mu_{(x)}$ can be obtained from the conditions
\be
g_{\mu\nu} E^\mu_{(x)} E^\nu_{(x)} = 1 \, , \quad 
g_{\mu\nu} E^\mu_{(x)} E^\nu_{(t)} = 0 \, .
\ee
For a static corona, we find
\be
E^\mu_{(x)} = \frac{1}{\sqrt{ -g_{tt} \left( g_{t\phi}^2 - g_{tt} g_{\phi\phi} \right)}}
\left( g_{t\phi} , 0 , 0 , - g_{tt} \right) \, .
\ee
For a corotating corona, we have
\be
E^\mu_{(x)} = \frac{1}{\sqrt{ \left( - g_{tt} - 2 \Omega_{\rm K} g_{t\phi} - \Omega_{\rm K}^2 g_{\phi\phi} \right)
\left( g_{t\phi}^2 - g_{tt} g_{\phi\phi} \right)}}
\left( g_{t\phi} + \Omega_{\rm K} g_{\phi\phi} , 0 , 0 , - g_{tt} - \Omega_{\rm K} g_{t\phi} \right) \, .
\ee

\section{C. Ring-like coronal geometry}\label{app:ring}

Our model in {\tt relxill\_nk} is constructed calculating the emissivity profiles of ring-like coronae and we can then obtain the emissivity profile of a disk-like corona summing up the emissivity profiles of ring-like coronae. It is thus straightforward to calculate iron lines and full reflection spectra from accretion disks illuminated by ring-like coronae and the results are presented in this appendix.

\begin{figure*}[t]
\begin{center}
\includegraphics[width=0.95\textwidth,trim={0cm 0.5cm 0cm 0cm},clip]{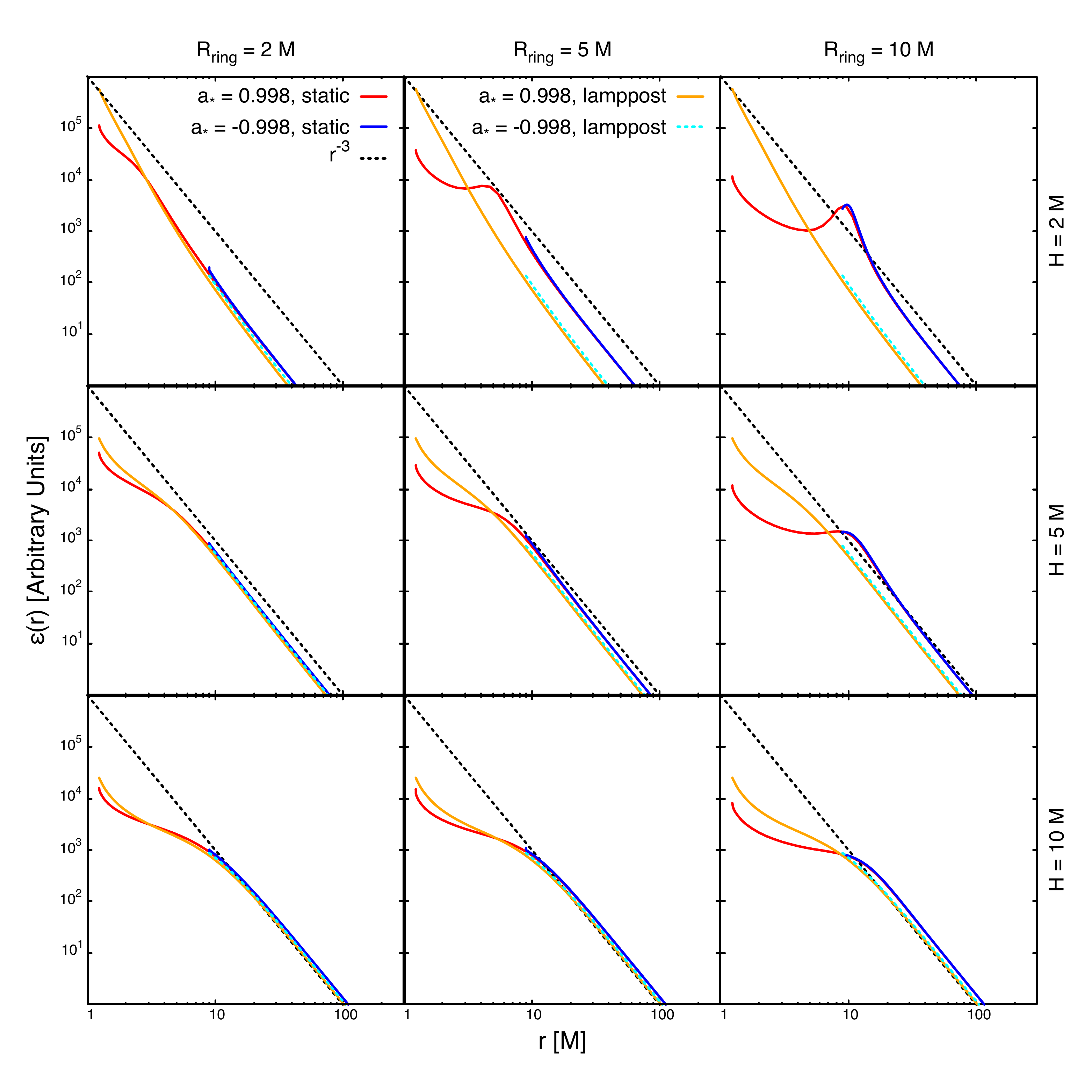}
\end{center}
\vspace{-0.5cm}
\caption{Emissivity profiles in the Kerr spacetime for different values of the corona radius $R_{\rm ring}$ and of the corona height $H$. Red curves are for static ring-like coronae and a black hole spin parameter $a_* = 0.998$. Blue curves are for static ring-like coronae and a black hole spin parameter $a_* = -0.998$. Black dashed lines are for the canonical emissivity profile $\varepsilon \propto r^{-3}$. Orange curves and cyan dashed curves are for the lamppost model with coronal height $h = H$ and a black hole spin parameter, respectively, $a_* = 0.998$ and $-0.998$. 
\label{f-emissivity-r}}
\end{figure*}

Fig.~\ref{f-emissivity-r} shows the emissivity profiles in the Kerr spacetime for different values of the static ring-like corona radius $R_{\rm ring}$ and height $H$ and can be compared with Fig.~\ref{f-emissivity} for the case of disk-like coronae. In these plots, we assume that the photon index of the coronal spectrum is $\Gamma = 1.7$.

\begin{figure*}[t]
\begin{center}
\includegraphics[width=0.90\textwidth,trim={0cm 0.5cm 0cm 0cm},clip]{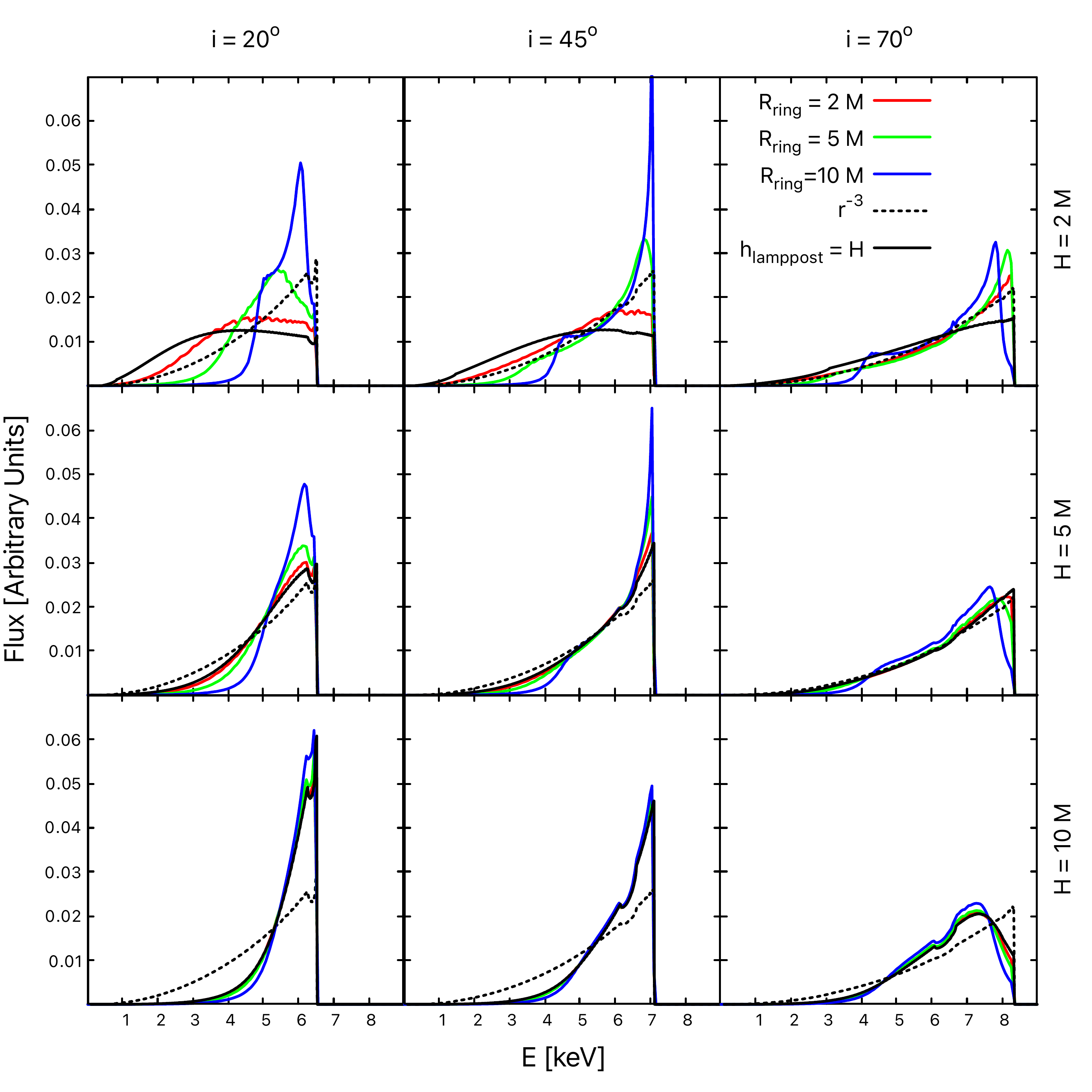}
\end{center}
\vspace{-0.4cm}
\caption{Static ring-like coronae. Iron line profiles in Kerr spacetime with $a_* = 0.998$. The radius of the corona is $R_{\rm ring} = 2$~$M$ (red profiles), 5~$M$ (green profiles), and 10~$M$ (blue profiles). In every panel we also show an iron line for a power-law emissivity profile with emissivity index $q = 3$ (black dotted profiles) and for a lamppost corona (black solid profiles). \label{f-lines-r}}
\end{figure*}

Fig.~\ref{f-lines-r} shows the iron line shapes from disks illuminated by static ring-like coronae. Even in these plots we assume that the spacetime is described by the Kerr metric and that the photon index of the coronal spectrum is $\Gamma = 1.7$. Fig.~\ref{f-lines-r} can be compared with Fig.~\ref{f-linesstat} for static disk-like coronae.

\bibliographystyle{aasjournal}
\bibliography{bbb}

\end{document}